\documentclass[12pt,reqno]{amsart}
\usepackage{setspace,multirow}
\usepackage{etoolbox}
\patchcmd{\section}{\normalfont}{\normalfont\bfseries\sffamily}{}{}
\patchcmd{\section}{\centering}{\flushleft}{}{}
\patchcmd{\subsection}{\bfseries}{\normalfont\itshape\sffamily}{}{}

\makeatletter
\renewcommand\@biblabel[1]{}
\patchcmd{\@settitle}{\uppercasenonmath\@title}{\Large\sffamily}{}{}
\patchcmd{\@settitle}{center}{flushleft}{}{}
\patchcmd{\@setauthors}{\MakeUppercase}{\large\sffamily}{}{}	
\patchcmd{\@setauthors}{\centering}{}{}{}
\makeatother

\renewcommand{\abstractname}{Summary}

\usepackage{amsmath}
\usepackage{amssymb}
\usepackage[authoryear,round]{natbib}
\usepackage{enumerate}
\usepackage{graphicx, booktabs,subcaption}
\usepackage{hyperref}
\hypersetup{
    colorlinks=true,
    linkcolor=blue,
    citecolor=blue,
    filecolor=magenta,      
    urlcolor=cyan,
    pdftitle={Overleaf Example},
    pdfpagemode=FullScreen,
    }

\let\OLDthebibliography\thebibliography
\renewcommand\thebibliography[1]{
  \OLDthebibliography{#1}
  \setlength{\parskip}{0pt}
  \setlength{\itemsep}{0pt plus 0.3ex}
}

\usepackage[export]{adjustbox}

\newtheorem{theorem}{Theorem}

\newtheorem{assumption}{Assumption}

\newtheorem{corollary}{Corollary}

\newtheorem{example}{Example}

\newtheorem{lemma}{Lemma}

\newtheorem{remark}{Remark}

\usepackage{xcolor}

\begin{document}

%TCIMACRO{\TeXButton{Section}{\sectionfont{\bfseries\large\sffamily}}}%
%BeginExpansion
%\sectionfont{\bfseries\large\sffamily}%
%EndExpansion

%TCIMACRO{%
%\TeXButton{Subsection}{\subsectionfont{\bfseries\sffamily\normalsize}}}%
%BeginExpansion
%\subsectionfont{\bfseries\sffamily\normalsize}%
%EndExpansion

%TCIMACRO{\TeXButton{noindent}{\noindent}}%
%BeginExpansion
\noindent%
%EndExpansion
%TCIMACRO{%
%\TeXButton{title}{{\sffamily\bfseries\Large Using Evidence Factors to Clarify Exposure Biomarkers}}}%
%BeginExpansion
% {\sffamily\bfseries\Large %Sridharacharyan 
% Inferring the Effect of a Confounded Treatment 
% by Calibrating Resistant Population's Variance}
% \medskip
%
%EndExpansion

\title[\Blocking \Method]{Inferring the Effect of a Confounded Treatment by Calibrating Resistant Population's Variance}
\author[Z.~Qin and B.~Karmakar]{Z.~Qin and B.~Karmakar \smallskip \newline 
\textit{Department of Statistics, University of Florida, 102 Griffin-Floyd Hall, Gainesville, FL 32611 USA.}}

%TCIMACRO{\TeXButton{noindent}{\noindent}}%
%BeginExpansion
\noindent%
%EndExpansion
%
%
%
%
%
%
%

% \textsf{Z.~Qin and B.~Karmakar}\footnote{%
% Department of Statistics, University of Florida, 102 Griffin-Floyd Hall, Gainesville, FL 32611 USA. \ \textsf{E-mail:} \textsf{bkarmakar@ufl.edu}.}
% \smallskip
%
%%TCIMACRO{\TeXButton{noindent}{\noindent}}%
%%BeginExpansion
%\noindent%
%%EndExpansion

% \textsf{University of Florida, Gainesville}
% \vspace{-30pt}

%TCIMACRO{\TeXButton{noindent}{\noindent}}%
%BeginExpansion
\noindent%
%EndExpansion
\begin{abstract}
    In a general set-up that allows unmeasured confounding, we show that the conditional average treatment effect on the treated can be identified as one of two possible values. Unlike existing causal inference methods, we do not require an exogenous source of variability in the treatment, e.g., an instrument or another outcome unaffected by the treatment. Instead, we require (a) a nondeterministic treatment assignment, (b) that conditional variances of the two potential outcomes are equal in the treatment group, and (c) a resistant population that was not exposed to the treatment or, if exposed, is unaffected by the treatment. Assumption (a) is commonly assumed in theoretical work, while (b) holds under fairly general outcome models. For (c), which is a new assumption, we show that a resistant population is often available in practice. We develop a large sample inference methodology and demonstrate our proposed method in a study of the effect of surface mining in central Appalachia on birth weight that finds a harmful effect.
% \end{abstract}
\medskip
%TCIMACRO{\TeXButton{noindent}{\noindent}}%
%BeginExpansion

\noindent%
%EndExpansion

\end{abstract}

\maketitle
\section{Introduction}
Although the no unmeasured confounders assumption \citep{Barnow1980, Rosenbaum1983, Greenland1986} is central to the current research on causal inference from observational studies, the statistical methodology and empirical literature show diverging attitudes toward this assumption. A significant section of current methodological development assumes no unmeasured confounders and aims to find fine-tuned methods to estimate the overall or more specific effects. In contrast, nearly all published observational study discusses the possibility of bias in their inference due to unmeasured confounders. These discussions often follow standard analyses using linear and generalized linear models after matching on measured covariates or propensity weighting. The penultimate sentence of Rosenbaum and Rubin's commentary \citep{Rosenbaum2022} on the 40th anniversary of their famous 1983 paper \citep{Rosenbaum1983} reads, ``{\it It is important to control for measured covariates, but in any nonrandomized study for causal effects the key activity is the step from association to causation, where bias from unmeasured covariates remains a possibility.}" When the no unmeasured confounders assumption is inaccurate, inferences from methods that work well under the assumption are generally unreliable and may have varying levels of sensitivity to potential unmeasured confounders.

The no unmeasured confounders assumption, also called the ignorability assumption, states that the treatment and potential outcomes are conditionally independent given the measured confounders. The term {\it conditional independence} and the sentiment behind it are easily understood in English, which encourages arguments for whether the collected confounders are sufficient for an ignorable treatment assignment. However, a probabilist can caution us regarding the nuances of independence and conditional independence. For example, the assumption is often justified by collecting a large number of covariates. But, in theory, conditioning on a subset of these covariates may make the treatment assignment ignorable, i.e., all confounders are measured, yet conditioning on all of those covariates may result in conditional dependence (see illustrative cases in \cite{Stoyanov2013}, problem 7.14). In the end, the ignorability assumption is usually untestable. In summary, absent detailed knowledge of the treatment assignment process, we cannot justify the ignorability assumption.

Alternatives to the no unmeasured confounders assumption have been developed. Among them are sensitivity analysis methods assuming a bounded influence of unmeasured confounders \citep{cornfield1959smoking, Rosenbaum2002book, Imbens2003, Tan2006, ding2016sensitivity, Zhao2019} and instrumental variable (IV) methods using exogenous variability in treatment assignment \citep{White1982, Angrist1996, Deaton2010, Baiocchi2016, Rudolph2017, Kennedy2019}. Despite its growing literature, sensitivity analysis methods have seen limited adoption in empirical applications. This could be because sensitivity analysis procedures typically need to be derived separately for each inference method, and we need a better understanding of their relative performance across inference methods.  A primary challenge in IV methods is finding a variable that qualifies as a good IV based on scientific understanding and statistical evidence. For specific problems, researchers have proposed instruments that are likely to be valid \citep{Baiocchi2016}. Note, however, that a valid instrument must also satisfy a conditional independence condition, among other conditions. 

We propose a new method for heterogeneous effect estimation from nonrandomized studies under the possibility of an unbounded amount of unmeasured confounding. Our method assumes much less and avoids any ignorability or exogenous variability conditions while costing us a bit of ambiguity in the inference. 
Specifically, we establish a two-point identification of the conditional average treatment effect on the treated (CATT) rather than a point identification.
Thus, our method, when given an infinite amount of data, even with a biased treatment assignment,  will give at most two possible values for the treatment effect. So, when given finite data, the method gives at most two possible confidence intervals, which may be combined into one confidence set, for the heterogeneous effect. 

The proposed method requires that there be a {\it resistant population}. This population is spared exposure to the treatment or, if exposed to it, is not affected by the treatment. The resistant population is used to calibrate the variability of control potential outcomes. Mathematically, the method needs an estimate of the conditional variance of the control potential outcomes given covariates. We show that resistant populations are available in various observational studies.

\subsection{Summary of our contributions}
\label{sec_contributions}

Section \ref{sec_cte} gives the identification result in Theorem \ref{thm_main} using Assumption \ref{assump_main}. These assumptions are discussed and examined in Section \ref{sec_discuss_assump}. There is some precedence to the idea of two-point identification. In a classification problem, \cite{shaham2016deep} propose a flexible model that identifies either the probability of having the target label or the probability of not having the target label. In determining the optimal experimental design from a large class, \cite{cai2022optimizing} show that at least one of the stratified designs and the cluster-based design is optimal when the units do not interfere. Further, sensitivity analysis to unmeasured confounders identifies the treatment effect in an interval of values \citep{Rosenbaum2002book}.

Our two-point identification proposes two treatment effect estimands. Section \ref{sec_estimation} provides an algorithm, which involves several steps and numerical optimizations, for non-parametric estimation of our treatment effect estimands. Our final estimator is based on a few (constrained) local linear regression estimators. While the proposed framework in Section \ref{sec_cte} is open to other estimation methods,  our local linear regression-based algorithm facilitates the derivation of the asymptotic distribution of the estimators and, hence, the inference of our treatment effect estimands. Section \ref{sec_inference} develops large sample theory and confidence intervals under technical Assumption \ref{assm_technical}. Interpreting the inference from the proposed method may require some change in perspective since it gives {\it two} estimates and {\it two} confidence intervals. The cost is that we do not know which of the two estimates is consistent, at least one of them is, or which of the two confidence intervals covers the treatment effect, at least one of them does. The benefit is we do not need to assume that all confounders are available during the analysis. Still, additional knowledge may be used to determine the right choice. For example, researchers routinely discuss the direction of possible bias due to unmeasured confounders in published observational studies. In practice, those discussions can inform to pick the smaller of the two estimates, with the corresponding confidence interval, when that bias is thought to be positive and vice versa.
Section \ref{sec_remarks} provides further remarks for practice.

Section \ref{sec_ate} provides an estimation process for the average treatment effect on the treated (ATT) that removes the no unmeasured confounders assumption. This process requires an additional Assumption \ref{assump_ateiden}, but estimation accuracy for the ATT is much better than for the heterogeneous treatment effect function. Section \ref{sec_simulation} validates our method and theory in simulation. Section \ref{sec_mining_birthweight} presents an empirical study of the effect of surface mining in central Appalachia counties on infant birth weight using our proposed method. Section \ref{sec_discussion} has further discussion. 
Code implementing our proposed methods and all data sets except the birth data for our empirical study are available at \url{https://github.com/bikram12345k/RPCOVA}.

\section{Identification of the conditional average treatment effect on the treated}\label{sec_cte}
Suppose $(Y_{i}(1), Y_{i}(0), Z_i, X_i)$, for $n$ units $i=1,\ldots,n$, are drawn independently from a probability distribution,
called the {\it full data distribution}. Potential outcomes $Y_i(1)$ and $Y_i(0)$ are univariate \citep{Neyman1990, rubin1974estimating}, 
$Z_i$ is a binary indicator for the treatment, and $X_i$ is possibly multidimensional. We assume that SUTVA holds and we observe $(Y_i, Z_i, X_i)$, with $Y_i=Z_iY_{i}(1)+(1-Z_i)Y_i(0)$. The distribution of $(Y_i, Z_i, X_i)$ is the {\it observed data distribution}. In an observational study, a parameter, perhaps of the full data distribution, is identifiable if it can be written as a function of the observed data distribution; any such function will be called an estimable function.

Here we study the identification of the conditional average treatment effect on the treated using the observed data distribution without assuming an ignorable treatment assignment or any other exogenous variability. 
Specifically, we aim to estimate 
\begin{equation}\label{eq_catt}
\tau(x) := E(Y_i(1) - Y_i(0)\mid Z_i = 1, X_i = x),    
\end{equation}
the conditional average treatment effect on the treated (CATT), under the following assumptions that we elaborate on in Section \ref{sec_discuss_assump} below.

\begin{assumption}\label{assump_main}  (Identification assumptions)
\begin{enumerate}
	\item[(a)]  $\pi(x) := \Pr(Z_i=1\mid X_i=x) \in (0,1)$.
	\item[(b)]  ${Var}(Y_i(0)\mid Z_i = 1, X_i=x) = {Var}(Y_i(1)\mid Z_i = 1, X_i=x)$.
	\item[(c)]  $\sigma_0^2(x) := {Var}(Y_i(0)\mid X_i=x)$ is known (later, assumed to be estimable).
\end{enumerate}
\end{assumption}

Let $\beta(x) = E(Y_i\mid Z_i=1, X_i = x) - 
E(Y_i \mid Z_i=0, X_i = x)$ and 
$\sigma^2(x) = {Var}(Y_i\mid X_i=x)$. Write
$\Delta(\beta, \tau)(x) := \beta(x) - \tau(x),$
for the {\it identification bias} of the target function $\tau$ using 
estimable function $\beta$.%; this bias is equal to $E(Y_i(0)\mid  Z_i=1, X_i=x) - E(Y_i(0)\mid Z_i=0, X_i=x)$. 

\begin{theorem}\label{thm_main}
Under Assumption \ref{assump_main},
\begin{equation}\label{eq_ind}
\left\{\Delta(\beta, \tau)(x)\right\}^2 =  \beta(x)^2 - \frac{\sigma^2(x) - \sigma_0^2(x)}{\pi(x)(1-\pi(x))}.
\end{equation}
\end{theorem}

 Since the observed data distribution or 
Assumption \ref{assump_main} does not identify the sign of $\Delta(\beta, \tau)(x)$, Theorem \ref{thm_main}
provides a `two-point identification' of the conditional average treatment effect on the treated in the following sense. Let us abbreviate $\Delta(x)\equiv \Delta(\beta, \tau)(x)$. Define
$$\tau_{-}(x) = \beta(x) - |\Delta(x)| \quad \text{ and } \quad \tau_{+}(x) = \beta(x) + |\Delta(x)|.$$
Then, under our identification assumptions, for a given $x$, $\tau(x) = \tau_-(x)$ or $\tau(x) = \tau_+(x)$ and
both are estimable functions from the observed data distribution. We refer to this property as two-point identification of $\tau(x)$.

\begin{figure}[htbp]
\centering
\includegraphics[width=.7\textwidth]{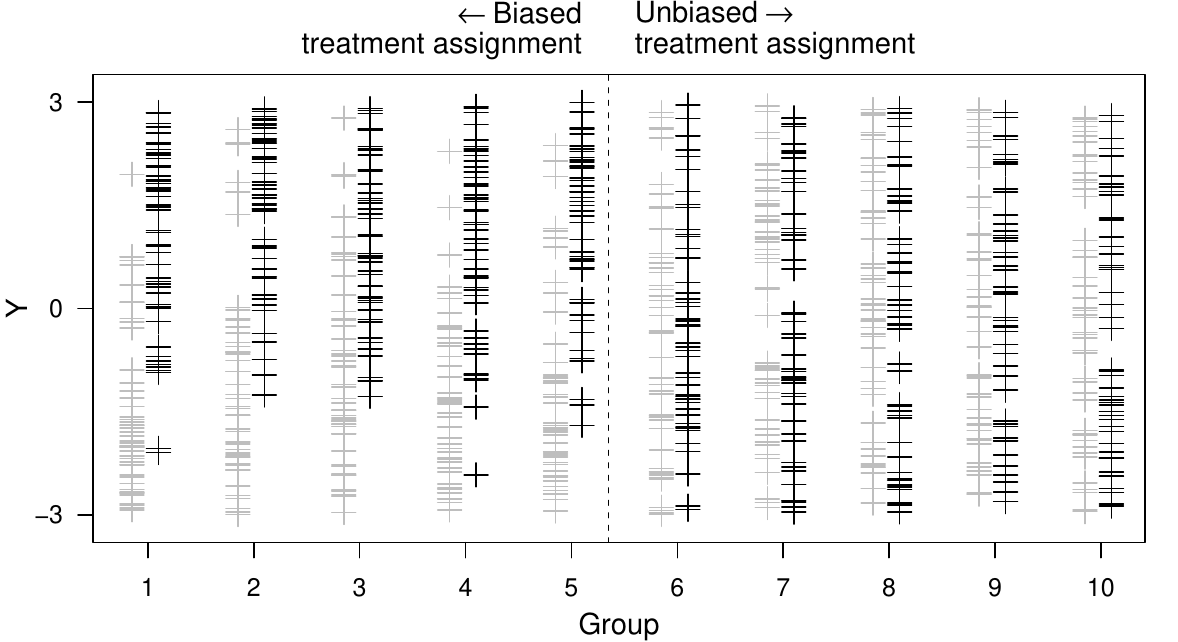}
\caption{Illustration of detecting biased treatment assignment by comparing variances. In truth, there is no treatment effect. Grey points show the control units and black points show the treated units. In each group, the grey and black points have been slightly separated horizontally to improve clarity.}
\label{fig_exam_illu}
\end{figure}

% \vspace{-20pt}
\begin{remark}\label{exam_illu}  Building intuition. \normalfont
Without delving into technical details, we give an illustration here to build intuition for why the squared bias can be identified as in Theorem  \ref{thm_main}. Figure \ref{fig_exam_illu} shows data from a simulated experiment where the $Y_i(1)=Y_i(0)$ is uniformly distributed over $(-3,3)$. Hence, there is no treatment effect; $\tau(x)\equiv 0$. There are ten groups with identically distributed potential outcomes. Exactly half of the units in each group are assigned to the treatment. However, the treatment assignment is biased -- assignment probabilities are proportional to the square of three plus potential outcomes ($Pr(Z_i = 1 \mid Y_j(1), Y_j(0), j=1,\ldots,50) = (3+Y_i(1))^{2} / (\sum_j (3+Y_j(1))^{2}$) -- in the first five groups, and is completely randomized in the last five groups. Thus, treatment has a spurious effect in the first five groups; $\beta(x)> 0$. Notice now, in each group, the variability of the black points, i.e., the treated units' outcomes, and that of the grey points, i.e., the control units' outcomes. In the unbiased groups, these two sets of points mix with each other; in fact, their theoretical variances are equal. Compared to this common variance, the variabilities of the points within both the treated and controls seem smaller in the biased groups because these groups prefer the treatment when outcomes are higher.  This phenomenon holds broadly, and Theorem \ref{thm_main} mathematically connects the differences in the variabilities between the unbiased and biased cases to identify the magnitude of bias in effect estimation.
\end{remark}

\vspace{-20pt}
\begin{figure}[!htbp]
\centering
\includegraphics[width=.9\textwidth]{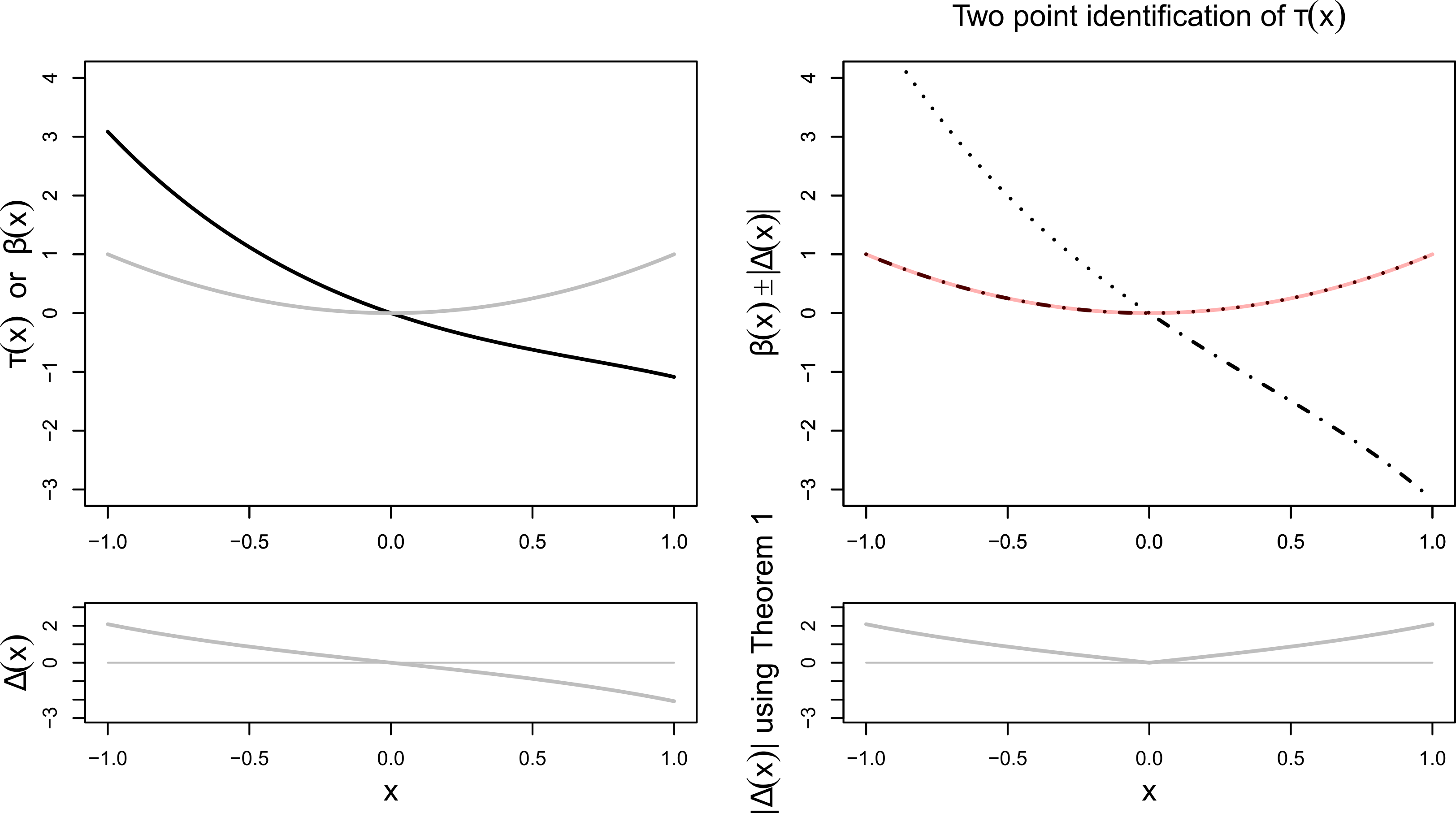}
\caption{Illustration of Example \ref{exam_1}. Top-left: conditional treatment 
effect on the treated $\tau(x)$ in `grey', and biased identification $\beta(x)$ in `black';
bottom-left: identification bias $\Delta(\beta, \tau)(x)$; bottom-right: absolute identification 
bias using \eqref{eq_ind}; top-right: two-point identification using the observed
data under Assumption \ref{assump_main}; the dotted line is for $\tau_{+}(x)$, dash-dotted line is for $\tau_{-}(x)$ and the `pink' line is for the true $\tau(x)$. All quantities are calculated analytically; see the online supplement. }
\label{fig_exam}
\end{figure}

\begin{example}\label{exam_1}\normalfont 
Consider a technical illustration for two-point identification.
Let $(X_i, U_i, \epsilon_i)$ be independent Normal(0, 1) random variables.
Let $Y_i(0) = X_i + |U_i| + \epsilon_i$ and $\Pr(Z_i=1\mid X_i+U_i) = \Phi(X_i+U_i)$
where $\Phi(\cdot)$ is the standard normal distribution function. Finally, let $Y_i(1) = X_i^2 + X_i + |U_i| + \epsilon_i$ so that $\tau(x) = E( Y_i(1) - Y_i(0) \mid Z_i=1, X_i=x) = E( X_i^2 \mid Z_i=1, X_i=x)=x^2$. Our observed data for unit $i$ are $(Y_i,Z_i,X_i)$.

Figure \ref{fig_exam} shows the implication of Theorem \ref{thm_main} in 
the identification of $\tau(\cdot)$ as one of two possible curves. When  $x=0$, 
$\beta(0) = E(\lvert U_i \rvert \mid Z_i = 1, X_i = 0) - E(\lvert U_i \rvert \mid Z_i = 0, X_i = 0)$ is equal to $\tau(0)$. The identification bias, $\Delta(x)=E(\lvert U_i \rvert \mid Z_i = 1, X_i = x) - E(\lvert U_i \rvert \mid Z_i = 0, X_i = x)$,  is negative for $x>0$ and 
positive for $x<0$; see the supplement for a proof. The bias occurs because of the unobserved confounder $U_i$ in this simulation.
The bottom-right panel of the figure calculates the 
absolute identification bias, $|\Delta(x)|$, by taking the positive square
root of the right-hand side of \eqref{eq_ind}. Consequently, we identify the
CATT $\tau(x)$ by $\tau_+(x)$ when $x>0$ and 
by $\tau_-(x)$ when $x<0$. At the center, $\tau(0)=\tau_+(0)=\tau_-(0)$.

\end{example}

\subsection{Discussion of the identification assumptions}\label{sec_discuss_assump}
Assumption \ref{assump_main}(a) is the well-known {\it overlap} assumption in the causal inference literature \citep{Rosenbaum1983}. It is clear that when Assumption 
\ref{assump_main}(a) fails, without additional assumptions,  $\tau(x)$ cannot be partially
identified in any bounded set. 

Assumption \ref{assump_main}(b) says that, within the treated group, two potential outcomes have the same conditional variances. 
Specifically, Assumption \ref{assump_main}(b) holds if we assume $Y_i(1) - Y_i(0) = f(X_i)$ (or $Y_i(1) - Y_i(0) \mid \{Z_i = 1\} = f(X_i)$), which is saying that the treatment effect (on the treated) is a fixed effect and $X_i$ includes all the effect modifiers. The effect modifiers describe the variability in the treatment effect when the effect is heterogeneous. There are practical benefits to assuming that all effect modifiers are observed because with incomplete information on effect modifiers, inference for a heterogeneous treatment effect would also give incomplete knowledge of the effect. 

However, Assumption 1(b) does not rule out a random treatment effect or require that all effect modifiers are observed. To see this write $Y_i(z) = m_z(X_i) + \epsilon_{iz}$ where $m_z(X_i) = E(Y_i(z) \mid Z_i=1, X_i)$ and $\epsilon_{iz} = Y_i(z) - m_z(X_i)$ for $z=0,1$, so that the CATT $\tau(x) = m_1(x) - m_0(x)$. The equalities in the previous sentences are purely algebraic; they do not require any assumptions. In this case, Assumption \ref{assump_main}(b) is requiring
$Var(\epsilon_{i0} \mid Z_i = 1, X_i) = Var(\epsilon_{i1} \mid Z_i = 1, X_i)$, i.e., equality of the residual variances from the two potential outcomes conditional on $X_i$.

In some situations, domain knowledge does not allow Assumption \ref{assump_main}(b). 
For example, consider $Y_i(z) \mid \{Z_i=1, X_i=x\}\sim Poi(m_z(x))$ for $z=0, 1$, where the outcomes are counting data with means associated with $X_i$.
Under this model, the variances of $Y_i(0)$ and $Y_i(1)$ will not be equal unless $m_0(x)=m_1(x)$. However, one could apply a variance stabilizing transformation \citep[][Chapter 8]{Box2005}, a common tool of a statistician, to the data so that the assumption may be met. In this example, a square root transformation, i.e., $\sqrt{Y_i}$ becoming our outcome of interest, should make the variances approximately equal.

Like the conditional unconfoundedness assumption, Assumption \ref{assump_main}(b) is counterfactual and hence generally untestable. In general, Assumption 1(b) is neither stronger nor weaker than the conditional unconfoundedness assumption. But if further given that ${Var}(Y_i(0)\mid Z_i = 0, X_i=x) = {Var}(Y_i(1)\mid Z_i = 1, X_i=x)$, which is a factual and testable condition,  Assumption \ref{assump_main}(b) is equivalent to ${Var}(Y_i(0)\mid Z_i = 0, X_i=x) = {Var}(Y_i(0)\mid Z_i = 1, X_i=x)$.  The latter is saying that $Y_i(0)$ has equal conditional variance in both the treated and control groups. This is also implied by the assumption $Y_i(0) \perp Z_i \mid X_i$, which is a version of the conditional unconfoundedness commonly assumed for identification of the CATT.

The following example provides an illustration of when the no unmeasured confounders assumption does not hold yet Assumption 1(b) holds.

\begin{example}\normalfont 
    Suppose we have observed some covariates $X_i$ and there are unobserved covariates $U_{i1} \sim $Unif$(0,1)$ and $U_{i0} \sim$ Unif$(-1, 0)$ and let $U_i = U_{i0} + U_{i1}$.   
    Let the treatment be assigned (implicitly) according to $U$, such that
    $P(Z_i = 1 \mid U_i = u, X_i=x) = 1$ if $u \geq 0$ and $P(Z_i = 1 \mid U_i = u, X_i=x) = 0$ otherwise.
    Then, let  $Y_i(z) = m_z(X_i) + U_{iz} + \epsilon_{iz}$ for some functions $m_0(\cdot)$ and $m_1(\cdot)$ and mean-zero random errors $\epsilon_{i0}$, $\epsilon_{i1}$. Here $\epsilon_0, \epsilon_1$ are independent of $(X_i, U_{i0}, U_{i1}, Z_i)$. Note that, 
    the CATT $\tau(x) = m_1(x) - m_0(x) + E(U_{i1} - U_{i0}\mid Z_i=1, X_i=x)$, which depends on the conditional distribution of $(U_{i0}, U_{i1})$ given $X_i$.

    Since $(U_{i0}, U_{i1}) \not\perp Z_i \mid X_i$, the potential outcomes $Y_i(0), Y_i(1)$ are also dependent on $Z_i$ when given $X_i$. Consequently, the conditional unconfoundedness assumption does not hold. 
    The assumption that $Y_i(0) \perp Z_i \mid X_i$, usually assumed for identification of CATT, also does not hold.    Sometimes, researchers may assume the conditional mean independence, i.e., $E(Y_i(0)\mid X_i, Z_i=0)=E(Y_i(0)\mid X_i, Z_i=1)$, a milder assumption than the conditional independence. %, for identification of CATT (or CATE, etc.).
    This assumption does not hold in this example either unless for particular choices of $m_0(\cdot)$ and $m_1(\cdot)$. On the other hand, the Assumption \ref{assump_main}(b) holds in this example as long as $Var(\epsilon_{i0}) = Var(\epsilon_{i1})$. %, even though the confounder $(U_0, U_1)$ is not observed.

\end{example}

Assumption \ref{assump_main}(c) is unique to our method and, to the best of 
our knowledge, has not been made in the literature. Here we discuss many common
situations where this assumption is fair. To guide practitioners, we also 
point to a few cases where this assumption might fail.
\begin{enumerate}
	\item[(a)] {\it Ignorable treatment assignment.} For illustrative purposes, we start 
	with a theoretically convenient but practically uninteresting case.
	Suppose $Y_i(0)\perp Z_i \mid X_i$, which is 
	part of the ignorable treatment assignment assumption \citep{Rosenbaum1983}. Then $\sigma_0^2(x)
 = Var( Y_i \mid Z_i=0, X_i=x)$ is estimable. Of course, two-point identification is uninteresting in this scenario because $Y_i(0)\perp Z_i \mid X_i$ along with assumption 1(a) and (b) implies point identification of $\tau(x)$ with $\beta(x)$. However, nothing is lost by using Theorem \ref{thm_main} because it gives $\Delta^2(x)=0$; hence $\tau_\pm(x)=\beta(x)$.
    
	\item[(b)] {\it Temporal variance stationarity.} Suppose the treatment is introduced to the population at a particular point in time and the outcome is measured not too long after that. A big class of observational studies is of this form. Consider two illustrative examples: the effect of a get-out-the-vote campaign in a statewide election on voter turnout, and, the effect of a pain drug on productivity. 
	
	A get-out-the-vote campaign is likely to be focused on districts where a higher benefit is expected. Thus, there is clear suspicion of confounding, which is difficult to remove by precisely measuring all the confounders. In such a study, we can estimate the conditional variance $\sigma_0^2(x)$ from the district-level voter turnout data from the previous election(s) when there was no such campaign. Since the last election, there could be a general change in the rate of voter turnout, e.g., an increase in the average voter turnout. This would not affect our identification. We only need the variance of the voter turnout to remain the same as in the last election if the campaign had not been used in the current election in a counterfactual situation.
	
	Consider the second example. Doctors prescribe pain medication to patients who need them. Further, a patient decides to visit the doctor when they are conscious of the pain and feel a doctor's input will be beneficial. Thus, the confounding effect of these complex decisions may be difficult to untangle. Nevertheless, in this case, we can estimate $\sigma^2_0(x)$, the conditional variance of individuals' productivity in the control group, based on their productivity on the day or week before they visited the doctors.
	
	Under temporal variance stationarity, the variance of the potential outcome in the control group is stationary before and after the treatment is introduced. Mathematically, this property is satisfied if $Y_{i}(0)$ is a second-order stationary random process across a period around the time of exposure and outcome measurement \citep[see][Chapter 2]{Brockwell2016}. However, second-order stationary processes assume a constant mean and a stationary auto-covariance over time. Temporal variance stationarity is thus more general than second-order stationarity.
    
	\item[(c)] {\it Geographical variance stationarity.} When the treatment pertains to long-term exposure or the outcome is measured a significant period after the exposure, temporal variance stationarity could be questionable. In some of those situations, the alternative may be geographical variance stationarity. Consider two examples: the effect of a voluntary professional development program on income after 5 years, and, the effect of surface mining for coal on the birth weight of babies \citep{Hendryx2016, Small2021}. 
	
	Many employers offer voluntary professional development programs. We might be interested in the effect of professional development programs for early career individuals on an individual's income at age 45. We can estimate $\sigma^2_0(x)$ from income data at age 45 of individuals who could not participate in such a program but are from the same locality and profession as those in the observational study.
	
	Surface mining in the Central Appalachia region of the eastern United States increased after 1989, partly resulting from the Clean Air Act Amendments of 1990, which made surface mining financially attractive \citep{copeland2012mountaintop}. We might be interested in estimating the effect on the birth weight of babies after a long period of mining activity in the region, say in 2010. In that case, birth weight data in 2010 from regions outside of the Appalachian Counties with mining permits can be used to estimate $\sigma_0^2(x)$. We present our analysis for this study in Section \ref{sec_mining_birthweight}.
	
	Similar to temporal variance stationarity, a second-order stationary process model for the control potential outcome over a geographical region, which includes the observational study as a smaller region, will justify our estimation of $\sigma_0^2(x)$ under geographical variance stationarity.
    
	\item[(d)] {\it Regression discontinuity.} In a regression discontinuity design, the treatment assignment probability jumps at a threshold of a running variable \citep{thistlethwaite1960regression, Hahn2001}. Consider a fuzzy regression discontinuity design where the jump in the probability is less than 1, i.e., the treatment is not impossible before the threshold or assured after the threshold. We can estimate the treatment effect in such a design if the treatment assignment was ignorable near the threshold \citep{Hahn2001}. But this unconfoundedness assumption may fail.
	
	For example, consider studying the effect of a college scholarship on starting salary after graduation. A scholarship-granting institution may have a threshold on a college applicant's standardized test score that increases their scholarship chances. However, these institutions typically follow other discretionary practices, perhaps using their subjective judgment on applicants' motivations and abilities, in granting scholarships to certain applicants slightly below the threshold or refusing scholarships to some applicants slightly above the threshold of the test score. This process will create a fuzzy regression discontinuity but also potential confounding. 
	
	We can use the two-point identification strategy in such cases by estimating $\sigma^2_0(x)$ from the outcomes of the units whose running variable values make them ineligible for the treatment. In this example, college graduates from similar backgrounds who did not take a standardized test may be used for estimating $\sigma^2_0(x)$.
    
	\item[(e)] {\it Resistant population.} Generally, we can estimate $\sigma_0^2(x)$ if
	there is a `similar' population that was not exposed to the treatment. In the previous examples, we rationalized such populations. Note that we can also estimate $\sigma_0^2(x)$ if that similar population was exposed to the treatment, yet the treatment is not expected to have an effect on this population. In the surface mining example, families could have traveled out of the Central Appalachia region shortly after surface mining in the area started. Then, births in those families in later years are unlikely to be affected by the mining activity as they were not exposed to the treatment for any significant amount of time. Birth data for this population can be used to estimate $\sigma_0^2(x)$.

	We call a comparable population that was not exposed to the treatment or, if exposed, is not expected to have an effect, a {\it resistant
	population}. Notice that the observed and the resistant population need not have the same distribution of $X_i$, or even the same mean of $Y_i(0)$, for us to calibrate $\sigma^2_0(x)$ from the resistant population. 
    Explicitly, let $\{(\widetilde{Y}_i, \widetilde{X}_i)\}$, for $i=1,\ldots, m$, be a sample from the resistant population.
    Assumption \ref{assump_main}(c) requires that
    ${Var}(\widetilde{Y}_i \mid \widetilde{X}_i=x) = {Var}(Y_i(0)\mid X_i=x)$.\footnote{Alternatively, we can think the resistant population and the target population belonging to a single larger population. Then, Assumption \ref{assump_main}(c) requires that $Var[Y_i \mid X_i = x, S_i = 1] = Var[Y_i(0) \mid X_i = x, S_i=0]$, where $S_i$ is indicating the resistant population, while we understand all quantities calculated for the observational study conditions on $S_i=0$.}
    
\end{enumerate}

Now consider a couple of examples where one could make a mistake in the calibration of $\sigma^2_0(x)$. First, $\sigma^2_0(x)$ should not be estimated by an estimate of $Var(Y_i\mid Z_i=0,X_i=x)$ without making a strong assumption that 
$Var(Y_{i}(0)\mid Z_i=0,X_i)=Var(Y_{i}(0)\mid X_i)$. While this is a milder assumption than conditional ignorablility of the treatment given $X_i$, it is nonetheless an untestable assumption and requires justification. 
Second, consider the professional development example from above. Such programs are likely aimed at individuals on the lower income scale. While a resistant population here need not have low income, income distribution is typically more variable with a higher income scale \citep[][Figure 12]{nichols2008measuring}. Thus, a resistant population should have similar outcome values to the observed population in cases such as these, where the variability is related to the mean of the outcome.

\section{Inference for the conditional treatment effect}

\subsection{Nonpametric estimation}\label{sec_estimation}
Consider $(Y_i, Z_i, X_i)$, for $i=1,\ldots, n$, i.i.d.~from the 
observed data distribution. This and the following section provide consistent estimators of $\tau_{\pm}(x)$
and corresponding large sample confidence intervals. Our nonparametric estimation method 
uses local linear regression \citep{Fan1996} to estimate the various components of $\tau_{\pm}(x)$.
For simplicity, we assume here that $X_i$ is a $d$-dimensional continuous variable.  
Our methods can be easily extended to incorporate discrete/categorical data 
following the existing literature on local linear methods.

While standard local linear regression methods can give estimates of
$\beta(x)$ and $\sigma^2(x)$, we have to ensure that when these estimates are 
combined to calculate $\widehat{\Delta^2}(x)$, the value is non-negative. To ensure
this, we impose a constraint on the least squares problem in the local linear regression for 
calculating $\widehat{\beta}(x)$. Apart from ensuring the non-negativity of $\widehat{\Delta^2}(x)$, 
this has additional implications. First, the resultant estimate $\widehat{\Delta^2}(x)$ behaves 
differently when $\Delta(x)=0$ v.s. when $\Delta(x)>0$. We provide asymptotic analysis under both these conditions. To give some details, while $\widehat{\Delta^2}(x)$
(after appropriate centering and scaling) is asymptotically normal when $\Delta(x)>0$, it is not normal 
in the case $\Delta(x)=0$. We use these distributions to provide asymptotically valid 
confidence intervals for $\tau_{\pm}(x)$ in the next section. Second, the corresponding constrained optimization 
problem is non-convex. But it can be rewritten in a way that modern 
mathematical optimizers can solve the problem easily; we implemented the optimization using the Gurobi optimizer \citep{gurobi}.

Suppose $\widehat{\sigma_0^2}(x)$  and $\widehat{\pi}(x)$ are estimators of $\sigma_0^2(x)$ and $\pi(x)$, respectively. Let $K(\cdot)$ denote a kernel function, a $d$-variate real-valued function, and $H_1,H_2,H_3$ and $H_4$ are diagonal bandwidth matrices of size $d$. Our estimation steps are as follows.
\begin{enumerate}
	\item[(I)] Estimate $m(x) = E( Y_i \mid X_i=x)$ as $\widehat{m}(x)=\widehat{\mu}_m$ where 
\begin{align}\label{eq_mx}
    (\widehat{\mu}_m, \widehat{{\zeta}}_m) = 
    \arg\min_{\mu, {\zeta}} \sum_{i=1}^n 
(Y_i - \mu - {\zeta}^\top (X_i-x))^2 K\left( H_1^{-1}(X_i-x)\right). 
\end{align}

\item[(II)] Using $\widehat{m}(x)$ from (I), let $e_i = Y_i-\widehat{m}(X_i)$. Now estimate $\sigma^2(x)$ as $\widehat{\sigma^2}(x)=\widehat{\mu}_v$ 
where 
\begin{align}\label{eq_sigma2x}
    (\widehat{\mu}_v, \widehat{{\zeta}}_v) = \arg\min_{\mu, {\zeta}} \sum_{i=1}^n (e_i^2 - \mu - {\zeta}^\top (X_i-x))^2 K\left(H_2^{-1}(X_i-x)\right).
\end{align}

\item[(III)] Finally, estimate $\beta(x)$ as $\widehat{\beta}_C(x)=\widehat{\mu}_1-\widehat{\mu}_0$ where 
$\widehat{\mu}_z$, $z=0,1$ come from the constrained optimization problem

\begin{align}\label{eq_betax}
(\widehat{\mu}_0, \widehat{{\zeta}}_0, \widehat{\mu}_1,\widehat{{\zeta}}_1) = &
\arg\min_{\mu_0,{\zeta}_0,\mu_1,{\zeta}_1}
\sum_{i=1}^n Z_i(Y_i - \mu_1-{\zeta}_1^\top(X_i-x))^2 K\left(H_3^{-1}(X_i-x)\right) \nonumber\\
& \qquad+ 
\sum_{i=1}^n (1-Z_i)(Y_i - \mu_0-{\zeta}_0^\top(X_i-x))^2 K\left(H_4^{-1}(X_i-x)\right) \nonumber\\
& \text{subject to }  (\mu_1-\mu_0)^2 \geq \frac{1}{\widehat{\pi}(x)(1-\widehat{\pi}(x))}( \widehat{\sigma^2}(x) - \widehat{\sigma_0^2}(x)).
\end{align}

\item[(IV)] Calculate 
$$\widehat{\Delta^2}(x) = \{\widehat{\beta}_C(x)\}^2 - \frac{\widehat{\sigma^2}(x) - \widehat{\sigma_0^2}(x)}{\widehat{\pi}(x)(1-\widehat{\pi}(x))},
\text{ and }
\widehat{\tau_\pm}(x) = \widehat{\beta}_C(x) \pm \sqrt{\widehat{\Delta^2}(x)}.$$
\end{enumerate}

Steps (I) and (II) are immediate generalizations of \cite{Fan1998} on estimating univariate conditional variance function to the multivariate case. As \eqref{eq_mx} and 
\eqref{eq_sigma2x} are straightforward weighted least squares regressions, $\widehat{m}(x)$ and $\widehat{\sigma^2}(x)$ have explicit expressions in the vector-matrix notation. However, $\widehat{\beta}(x)$ in  (III) does not have an explicit expression because of the constraint in \eqref{eq_betax}. The subscript $C$ in $\widehat{\beta}_C(x)$ is used to emphasize that the 
estimator is from the constrained optimization. We will call $\widehat{\beta}_U(x)$ the 
estimate of $\beta(x)$ from the corresponding unconstrained optimization.

\begin{remark}\label{rem_sigma0}
{Estimation of $\sigma^2_0(x)$} \normalfont can follow steps (I) and (II) but with the resistant population data and their own bandwidths. Typically, $\sigma^2_0(x)$ will be a simpler function than $\sigma^2(x)$ since the latter includes additional variability in the treatment assignment. In the simplest case, $\sigma^2_0(x)$ is a constant, i.e., the control potential outcomes are homoscedastic,  which is a common model assumption in the literature. This common variance can be better estimated as the sample variance of the residuals from a regression of the resistant population's outcome data on $x$. The resistant population data may also be much larger than the observational study data. In that case, again, the error in estimating $\sigma^2_0(x)$ using Steps (I) and (II) will be much smaller than that in estimating $\sigma^2(x)$.
\end{remark}
\subsection{Large sample confidence interval}\label{sec_inference}
We provide large sample guarantees of the proposed estimator and then provide a method to calculate asymptotically valid confidence intervals. We start with our technical assumptions.
Let $m_z({x}) = E \left( Y_i \mid  Z_i = z, {X}_i = {x} \right)$, $z=0,1$, the conditional means of $Y_i$ of the treated and control group, respectively.
%{\color{red} (Going to use this notation in the assumptions.)}

\begin{assumption}\label{assm_technical}  (Technical assumptions)
\begin{enumerate}
    \item[(a)] $K(u_1,\ldots,u_d) = \prod_{i=1}^dk(u_i)$, where $k$ is a density function symmetric about zero with bounded support on the real line and finite fourth moment.
    $H_i = h_i I_d, i=1,2,3,4$ with
    $h_1 \asymp h_2\asymp h_3\asymp h_4 \asymp h $, for some $h$ such that $h\rightarrow 0$ and $nh^d\rightarrow \infty$ as $n\rightarrow\infty$.
    \item[(b)] $f(x)$, the density function of $X_i$, $\pi(x)$ and $\sigma^2(x)$ are positive and continuous. $E(Y^r \mid X=x)$ and $E(Y^r \mid X=x, Z=z)$ are continuous for $r=3,4$ and $z=0,1$. The second derivatives of $m(x)$, $m_0(x)$, $m_1(x)$ and $\sigma^2(x)$ are uniformly continuous on an open set containing $x$. $E(Y^4) <\infty$. 
    \item[(c)] $\widehat{\pi}(x) \stackrel{p}{\rightarrow} \pi(x)$ and $(nh^d)^{-1/2}\,\widehat{\sigma^2_0}(x) \stackrel{p}{\rightarrow} 0$ as $n\rightarrow\infty$.
\end{enumerate}
\end{assumption}

Among these assumptions, (a) and (b) are standard assumptions adapted from the local linear regression literature. Regarding (c), any reasonable estimator $\widehat{\pi}(x)$ should be consistent for $\pi(x)$, while its second part is justified by Remark \ref{rem_sigma0}.

Before stating our distributional convergence result, we introduce some notation. Let $\epsilon_i = \left( Y_i - m({X}_i) \right)/\sigma({X}_i)$ for $i=1,\ldots, n$ and $\lambda^2({x}) = E\left\{ (\epsilon_i^2-1)^2 \mid {X}_i={x}\right\}$. 
Also, denote 
$\nu_z^2({x}) = Var(Y_i \mid  Z_i = z, {X}_i = {x})$, and let $\xi_{z,i}= \left(Y_i-m_z({X}_i) \right)/\nu_z({X}_i)$, and 
$\eta_z({x}) = E\left\{ \xi_{z,i} \epsilon_i^2 \mid  Z_i = z, {X}_i = {x}\right\}$ 
 for $z=0,1$ and $i=1,\ldots, n$. 
 
Write $\theta_K^d = \int K(u)^2 du$. Finally, with $h^d =(\sum_{j=2}^4 |H_j|)/3$, let $|H_j|/h^d\rightarrow \alpha_j^d$ as $n \rightarrow \infty$ for $j=2,3,4$. 
Define 
%{\tiny \it [$pf_1(x)=\pi(x)f(x)$ and $(1-p)f_0(x)=(1-\pi(x))f(x)$ I HAVE CORRECTED THE FORMULAS]}
\begin{align}
v_\Delta^2:= & 4 \theta_K^d \beta({x})^2 \Big( \frac{\nu_1^2({x})}{\alpha_3^d \pi(x)f({x})}  
+ \frac{\nu_0^2({x})}{\alpha_4^d (1-\pi(x))f({x})} \Big) 
+ \frac{4 \theta_K^d \beta({x}) \sigma^2({x})}{\pi({x})(1-\pi({x})) f({x})} \Big( \frac{\nu_0({x}) \eta_0({x})}{\alpha_2^{d/2}\alpha_4^{d/2}}  \nonumber\\
 &\qquad - \frac{\nu_1({x}) \eta_1({x})}{\alpha_2^{d/2}\alpha_3^{d/2}} \Big) + \frac{\theta_K^d \sigma^4({x}) \lambda^2({x})}{\alpha_2^d \pi({x})^2(1-\pi({x}))^2 f({x})},\label{eq_varDelta}
\end{align}
and 
\begin{align}
v_{\tau,\pm}^2:= & \ \frac{\theta_K^d \tau_\pm({x})^2}{\Delta^2({x})} \Big( \frac{\nu_1^2({x})}{\alpha_3^d \pi(x)f({x})} 
+ \frac{\nu_0^2({x})}{\alpha_4^d (1-\pi(x))f({x})} \Big) 
+ \frac{ \theta_K^d \tau_\pm({x}) \sigma^2({x})}{\pi({x})(1-\pi({x})) f({x}){\Delta^2({x})}} \times \nonumber \\
 &\qquad \Big( \frac{\nu_0({x}) \eta_0({x})}{\alpha_2^{d/2}\alpha_4^{d/2}}  - \frac{\nu_1({x}) \eta_1({x})}{\alpha_2^{d/2}\alpha_3^{d/2}} \Big) + \frac{\theta_K^d\sigma^4({x}) \lambda^2({x})}{4\alpha_2^d\Delta^2({x})\pi({x})^2(1-\pi({x}))^2 f({x})}.\label{eq_varTau}
\end{align}

\begin{theorem}
	\label{thm_largesample}  
Suppose that Assumptions \ref{assump_main} and  \ref{assm_technical} hold.
\begin{enumerate} 
    \item[(a)] When $E(Y_i(0)\mid Z_i=1, X_i=x) = E(Y_i(0)\mid Z_i=0, X_i=x)$, assuming that $nh^{d+4} \rightarrow 0$ as $n \rightarrow \infty$,
$$(nh^d)^{1/2}\left( \widehat{\Delta^2}(x) + O(h^2)\right) \stackrel{d}{\longrightarrow} \frac12 \delta_0 + \frac12 |N(0, v_\Delta^2)|, \text{ as $n \rightarrow \infty$,}$$
where $\delta_0$ is the degenerate distribution at 0 and $|N(0,v^2)|$ is the distribution of absolute value of a $N(0,v^2)$ random variable.	
    \item[(b)]	When $E(Y_i(0)\mid Z_i=1, X_i=x) \neq E(Y_i(0)\mid Z_i=0, X_i=x)$, 
$$(nh^d)^{1/2}\left( \widehat{\Delta^2}(x) -\Delta^2(x) + O(h^2) \right)\stackrel{d}{\longrightarrow} N(0, v_\Delta^2), \text{ as $n \rightarrow \infty$,}$$
and
$$(nh^d)^{1/2} \left( \widehat{\tau_{\pm}}(x) - \tau_{\pm}(x) + O(h^2)\right)
\stackrel{d}{\longrightarrow} N(0, v_{\tau,\pm}^2), \text{ as $n \rightarrow \infty$.}$$
\end{enumerate}
\end{theorem}

\begin{corollary}
Suppose that Assumptions \ref{assump_main} and  \ref{assm_technical} hold.	
Then, at least one of the following is true: 
\begin{enumerate}
    \item[(a)] $\widehat{\tau_+}(x)\stackrel{p}{\rightarrow} \tau(x)$, as $n \rightarrow \infty$, 
    \item[(b)] $\widehat{\tau_-}(x)\stackrel{p}{\rightarrow} \tau(x)$, as $n \rightarrow \infty$.
\end{enumerate}
% 1.~$\widehat{\tau_+}(x)\stackrel{p}{\rightarrow} \tau(x)$, 
% 2.~$\widehat{\tau_-}(x)\stackrel{p}{\rightarrow} \tau(x)$.
\end{corollary}

Next, to construct a confidence interval, note that one can construct 
consistent estimators of $v_\Delta^2$ and $v_{\tau,\pm}^2$ by plugging in consistent estimators of its components; see Remark \ref{rem_se} in Section \ref{sec_remarks} for details. 
So, let $\widehat{v_\Delta}(x)$ be a consistent estimator of $v_\Delta$ and
$\widehat{v_{\tau,\pm}}(x)$ be consistent estimators of $v_{\tau,\pm}(x)$. 
Then, using Theorem \ref{thm_largesample}, if we knew that $E(Y_i(0)\mid Z_i=1, X_i=x) \neq E(Y_i(0)\mid Z_i=0, X_i=x)$, we could create confidence 
intervals for $\tau_{\pm}(x)$ as
\begin{equation}\label{eq_ci_alt}
\left[\widehat{\tau_\pm} - \Phi^{-1}(1-\alpha/2)\times \frac{\widehat{v_{\tau,\pm}}(x)}{\sqrt{nh^d}},
\widehat{\tau_\pm} + \Phi^{-1}(1-\alpha/2)\times \frac{\widehat{v_{\tau,\pm}}(x)}{\sqrt{nh^d}}\right]
\end{equation}
which would have an asymptotic coverage rate of 100$(1-\alpha)\%$ by 
Theorem \ref{thm_largesample} (b).

Unfortunately, we do not know whether in fact $E(Y_i(0)\mid Z_i=1, X_i=x) = E(Y_i(0)\mid Z_i=0, X_i=x)$. Thus, we break down our confidence interval construction into multiple steps, where the first step uses an appropriate testing procedure to test for the null 
hypothesis $H_0: E(Y_i(0)\mid Z_i=1, X_i=x) = E(Y_i(0)\mid Z_i=0, X_i=x)$.
It uses the above interval when the hypothesis is rejected. Algorithmically, the 100$(1-\alpha)$\% confidence intervals for $\tau_\pm(x)$ are constructed using the three steps below. Theorem \ref{thm_ci} states the asymptotic validity of our confidence statements based on the resulting confidence intervals.

\begin{enumerate}
    \item[(I)]  Fix $\delta\in(0,1)$. Check for the inequality $\widehat{\Delta^2}(x) > \widehat{v_\Delta}(x) / (nh^d)^{.5(1-\delta)}$.
    \item[(II)]  If the inequality holds, use the confidence interval in \eqref{eq_ci_alt}.
    \item[(III)]  If the inequality fails to hold, calculate the confidence interval for $\tau_\pm(x)$ as 
    \begin{equation}\label{eq_betaU}
    \left[\widehat{\beta}_U(x) - \Phi^{-1}(1-\alpha/2)\times \frac{\widehat{v_{\beta,U}}(x)}{\sqrt{nh^d}},
    \widehat{\beta}_U(x) + \Phi^{-1}(1-\alpha/2)\times \frac{\widehat{v_{\beta,U}}(x)}{\sqrt{nh^d}}
    \right],\end{equation}
    where $\widehat{v_{\beta,U}}(x)$ is a consistent estimator of 
    $v_{\beta,U}^2(x)=\theta_K^d[ \nu_1^2(x)\{\alpha_3^d \pi(x)f(x)\}^{-1} 
+ \nu_0^2(x)\{\alpha_4^d(1-\pi(x))f(x)\}^{-1}]$; $\Phi$ denotes the standard normal cumulative distribution function.
\end{enumerate}
The confidence interval in \eqref{eq_betaU} can be replaced by any other confidence interval for $\tau(x)$ under the no unmeasured confounders assumption. We took $\delta=1/3$ in our implementation.

\begin{theorem}\label{thm_ci}
Suppose that Assumptions \ref{assump_main} and  \ref{assm_technical} hold. Further assume that $nh^{d+4} \rightarrow 0$ as $n \rightarrow \infty$. Then, for the confidence intervals calculated following steps (I)--(III), at least one of the following is true:
\begin{enumerate}
    \item[(a)]  the confidence interval for $\tau_+(x)$ has an asymptotic 100$(1-\alpha)$\% coverage for $\tau(x)$,
    \item[(b)]  the confidence interval for $\tau_-(x)$ has an asymptotic 100$(1-\alpha)$\% coverage for $\tau(x)$.
\end{enumerate}
\end{theorem}

\begin{remark} Simultaneous coverage.
\normalfont 
Above, we developed a method for estimation and confidence interval construction at a given $x$. One might want simultaneous confidence intervals for a range of values of $x$. Under assumptions stronger than our Assumption \ref{assm_technical}, one can derive convergence results similar to Theorem \ref{thm_largesample}  uniformly on a range of $x$; see, e.g., \cite{Lee2017}. Then, following the above steps, one can provide uniform inference. We do not pursue this exercise in this paper.
\end{remark}

\subsection{Some practical remarks}\label{sec_remarks}
\begin{remark} {\it Constant treatment effect.} \normalfont There are many practical benefits to a constant
additive treatment effect assumption \citep{Fisher1935, Rosenbaum2020}. A constant treatment effect is often a convenient starting point for establishing causality. Additionally, in many situations, identification of a constant treatment effect has immediate practical use \citep[][Section 2.4.5]{Rosenbaum2002book}.

Our identification and estimation methods using resistant population variance calibration simplify considerably under a constant treatment effect, leading to practically handy formulas. For this discussion let $\mu_t=E(Y_i\mid Z_i=1), \mu_c =E(Y_i\mid Z_i=0), \sigma_t^2=var(Y_i\mid Z_i=1), \sigma_c^2=var(Y_i\mid Z_i=0)$ and $p=\Pr(Z_i=1)$. Then, some calculations using \eqref{eq_ind} show that 
$$\Delta^2 = \frac{\sigma_0^2}{p(1-p)}-\left\{\frac{\sigma_t^2}{1-p}+\frac{\sigma_c^2}{p}\right\}.$$
Hence, $\tau_\pm=(\mu_t-\mu_c) \pm \{p(1-p)\}^{-1/2}\left[\sigma_0^2-\{p\sigma_t^2+(1-p)\sigma_c^2\}\right]^{1/2}$. We can estimate them by
\begin{equation}\label{eq_est_const_efft}
    \widehat{\tau_\pm} = (\overline{Y}_t-\overline{Y}_c) \pm \frac{n}{\sqrt{n_tn_c}}\sqrt{\widehat{\sigma_0^2} - S^2_{pooled}},
\end{equation}
where $(\overline{Y}_t-\overline{Y}_c)$ is the difference of the sample average outcomes between the treatment and control group of sample sizes $n_t$ and $n_c$ respectively, and $S^2_{pooled}$, defined as in classical statistics, is the pooled sampled variance of the two groups, i.e., $S^2_{pooled}=\{(n_t-1)S_t^2+(n_c-1)S_c^2\}/(n_t+n_c-2)$ with $S_t^2$ and $S_c^2$, the corresponding sample variances. Estimators \eqref{eq_est_const_efft} are valid irrespective of the observed or unobserved confounding as long as the treatment effect is constant. In practice, one might use $\max\{0,\widehat{\sigma_0^2} - S^2_{pooled}\}$ instead of $\widehat{\sigma_0^2} - S^2_{pooled}$ under the square root in \eqref{eq_est_const_efft}.

Next, notice that $S^2_{pooled}$ and $\overline{Y}_t-\overline{Y}_c$ are approximately independent for large samples. Hence, assuming $\widehat{\sigma_0^2}$ is calculated independently from the observational sample, the adjustment due to confounding in the observational study is approximately independent of the basic estimator $\overline{Y}_t-\overline{Y}_c$, which one could use if we had a completely randomized experiment. Further,  we can get the estimated large sample variance of $\widehat{\tau_\pm}$ as
\begin{align}\label{eq_var_const_efft}
\frac{S_t^2}{n_t}+\frac{S_c^2}{n_c}+\frac{n^2}{4n_tn_c(\widehat{\sigma_0^2} - S^2_{pooled})} \times \left[var(\widehat{\sigma_0^2}) + \frac{n_t}{n^2}(M_{4t}-S_t^4)+\frac{n_c}{n^2}(M_{4c}-S_c^4)\right], 
\end{align}
where $M_{4t}$ and $M_{4c}$ are the fourth central sample moments of the treatment and control group outcomes, respectively. Thus, the confounding in the study leads to an increase in the standard error over the standard error of a two-sample t-test, $\{{S_t^2}/{n_t}+{S_c^2}/{n_c}\}^{1/2}$.

While the above formulas do not use any covariates, additional covariate information can be used in estimation and inference. One way of covariate adjustment would be to use formulas \eqref{eq_est_const_efft} and \eqref{eq_var_const_efft} on residuals $Y_i-g(X_i)$, for some function $g(\cdot)$, instead of on raw outcomes $Y_i$. A common choice for $g$ is the linear function $a^\top X_i$.
\end{remark} 

\begin{remark}
 Two is less than infinity. \normalfont
There is a practical question of how one would use the inference resulting in two confidence intervals using the proposed method. We emphasize the benefit of the method that it liberates us from assuming treatment selection based on observables. A sensitivity analysis method also relaxes this assumption and, ideally, it gives bounds on the treatment effect estimates and confidence intervals \citep{cornfield1959smoking, Rosenbaum2002book, Imbens2003, Tan2006, ding2016sensitivity, Zhao2019} --- although, to our knowledge, no formal sensitivity analysis method has been developed for heterogeneous treatment effects. However, an increasing amount of data affects these two methods differently. For an externally set value for the maximum amount of bias from unmeasured confounding, a sensitivity analysis gives a bound on the effect estimate. This bound also grows wider, and eventually to infinity, the more we relax the ignorability assumption. The proposed method, on the other hand, with a larger amount of data, pinpoints the treatment effect to two numbers. 

 If no ex-ante knowledge can pick the correct number among the two, conservatively, the researcher can still estimate that the effect is between the interval of these two numbers. Similarly, being conservative, the two confidence intervals can be combined into a single interval, with the lower limit being the smallest of two lower limits and the upper limit being the largest of two upper limits. This will be a wider confidence interval but will not explode to infinity. Similar to sensitivity analysis, this combined interval will also become narrower the more likely the ignorability assumption is. Still, in contrast to sensitivity analysis, we would not need to know or specify the amount of bias from unmeasured confounding.
\end{remark}

\begin{remark}\label{rem_se}
Standard error calculation.  \normalfont
Constructions of confidence intervals, e.g., \eqref{eq_ci_alt} and \eqref{eq_betaU}, require estimation of the asymptotic variances, which contain quantities that can be or have been consistently estimated, such as $\pi({x})$, $f({x})$, $\beta({x})$ and $\sigma^2({x})$, as well as some quantities whose consistent estimation requires some more work, e.g., $\lambda({x})$, $\eta_0({x})$ and $\eta_1({x})$. 
Here, we propose a method to consistently estimate the latter group of quantities so that we can estimate the asymptotic variances using plug-in estimators.

Noting that $\lambda^2({x}) = E \left\{ (\epsilon_i^2 - 1)^2 \mid {X}_i = {x} \right\} = E \left\{  \epsilon_i^4 \mid {X}_i = {x} \right\} -1$, and that $\lambda({x})$ appears in the asymptotic variance formulas in terms of $\sigma^4({x}) \lambda^2({x})$, 
we can estimate the two components $E \left\{\sigma^4({X}_i) \epsilon_i^4 \mid {X}_i = {x} \right\}$ and 
$\sigma^4({x})$, separately. We already have an estimator $\widehat{\sigma^2}({x})$ for the latter.
Now we propose to estimate $E \left\{\sigma^4({X}_i) \epsilon_i^4 \mid {X}_i = {x} \right\}$ using an Nadaraya–Watson type estimator 
\begin{equation}\label{NW1}
    \frac{\frac{1}{nh_v^d}\sum_{i=1}^{n} {K}({H}_{v}^{-1}({X}_i - {x})) e_i^4}{\frac{1}{nh_v^d}\sum_{i=1}^{n} {K}({H}_{v}^{-1}({X}_i - {x}))},
\end{equation}
where $e_i = Y_i - \widehat{m}({X}_i)$, $i=1,\dots,n$ are the residuals as in step (II) of the main estimation procedure
and ${H}_{v} = h_v {I}_{d}$ for a bandwidth $h_v$. We show in the supplement material that the proposed estimator is consistent under certain regularity conditions.

For $z=0$ or $1$,
\begin{align}
    \sigma^2({x}) \nu_z({x}) \eta_z({x}) & = E(\sigma^2({X}_i) \nu_z({X}_i) \xi_{zi} (\epsilon_i^2 - 1) \mid  Z_i = z, {X}_i = {x})\nonumber\\
    & = E(\sigma^2({X}_i) \nu_z({X}_i) \xi_{zi} \epsilon_i^2 \mid  Z_i = z, {X}_i = {x}).\label{NW2}
\end{align}
Let $e_i^{(z)} = Y_i - \widehat{m}_z({X}_i)$, $i=1,\dots,n,$ be the residuals obtained in step (III).
Then we similarly propose to estimate $E(\sigma^2({X}_i) \nu_z({X}_i) \xi_{zi} \epsilon_i^2 \mid  Z_i = z, {X}_i = {x})$ using the Nadaraya–Watson type estimator 
\begin{equation*}
    \frac{\frac{1}{nh_v^d}\sum_{i: Z_i = z} {K}({H}_{v}^{-1}({X}_i - {x})) e_i^2 e_i^{(z)}}{\frac{1}{nh_v^d}\sum_{i:Z_i=z} {K}({H}_{v}^{-1}({X}_i - {x}))}.
\end{equation*}
The supplement shows that the resultant variance estimators are consistent under certain regularity conditions. Alternatively, other nonparametric methods that consistently estimate conditional expectation functions may be used to estimate these quantities. 
\end{remark}

\begin{remark}
 Bandwidth selection. \normalfont 
 There is a large literature on bandwidth selection in local linear regressions \citep{Fan1995,Ruppert1995,Ruppert1997, Yang1999, lee1999bandwidth}. In our numerical results for estimation, we calculate each of the bandwidths separately using a leave-one-out cross-validation method. We used the \url{np} package \citep{hayfield2008nonparametric} that implements the leave-one-out cross-validation to select the bandwidths.
 \citet{Li2004} showed that, under mild regularity assumptions, the relative difference between the leave-one-out cross-validation estimator of the bandwidth and the optimal bandwidth goes to zero,
 %difference between the leave-one-out cross-validation estimator of the bandwidth and the optimal bandwidth is of a smaller order compared to the optimal bandwidth {\color{red}(Is this statement ok?)}, 
 and that the local linear regression estimator using the cross-validated bandwidth attains the same asymptotic normality as the estimator using the oracle bandwidth. 
 Although it is a two-step procedure to estimate the conditional variance $\sigma^2({x})$, \citet{Fan1998} claimed that the bias contributed by the error in the estimation of $\widehat{m}({x})$ is of a higher order of infinitesimal compared to the bias of $\widehat{m}({x})$ itself. This fact enables us to use the cross-validated bandwidth in the first step without introducing any significant bias to $\widehat{\sigma^2}(x)$, as well as to cross-validate the bandwidth in the second step to consistently estimate the optimal bandwidth.  Lastly, to enforce undersmoothing in our confidence interval calculations as required in Theorem \ref{thm_ci}, we crudely take a power, with an exponent slightly larger (or smaller, if the bandwidth was greater than one) than 1, of the cross-validation based bandwidths, $h_1,h_2,h_3$ and $h_4$. This method works well, as seen in our numerical results in Section \ref{sec_sim_CI}.
\end{remark}

\section{Two-point identification and inference for the ATE}\label{sec_ate}
The method developed above gives a two-point inference for the conditional average treatment effect on the treated, $\tau(x)$. But we need to be aware of some practical challenges associated with any nonparametric inference method for $\tau(x)$. First, it is difficult to summarize and visually inspect the effect estimates and confidence intervals when the dimension of $x$ is greater than two. Second, the methods are cumbersome in higher dimensions --- they take longer time to compute and they are slow to converge. Many of these challenges are unavoidable in nonparametric inference for the CATT. To avoid these challenges,
now we consider inference for the average treatment effect on the treated (ATT), which is a single number defined as $ATT=E(Y_i(1)-Y_i(0) \mid Z_i = 1)=E(\tau(X_i))$. 
When the treatment has a highly heterogeneous effect, ATT hides a lot of information about the effect and can vary quickly when the distribution of $X_i$ changes. Still, the ATT has been studied extensively in the causal inference literature.

Our Assumption \ref{assump_main} is not sufficient for two-point identification of the ATT because the bias can change sign with $x$. Recall that we cannot identify the direction of the bias $\Delta(\beta, \tau)(x)$ just by Assumption \ref{assump_main}. Instead, we provide a sufficient condition, Assumption \ref{assump_ateiden}, below that excludes the possibility of an interaction between the sign of $\Delta(\beta,\tau)(x)$ and $x$ and thus allows two-point identification of the ATT. Without this assumption, researchers could use domain knowledge to justify that the sign of $\Delta(\beta,\tau)(x)$, i.e., the direction of the bias in estimating the conditional average treatment effect using the difference of two regression functions ($\beta(x)$), does not change with $x$ before using the method proposed below for the ATT.

\begin{assumption}\label{assump_ateiden} (Identification assumptions for the ATT)
\begin{enumerate}
    \item[(a)] The support of $X$ is a connected set.
    \item[(b)] $\Delta(\beta, \tau)^2(x)$ is continuous on its support.
    \item[(c)] For all $j=1,\ldots,d$, the function 
    $x_j\mapsto E [ \Delta(\beta,\tau)^2(X_1,\ldots,X_{j-1},x_j, X_{j+1},\ldots, X_d) ]$
    is positive on the support of the $j$th coordinate of $X$.
\end{enumerate}
\end{assumption}

Assumption \ref{assump_ateiden} undoubtedly puts restrictions on the observational study. But, on the positive side, this assumption is checkable in the sense that the assumption is on the observed data distribution.

\begin{theorem}
\label{thm_ateiden}
    Assumptions \ref{assump_main} and \ref{assump_ateiden} entail the identification of the magnitude of the bias as
    \begin{equation}|ATT - E\beta(X)| = E |\Delta(\beta,\tau)(X)|.
    \end{equation}
\end{theorem}

Thus, define
$$ATT_- = E\beta(X)  - E \sqrt{\Delta(\beta,\tau)^2(X)},
\text{ and } 
ATT_+ = E\beta(X)  + E \sqrt{\Delta(\beta,\tau)^2(X)}.$$
By Theorem \ref{thm_ateiden},  $ATT$ is equal to either $ATT_-$ or $ATT_+$. We can estimate $ATT_\pm$ by separately estimating $E\beta(X)$ and $E |\Delta(\beta,\tau)(X)|$. Various methods are available for the estimation of $E\beta(X)$, including the inverse probability weighted estimator, augmented inverse probability weighted estimator,  and nonparametric regression-based estimators. 
Particularly, $n^{-1}\sum_{i=1}^n\widehat{\beta}_C(X_i)$ and $n^{-1}\sum_{i=1}^n\widehat{\beta}_U(X_i)$ are nonparametric regression based estimators of $E\beta(X)$. The same way, we can estimate $E |\Delta(\beta,\tau)(X)|$ by $n^{-1}\sum_{i=1}^n\{\widehat{\Delta^2}(X_i)\}^{1/2}$. 

\begin{remark} Asymptotic properties of the estimator. \normalfont
Estimators of ${ATT}_\pm$ that are immediately available from our CATT estimation method are $n^{-1}\sum_{i=1}^n\widehat{\tau_\pm}(X_i)$. The mean squared errors of $n^{-1}\sum_{i=1}^n\widehat{\tau_\pm}(X_i)$ can be shown to decrease at the rate $n^{-1}$, which does not depend on $d$, under reasonable regularity conditions. 
Theorem S1 in the supplement gives a formal result for asymptotic normality of $\sqrt{n}(n^{-1}\sum_{i=1}^n\widehat{\tau_\pm}(X_i)-ATT_\pm)$.
The parametric rate of convergence of $n^{-1}\sum_{i=1}^n\widehat{\tau_\pm}(X_i)$ is demonstrated by simulation later in Section \ref{sec_ate_sim}.
\end{remark}

\section{Simulation: demonstration and comparison}\label{sec_simulation}

\subsection{Estimation of the conditional average treatment effect on the treated}\label{sec_simulation_est}

We demonstrate the estimation of the conditional average treatment effect on the treated based on two-point identification using simulation. We consider three data-generating processes that vary in their specification of the CATT function and confounding effect. Throughout this subsection, we have $n=3000$ and $d=1$, where the univariate $X_i$ is drawn from a standard normal distribution. In all three simulation models, let $U_i$ be an unmeasured confounder distributed as Unif(0,1) and let $\pi(x,u)=\Pr(Z=1\mid X_i=x,U_i=u)$.

{\it Simulation model 1 (Linear effect modification)} \ Set $\log\{\pi(x,u)/(1-\pi(x,u))\} = xu+x/2+3$, $Y_{i0}$ is normal with mean $4-6U_i+X_i$ and standard deviation $0.5$, and $\tau(x)=x/2+3/2$.

{\it Simulation model 2 (Quadratic effect modification)} \ Set $\log\{\pi(x,u)/(1-\pi(x,u))\} = u+4u^2+x/2$, $Y_{i0}$ is normal with mean $1+6U_i+X_i$ and standard deviation $0.5$, and $\tau(x)=x^2-2x+1/2$.

{\it Simulation model 3 (Cubic effect modification and heteroscedastic $Y_{i0}$)} \ Set $\log\{\pi(x,u)/(1-\pi(x,u))\} = u+4u^2+x/2$. As before, $Y_{i0}$ is normal with mean $1+4U_i\sqrt{|X_i|}+X_i$ and standard deviation $0.5$, hence $\sigma_0^2(x)=1/4+4|x|/3$, and $\tau(x)=x^3/2$.

\begin{figure}[htpb]
    \centering
    \includegraphics[width=.85\linewidth]{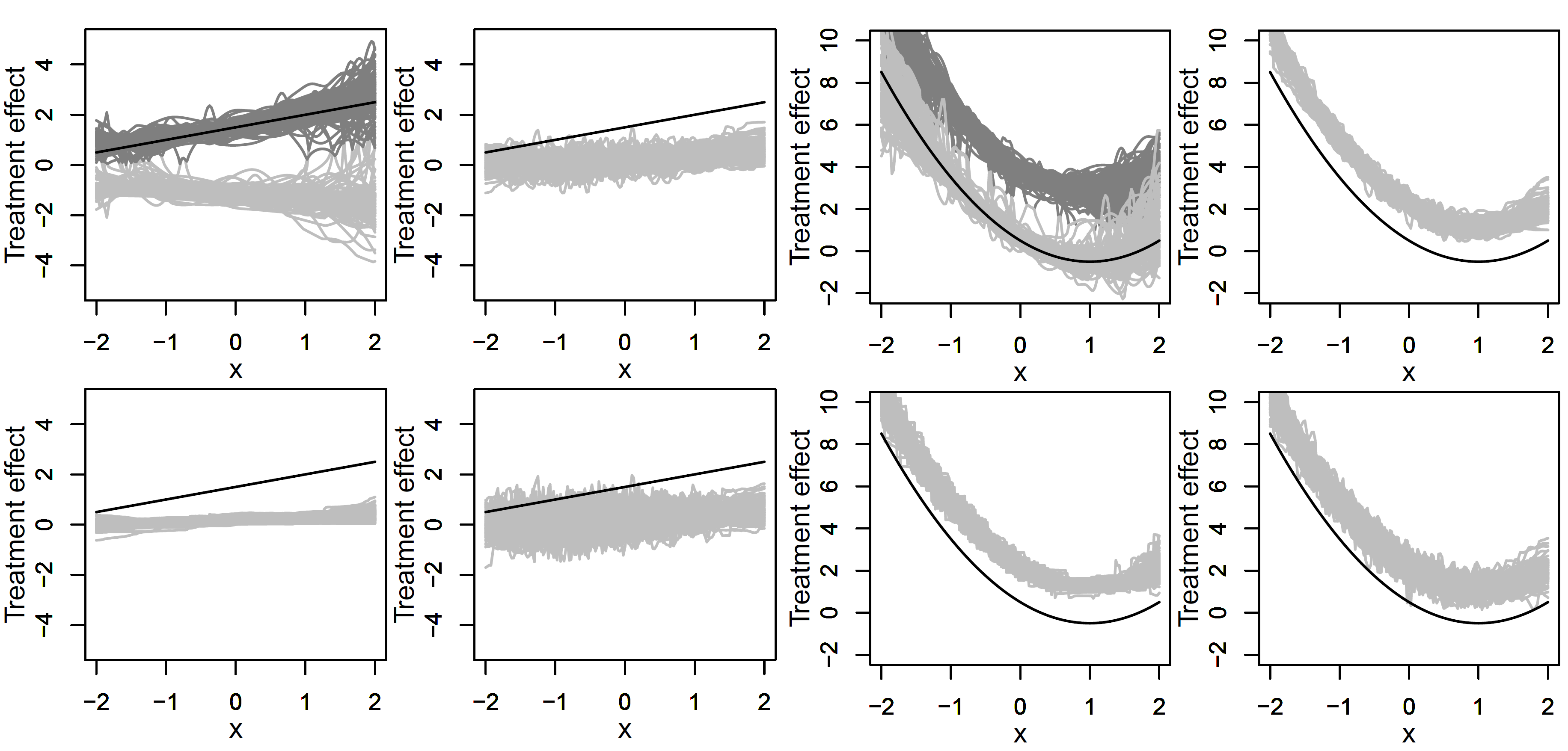}
    \caption{Results from two simulation models: model 1 in the left two columns and model 2 in the right two columns.
    The dark line in the plot shows the true effect. For each setting, the four plots are for: the proposed method (in the top-left, $\widehat{\tau_+}$ in darker grey,
        $\widehat{\tau_-}$ in lighter grey), causal forest (top-right), Bayesian causal tree (bottom-left), and X-learner (bottom-right).}
    \label{fig_linear}
\end{figure}

\begin{figure}[htpb]
    \centering
    \includegraphics[width=.5\linewidth]{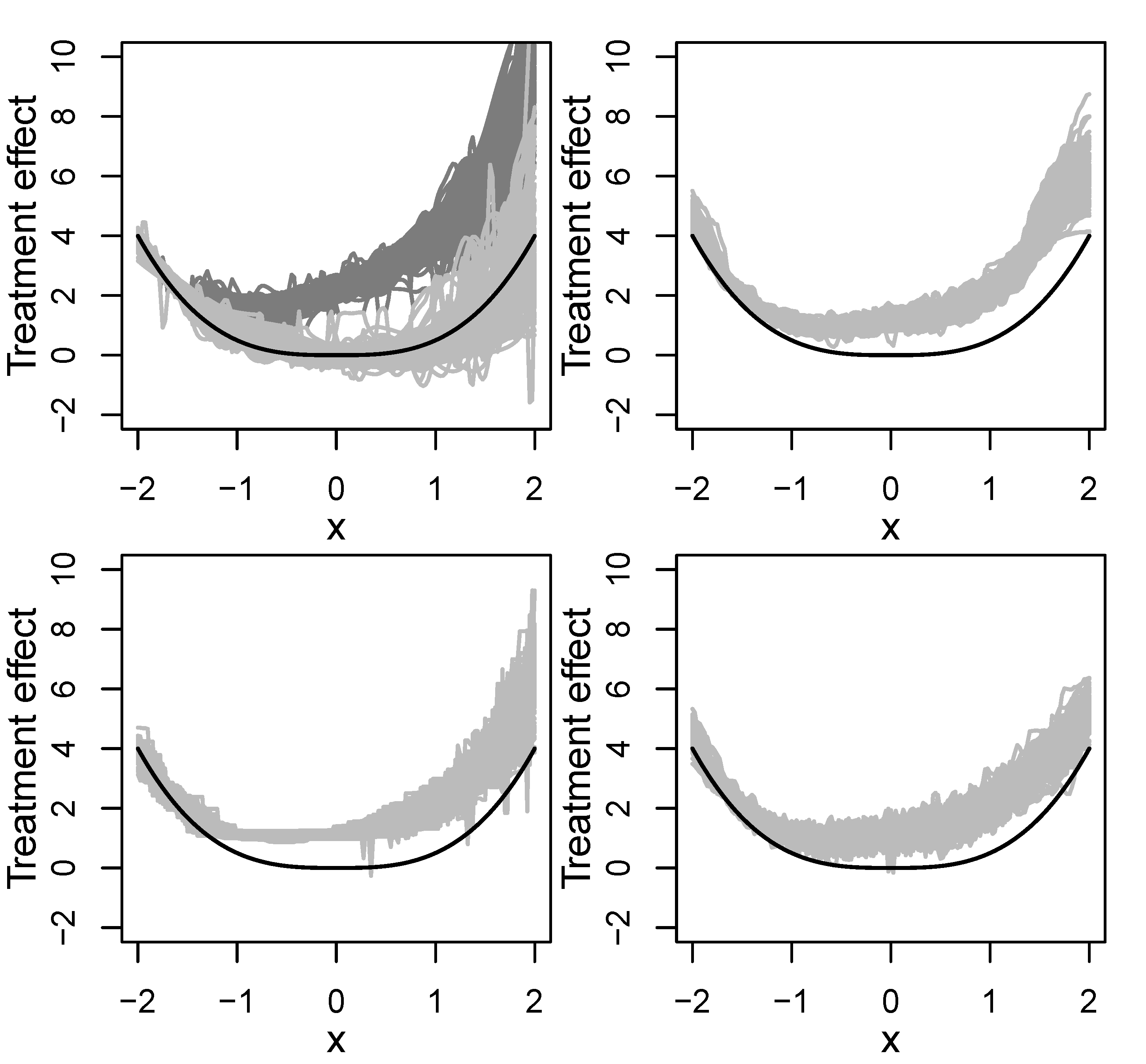}
    \caption{Results from simulation model 3. The dark line in the plot shows the true effect. The proposed method (in the top-left, $\widehat{\tau_+}$ in darker grey,
        $\widehat{\tau_-}$ in lighter grey), causal forest (top-right), Bayesian causal tree (bottom-left), and X-learner (bottom-right).}
    \label{fig_cubic_varysigma0}
\end{figure}

We compared our estimation method, as described in the four steps in Section \ref{sec_estimation}, with state-of-the-art heterogeneous treatment effect estimation methods: causal forest \citep{Wager2018}, Bayesian causal tree \citep{Hahn2020} and X-learner \citep{metalearners2019}. We estimated $\pi(x)$ in our method using a one-layer feed-forward neural network with 8 hidden neurons. The results of the simulation study are presented in Figure \ref{fig_linear} for simulation models 1 and 2 and Figure \ref{fig_cubic_varysigma0} for simulation model 3. 

While there are differences in the results of the three competing methods, in Figure \ref{fig_linear}, all of these methods give biased estimates of $\tau(x)$ because of the presence of the confounder $U_i$ which is highly correlated with the outcome. The confounding effect is also visible in the estimates $\widehat{\tau_+}(x)$ and $\widehat{\tau_-}(x)$ as they are two distinct sets of estimated curves. The ambiguity between these two estimates is also visible from comparing the two models. In simulation model 1, $\widehat{\tau_+}(x)$ covers the true CATT while in model 2, $\widehat{\tau_-}(x)$ covers the true $\tau(x)$.

Figure \ref{fig_cubic_varysigma0}, which corresponds to simulation model 3, shows interesting effects of the unmeasured confounder. The identification bias $|\Delta(\beta,\tau)(x)|$ is smaller for negative values of $x$ and gets larger with larger, positive values of $x$. Thus, $\widehat{\tau_-}(x)$ and $\widehat{\tau_+}(x)$ are close for $x$ values closer to $-2$ and are different for larger values of $x$. The confounding also affects the variability of the estimates $\widehat{\tau_\pm}(x)$, which is comparable to those of the competing methods when the confounding effect is small and is larger than those of the competing methods for larger confounding biases. However, the competing methods give biased estimates in the latter situation.

\subsection{Inference for the conditional average treatment effect on the treated}\label{sec_sim_CI}

We next assess the coverage of the proposed inference method and compare it against that of two frequentist methods, causal forest, and X-learner, using simulation. Our data-generating process generates (i) an unmeasured confounder $U_i$ distributed as Unif$(0,1)$, (ii) $X_i$ as a $5$-dimensional normal, independent of $U_i$, with mean $\mathbf{0}$ and variance $I_5$, 
(iii) $Y_{i0} = 1+2U_i + 5\overline{X}_i + \epsilon_iX_{i1}/2+\eta_i$, where $\epsilon_i$ and $\eta_i$ are independent standard normal random variables and independent of $X_i, U_i$, (iv) $Y_{i1} = Y_{i0} + \tau(X_i)$ where $\tau(X_i)=1+5(\overline{X}_i+X_{i1})/12$, and (iv) $Z_i$ based on $\Pr(Z_i=1\mid X_i, U_i, \epsilon_i, \eta_i) =$ $[1+\exp\{-(\overline{X}_i/2 + 4U_i -1.5)\}]^{-1}$. This logistic model for the treatment assignment ensures that the $\pi(X_i)$ is reasonably away from 0 and 1. Additionally,  $\sigma^2_0(X_i) = X_{i1}^2/4$ varies with the covariate.

\begin{figure}[htbp]
    \centering
    \includegraphics[width=1\linewidth]{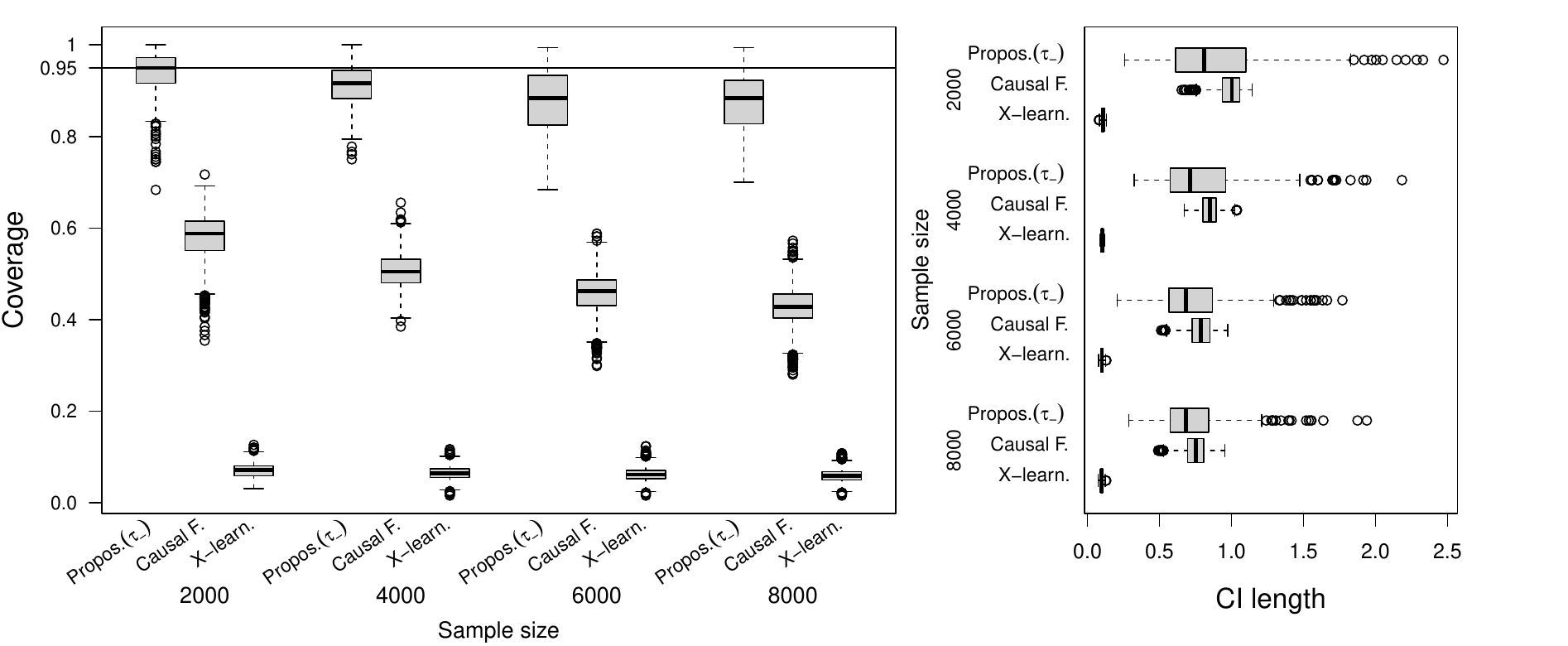}
    \caption{Coverage and length of confidence intervals for $\tau(x)$. Boxplots in the left two panels show the empirical coverage of $\tau(x)$ for a grid of $x$ values. Boxplots in the right two panels show the empirical average length of the confidence intervals for a grid of $x$ values. Based on 200 simulations.}
    \label{fig_ci}
\end{figure}

We evaluate the confidence intervals provided by the methods for $\tau(x)$ on a grid of points $400$ fixed $x$ values. In particular, we calculate the empirical coverage for $\tau(x)$ for each $x$ on this grid using 200 simulated data sets. To implement our method, we use the Bayesian Additive Regression Tree model \citep{Chipman2010} with its default specifications to estimate $\pi(x)$, $\sigma^2_0(x)$ and all the unknown functions in our asymptotic covariance formula. We use a resistant population of size $n^{1.1}$ to estimate $\sigma^2_0(x)$ when our observational study sample size is $n$.

The simulation results for the coverage rates for $\tau(x)$ and average CI lengths are given in Figure \ref{fig_ci}.
The figure shows that the empirical coverage rate for the proposed estimator $\widehat{\tau_-}(x)$ is close to the target $0.95$. However, there is a clear finite sample undercoverage. The other methods have very poor coverage. The lengths of the intervals are sometimes comparatively larger for the proposed estimator than the others. For example, when $n=8000$, the averages of the length of all the confidence intervals are $0.825$ and $0.717$ for the proposed method and causal forest, respectively, which are comparable; the average length is $0.098$ for the X-learner. As noted in Remark 4, albeit under the no covariates case,  wider intervals are expected for $\widehat{\tau_-}(x)$ compared to intervals for the estimators for the $\beta(x)$.  X-learner has poor coverage, and thus, its short intervals are misleading.

\subsection{Estimation of the average treatment effect on the treated}\label{sec_ate_sim}

Section \ref{sec_ate} showed that we can provide two-point identification of the average treatment effect on the treated under Assumption \ref{assump_main} when the sign of the bias does not interact with the effect modifiers. Thus, we have two-point estimates of the ATT as $\widehat{ATT_\pm} = n^{-1}\sum_{i=1}^n \widehat{\tau_\pm}(X_i)$. 
		
We compare the mean squared errors (MSEs) of $\widehat{ATT_\pm}$ and two popular competing estimators across different sample sizes and dimensions of $x$ in two data-generating models. These models specify (i) the unmeasured confounder $U$ distributed as Unif$(0,1)$, (ii) $X_i$ is $d$-dimensional normal, independent of $U_i$, with mean $\mathbf{0}$ and variance $I_d$.
In model 1, that has a relatively strong confounding effect, (iii) $Y_{i0}$ is normal with mean $4+4U_i^2+\overline{X}_i$ and variance 1, (iv) $\Pr(Z_i=1\mid X_i, U_i) =$ $[1+\exp\{-(2\overline{X}_iU_i+6U_i+.5\overline{X}_i+1.5)\}]^{-1}$, and $Y_{i}(1) = Y_{i}(0) + \theta(X_i)$. In model 2, (iii$'$) $Y_{i}(0) = m_i+0.25\eta_{i0}$ and $Y_{i}(1) = m_i + \theta(X_i)+0.25\eta_{i1}$, where $m_i=4+4U_i^2+d\overline{X}_i + \epsilon_i\overline{X}_i/6$ with $\epsilon_i$ and $\eta_i$ are independent standard normal variables and independent of  $X_i$ and $U_i$, and (iv$'$) $\Pr(Z_i=1\mid X_i, U_i) =$ $[1+\exp\{-(U_i+\overline{X}_i/2)\}]^{-1}$. Finally, (v) $\theta(X_i)=X_{i1}/2+3/2$ for $d=1$ and 
$\theta(X_i)=X_{i1}(.5-1/d)+\overline{X}_i/3+5/2$ for $d>1$. 

Some comments regarding the models are in order. To understand the amount of unmeasured confounding, note that Kendall's partial correlation coefficient between $Y_{i0}$ and $U_i$ given $X_i$ is about 0.5 and between $Z_{i}$ and $U_i$ given $X_i$ is about 0.3 for all $d$ for model 1; these correlations are respectively .4 and .1 for model 2. Next, model 1 has a constant additive treatment effect, so the average treatment effect, $ATE = E(Y_i(1)-Y_i(0))  = ATT$, while model 2 has different ATE and ATT, but our identification assumption holds. Finally, Model 1 has a constant value for $\sigma^2_0(x)$, while in model 2, $\sigma^2_0(x)=1.48+\bar{x}^2/36$.

\begin{table}[htpb]
\centering
\caption{Empirical mean squared errors for estimating ATT in simulation model 1}
\begin{tabular}{r|rrrr||rrrr}
\hline
 & \multicolumn{4}{|c||}{$d=7$} &   \multicolumn{4}{|c}{$d=8$}\\
\hline
 &  &  & Linear & Doubly   & &    & Linear & Doubly  \\
n & $\widehat{ATT_-}$ & $\widehat{ATT_+}$  &  regression &   robust &   $\widehat{ATT_-}$ & $\widehat{ATT_+}$ &  regression &   robust\\
\hline
500 & 2.62 & 16.97 & \textbf{1.25} & 1.37 &  3.38 & 19.31 & \textbf{1.30} & 1.43\\
1000 & \textbf{1.21} & 13.92 & 1.25 & 1.38 &  1.65 &  14.03 & \textbf{1.27} & 1.37\\
1500 & \textbf{0.71} & 12.67 & 1.24 & 1.36 &   \textbf{0.81} & 12.84 & 1.29 & 1.38\\
2000 &  \textbf{0.41} & 12.23 & 1.24 & 1.36 &  \textbf{0.55} & 12.41 & 1.26 & 1.36\\
\hline
\end{tabular}
\label{tab_ate}

\end{table}

\begin{table}[htbp]
\small
 \centering
  \caption{Empirical mean squared errors for estimating ATT in simulation model 2. We also report the empirical mean squared errors of the competing methods for estimating ATE because it is unclear what is their appropriate estimand}
   \begin{tabular}
   {c|p{.4in}p{.3in}p{.3in}p{.3in}p{.3in}||p{.4in}p{.3in}p{.3in}p{.3in}p{.3in}}
    \hline
    $n$     & \multicolumn{1}{p{.4in}}{} & \multicolumn{2}{p{.6in}}{Linear regression} & \multicolumn{2}{p{.6in}||}{Doubly robust}      &  & \multicolumn{2}{p{.6in}}{Linear regression} & \multicolumn{2}{p{.6in}}{Doubly robust} \\
        \cline{3-4}\cline{5-6}\cline{8-9}\cline{10-11}
          &    $\widehat{ATT_-}$   & \multicolumn{1}{p{.3in}}{for ATT} & \multicolumn{1}{p{.3in}}{for ATE} & \multicolumn{1}{p{.3in}}{for ATT} & \multicolumn{1}{p{.3in}||}{for ATE}    &    $\widehat{ATT_-}$   & for ATT & for ATE & for ATT & for ATE \\
    \hline
          &       &       & \multicolumn{1}{l}{d = 4}   &       &       &       &       & \multicolumn{1}{l}{d = 5} &       &  \\
    \hline
    500   & 0.48  & 0.08  & 0.11  & 0.11  & 0.13     & 0.28  & 0.10  & 0.12  & 0.12  & 0.13 \\
    1000  & 0.31  & 0.07  & 0.09  & 0.08  & 0.11       & 0.17  & 0.08  & 0.10  & 0.09  & 0.11 \\
    1500  & 0.29  & 0.07  & 0.09  & 0.08  & 0.11       & 0.15  & 0.08  & 0.10  & 0.09  & 0.11 \\
    2000  & 0.24  & 0.07  & 0.09  & 0.08  & 0.11      & 0.13  & 0.08  & 0.10  & 0.09  & 0.11 \\
    \hline
    \end{tabular}%
  \label{tab_ate_model2}%
\end{table}%

\begin{figure}[htpb]
    \centering
    \includegraphics[width=.8\linewidth]{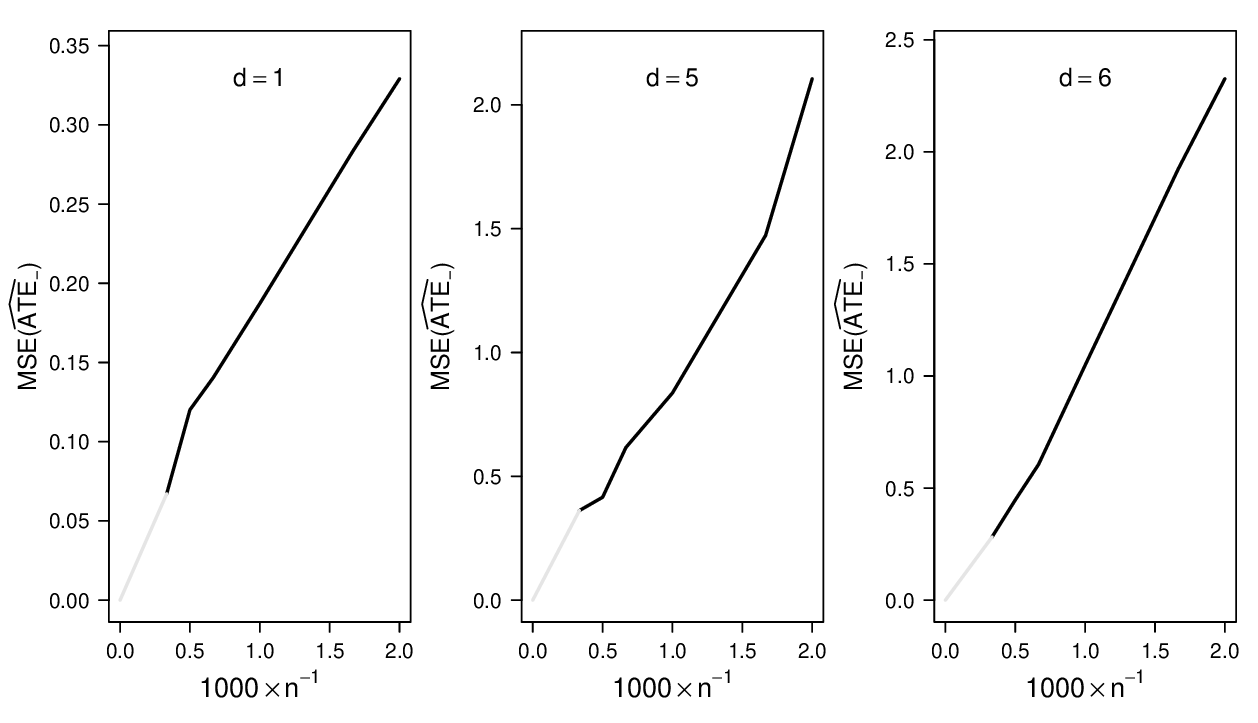}
    \caption{Simulated Mean Squared Error for estimating ATT with $\widehat{ATT_-}$ in model 1.}
    \label{fig_ate_rate}
\end{figure}

The estimators $\widehat{ATT_\pm}$ are suitable in these simulation models since the identification bias does not change sign; it is always positive. Our first competing estimator fits a linear regression to the outcome on $Z_i$ and $X_i$ and estimates the ATT as the coefficient of $Z_i$. This fairly simple estimator is arguably the most popular choice of practitioners, even though it is not recommended by recent statistics literature because of its strong reliance on the linear model specification.  Our second competing method is arguably the most flexible in terms of model specification. This is the doubly robust estimator using the augmented inverse probability weighted estimator that uses super learners for the outcome and assignment model \citep{vanderLaan2007}. We include a generalized additive model \citep{Hastie1990} and XGBoost, a nonparametric model based on generalized tree boosting \citep{Friedman2001, chen2015xgboost}.

Tables \ref{tab_ate} and \ref{tab_ate_model2} report the results of our comparative study. Table \ref{tab_ate} shows that for model 1,  $\widehat{ATT_-}$ provides the most precise estimate when  $n\geq 1000$ and $d=7$ or $n\geq 1500$ and $d=8$. Further, in both tables, we see that the linear regression and doubly robust method are asymptotically biased because of the unmeasured confounding. Notice that MSEs of the competing estimators in these tables, and to some extent of $\widehat{ATT_+}$ in Table \ref{tab_ate}, are relatively constant across $n$ as the variance is much smaller than the squared bias in these sample sizes. In contrast, the MSE of $\widehat{ATT_-}$ decreases with increasing sample size. However, in Model 2, the MSEs of $\widehat{ATT_-}$ are still relatively large because our nonparametric estimation method plugs in several nonparametric estimators. Even though consistent, some components have finite sample bias. Consequently, with a finite sample size, the plug-in estimator tends to exhibit a relatively higher MSE. %We thank a reviewer for the suggestion to explore this aspect of the proposed estimator in simulation.

Figure \ref{fig_ate_rate} investigates the rate of the decrease of the MSE of $\widehat{ATT_-}$ with the sample size. The figure shows that across different dimensions $d=1,5$ and $d=6$, the rate of MSE is proportional to $n^{-1}$. 

\section{Effect of surface mining on birth weight}\label{sec_mining_birthweight}

We study the effect of substantial surface mining in the central Appalachian region on infant birth weight. Surface mining activities became prominent in some areas of central Appalachia starting in the 1990s due to technological developments in large-scale surface mining and amendments to the Clean Air Act in 1990 which made coal reserves in the area, which are low in sulfur content, financially attractive. Mountaintop removal mining, which involves mining coal upon steep terrains, often near residential areas, using explosives, is a majority among surface mining activities. 

The central Appalachian region is contained inside four states -- Kentucky, Tennessee, Virginia, and West Virginia. Regulations only provided permissions for mining in 91 of the 181 central Appalachian counties. Births in those 91 counties form our observational study population, where 23 of those counties saw substantial surface mining. Births in the remaining 298 counties, in those four states, without mining permits, will provide data on our resistant population. We use data from the U.S.~National Center for Health Statistics and combine it with U.S.~census data. We include all data sets except the Birth data used in this study in the supplement to this paper; the Birth data is available upon request from \url{https://www.cdc.gov/nchs/nvss/dvs_data_release.htm}.

Our outcome is birth weight in grams. If birth weight is affected by surface mining, we may see variability in this effect because of biological factors associated with birth. Thus, we use covariates: mother's age, number of prenatal visits, and number of months of prenatal care. The first variable accounts for the mother's biology and the latter two account for the quality of care during the pregnancy. Of course, we assume that surface mining does not affect these variables.

Socioeconomic variables, e.g., income and father's education, may be associated with birth weight and thus are likely confounders. However, in the current analysis, we are not interested in how they might influence the possible effect of surface mining on infant birth weight.
Thus, we do not include these socioeconomic variables in our analysis.

Variability in birth weight in the areas of the same states with no mining permits can be argued to be a good choice for our resistant population variance. We reinforce this argument by matching our 91 observational study counties with the remaining 298 counties in a one-to-one match on several socioeconomic variables. The 28850 births in 2010 in these counties form our resistant population data. Table \ref{tab_resistantmatch} shows that this match makes the matched resistant counties similar to our observational study counties.

\begin{table}[htbp]
  \centering
  \caption{Matching counties in central Appalachia with mining permit to counties of KY, TN, VA, and WV without mining permit on socioeconomic variables in 1990}
    \begin{tabular}{lcccc}
    \hline
          & \multicolumn{2}{c}{Before matching} & \multicolumn{2}{c}{After matching} \\
          & Mining  & Mining & Mining & Mining \\
            &  permitted &  not permitted &  permitted & not permitted \\
    \hline
    No.~of counties & 91    & 298   & 91    & 91 \\
    Poverty rate (\%) & 26.031 & 15.839 & 26.031 & 22.403 \\
    Median income (\$) & 8622 & 10562 & 8622 & 9237 \\
    Not high-school graduate (\%) & 45.390 & 36.928 & 45.390 & 43.848 \\
    Proportion white & 0.977 & 0.863 & 0.977 & 0.900 \\
    Proportion black  & 0.016 & 0.122 & 0.016 & 0.091 \\
    Maternal smoking rate & 0.289 & 0.220  & 0.289 & 0.262 \\\hline
    No.~of births in 2010 & 34173    & 215182   & 34173   & 28850\\   
    \hline
    \end{tabular}%
  \label{tab_resistantmatch}%
\end{table}%

\begin{figure}[htbp]
    \centering
    \includegraphics[width=.95\linewidth]{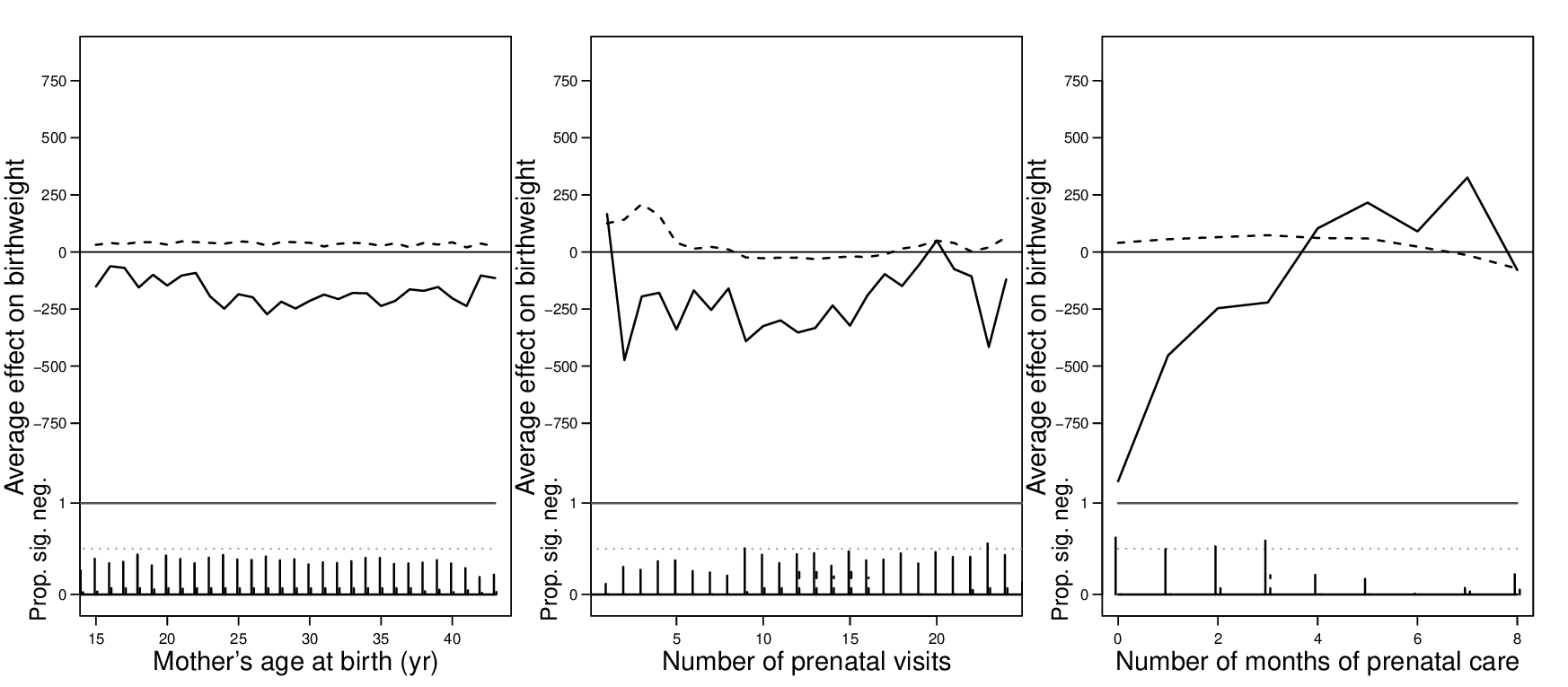}
    \caption{Estimated conditional average treatment effects on the treated (in gms) of surface mining, assuming a non-negative bias (solid line) or no bias (dotted line) from unmeasured confounding. The bar plots summarize results of hypothesis tests of the form $H_0: \tau(x)=0$ vs $H_1:\tau(x)<0$, where $x$ is a vector of values of our three covariates. {The solid bars (the left ones in each pair) correspond to the tests assuming a non-negative bias and the dotted bars (the right ones in each pair) correspond to the tests assuming no bias.} The height of each bar in the first panel is the proportion of rejected tests, at 5\% significance level, in the collection of tests where the mother's age at birth in $x$ is fixed and the other two covariates are varied. Similarly, for the second and the third panels, the corresponding covariate values are fixed and the proportion is over the other two covariates.}
    \label{fig_birthweight1}
\end{figure}

We estimate the conditional effects and confidence intervals using the proposed methods in Sections \ref{sec_estimation} and  \ref{sec_inference}, respectively. Since our theory requires that the resistant population variance is estimated more precisely than the estimands in the study population, we took a 40\% random sub-sample of our 34173 births in mining counties for our calculations, giving us a bit less than half the number of births in the resistant population. In our secondary check, this subsampling did not affect our estimates.

The naive method that ignores any bias correction due to unmeasured confounding finds an estimated effect close to zero or positive of surface mining on birth weight; see Figure \ref{fig_birthweight1}. If we were to assume a negative bias in the naive estimator due to unmeasured confounding, it would suggest an increase in birth weight because of surface mining. However, there is no apparent scientific support for an increase in birth weight due to surface mining. Thus, assuming a non-negative bias in the naive estimator, i.e., taking $\widehat{\tau_-}(x)$, Figure \ref{fig_birthweight1} presents a summary of our bias-corrected estimation and inference. These results show often substantial estimated biases and typically negative estimated effects on birth weight. This finding of a harmful effect of surface mining complements \cite{Small2021} who use a binary indicator for low birth weight as the outcome, adjust for socioeconomic variables, and do not study effect heterogeneity. Our results show that while the mother's age and the number of prenatal visits do not appear to be effect modifiers, births that followed less prenatal care are affected more; this effect mitigates with longer prenatal care. 

Note, in Figure \ref{fig_birthweight1}, the naive estimator, which estimates the treatment effect using the differences in the conditional means, sometimes being below the estimator that assumes a nonnegative bias is not unreasonable because the latter uses a separate penalized estimation method for the difference.

\section{Discussion}\label{sec_discussion}

Analyses of variances have been vital tools for statistical analysis of controlled experiments. Although one wants inference regarding the difference in means or the main effect, comparisons of variabilities in different parts of the data are critically used in the analyses of (co-)variance. Fundamentally, our method builds on this idea (cf.~Remark \ref{exam_illu}). The difference in the variance of the whole data to that of the resistant population turns out to be informative about the magnitude of the bias in the CATT due to unmeasured confounding. The proposed method may thus be termed Resistant Population Calibration Of VAriance, or RPCOVA, following suit of the classical abbreviations. Similar to the fact that the ANOVA table informs us of whether the treatment effect is nonzero but not the effect size, RPCOVA can inform us of the size of the bias but not the direction. 

RPCOVA gives two possible estimates of the treatment effect but fails to provide further information regarding which of these two is the right choice. In fortunate situations where there is no bias in the treatment assignment, these two estimates collapse into a single estimate, or either is a consistent estimate. In other situations, when there is consensus regarding the direction of bias from unmeasured confounders, we can choose the corresponding estimate -- the smaller of the two when the bias is positive and the larger of the two when it is negative. In our empirical study, we ruled out a negative bias on scientific grounds. Even in other situations, it may be sufficient to know the two estimates. If establishing causality consists of a connected mesh of coherent arguments (\cite{Peirce1868}, \cite{Karmakar2020}), then RPCOVA, even with its slight inconclusiveness, will substantially contribute to these arguments. %Finally, it is hard at this moment to elaborate on other possibilities and specific instances of how RPCOVA will be instrumental.

Beyond developing the notion of two-point identification of the effect of a confounded treatment, this paper proposed and implemented an algorithm for nonparametric estimation and then provided easy-to-compute large sample confidence intervals. However, we could also aim to estimate better and faster, leaving further inference to resampling-based tools, perhaps. This would invariably lead to less restrictive technical conditions than those in Assumption \ref{assm_technical}. 

\section{Supplementary material}
In the supplementary document, we present proofs of the theorems and some examples and remarks given in the
main paper, provide some auxiliary results to the proofs and give an additional
result regarding the parametric rate of convergence for ATT estimation.
Code implementing our proposed methods and all data sets except the birth data for our empirical study are available at \url{https://github.com/bikram12345k/RPCOVA}.

\section{Acknowledgments}
%The authors thank the anonymous reviewers for their valuable suggestions. 
This work is supported in part by funds from the U.S.~National Science Foundation.

\bibliographystyle{apa}
\bibliography{references}

\clearpage

\newpage
%\appendix

\setcounter{equation}{0}
\setcounter{figure}{0}
\setcounter{table}{0}
\setcounter{page}{1}
\setcounter{section}{0}
\setcounter{theorem}{0}
\renewcommand{\theequation}{S\arabic{equation}}
\renewcommand{\thetheorem}{S\arabic{theorem}}
\renewcommand{\thelemma}{S\arabic{lemma}}
\renewcommand{\thefigure}{S\arabic{figure}}
\renewcommand{\thesection}{S\arabic{section}}
\renewcommand{\thetable}{S\arabic{table}}
\renewcommand{\thefigure}{S\arabic{figure}}

\pagestyle{myheadings}
\markright{SUPPLEMENT}
\markleft{SUPPLEMENT}

\noindent\textsf{\large SUPPLEMENT TO ``Inferring the Effect of a Confounded Treatment by Calibrating Resistant Population's Variance''}

\begin{abstract}
In this document, we present proofs of the theorems and some examples and remarks in the main paper, provide some auxiliary results to the proofs and give an additional result regarding the parametric rate of convergence for ATT
estimation in Section \ref{sec:parametric.rate}.
\end{abstract}

\noindent%
% {\it Keywords:}  Asymptotic dependence; Central Limit Theorem; Dose-response relation with a downturn; Non-contiguous alternatives; Rank statistics; Sequential ranks.
% \vfill

% \newpage
% \spacingset{1.9} % DON'T change the spacing!

\section{Proofs of the identification results (Theorems \ref{thm_main} and \ref{thm_ateiden})}

\begin{proof}[Proof of Theorem \ref{thm_main}.]
First write
\begin{align*}
\sigma^2(x) &= Var(Y_i \mid X_i=x) \\
& = E\{Var(Y_i \mid Z_i, X_i=x)\} + Var\{E(Y_i \mid Z_i, X_i=x)\}\\
& = \pi(x) Var(Y_i(1) \mid Z_i=1, X_i=x) + (1-\pi(x)) Var(Y_i(0) \mid Z_i=0, X_i=x) \\
& + (\pi(x)(1-\pi(x))) \beta(x)^2,
\end{align*}
and
\begin{align*}
\sigma_0^2(x) &= Var(Y_i(0) \mid X_i=x) \\
& = E\{Var(Y_i(0) \mid Z_i, X_i=x)\} + Var\{E(Y_i(0) \mid Z_i, X_i=x)\}\\
& = \pi(x) Var(Y_i(0) \mid Z_i=1, X_i=x) + (1-\pi(x)) Var(Y_i(0) \mid Z_i=0, X_i=x) \\
& + (\pi(x)(1-\pi(x))) (\beta(x) - \tau(x))^2.
\end{align*}
In the first expression, $Var\{E(Y_i \mid Z_i, X_i=x) = (\pi(x)(1-\pi(x))) \beta(x)^2$ is true because
\begin{align*}
&Var\{E(Y_i \mid Z_i, X_i=x)\}\\
=& E[E(Y_i \mid Z_i, X_i=x)-\{\pi(x) E(Y_i \mid Z_i=1, X_i=x) + (1-\pi(x)) E(Y_i \mid Z_i=0, X_i=x)\}]^2 \\
=& \pi(x) \{(1-\pi(x)) E(Y_i \mid Z_i=1, X_i=x) - (1-\pi(x)) E(Y_i \mid Z_i=0, X_i=x)\}^2 \\
&+ (1-\pi(x)) \{\pi(x) E(Y_i \mid Z_i=1, X_i=x) - \pi(x) E(Y_i \mid Z_i=0, X_i=x)\}^2 \\
=& (\pi(x)(1-\pi(x))) \beta(x)^2.
\end{align*}
And $Var\{E(Y_i(0) \mid Z_i, X_i=x)\} = (\pi(x)(1-\pi(x))) (\beta(x) - \tau(x))^2$ can be proven similarly.

Under Assumption \ref{assump_main} (b),
\begin{align*}
\sigma^2(x) - \sigma_0^2(x) &= (\pi(x)(1-\pi(x))) \{\beta(x)^2 - (\beta(x) - \tau(x))^2\}\\
&= (\pi(x)(1-\pi(x))) [\beta(x)^2 - \left\{\Delta(\beta, \tau)(x)\right\}^2]
\end{align*}
Then, the proof is completed by rearranging.
% The proof then follows from the quadratic formula used to find $\tau(x)$. 
Specifically, by solving the above quadratic equation in $\tau(x)$ we get
$$\tau(x) = \beta(x) \pm \Big\{\beta(x)^2 - \frac{\sigma^2(x)-\sigma_0^2(x)}{\pi(x)(1-\pi(x))} \Big\}^{1/2}.$$
Hence, the proof is complete.% Thus, the proof is completed by rearranging.
\end{proof}

\begin{proof}[Proof of Theorem \ref{thm_ateiden}]
    The proof follows from Theorem \ref{thm_main} and the fact that under Assumption \ref{assump_ateiden} $\Delta(x)$ is either non-positive for all $x$ or non-negative for all $x$.
\end{proof}

\section{Proofs of the results in Section \ref{sec_inference}} \label{sec_proofs_asymp}
We first prove the following lemma regarding the asymptotic joint distribution of the estimators.
\begin{lemma}\label{lemma_jointCLT}
Under the technical assumptions \ref{assm_technical}, further assuming that $\sigma_k^2 = \int u^2 k(u) du > 0$ and that
$h_i / h \rightarrow \alpha_i > 0$
for some $\alpha_i>0$, $i=2,3,4$, then for the estimators $\widehat{m_1}({x})$ and $\widehat{m_0}({x})$ obtained in the unconstrained version of \eqref{eq_betax} and $\widehat{\sigma^2}({x})$ obtained in \eqref{eq_sigma2x}, we have the following asymptotic joint distribution:
\begin{equation}
    \begin{split}
        &\sqrt{n h^d}
        \left( \begin{array}{c}
         \widehat{m_1}({x}) - m_1({x}) - \frac{1}{2} \alpha_3^2 h^2 \sigma_k^2 tr(\Ddot{m}_1({x})) + o_p(h^2)  \\
         \widehat{m_0}({x}) - m_0({x}) - \frac{1}{2} \alpha_4^2 h^2 \sigma_k^2 tr(\Ddot{m}_0({x})) + o_p(h^2) \\
         \widehat{\sigma^2}({x}) - \sigma^2({x}) - \frac{1}{2} \alpha_2^2 h^2 \sigma_k^2 tr(\Ddot{(\sigma^2)}({x})) + o_p(h^2) 
    \end{array} \right) \\
    \overset{d}{\to}& {N}_3 
    \left( {0}_3, 
    \left(\begin{array}{ccc}
         \frac{\theta_{K}^d \nu_1^2({x})}{\alpha_3^{d} \pi(x) f_1({x})} & 0 & \frac{\theta_{K}^d \nu_1({x}) \sigma^2({x}) \eta_1({x}) }{\alpha_2^{d/2}\alpha_3^{d/2} f({x})} \\
         0 & \frac{\theta_{K}^d \nu_0^2({x})}{\alpha_4^{d} (1-\pi(x)) f_0({x})} & \frac{\theta_{K}^d \nu_0({x}) \sigma^2({x}) \eta_0({x}) }{\alpha_2^{d/2}\alpha_4^{d/2} f({x})} \\
         \frac{\theta_{K}^d \nu_1({x}) \sigma^2({x}) \eta_1({x}) }{\alpha_2^{d/2}\alpha_3^{d/2} f({x})} & \frac{\theta_{K}^d \nu_0({x}) \sigma^2({x}) \eta_0({x}) }{\alpha_2^{d/2}\alpha_4^{d/2} f({x})} & \frac{\theta_{K}^d \sigma^4({x}) \lambda^2({x}) }{\alpha_2^{d} f({x})}
    \end{array} \right)
    \right),
    \end{split}
\end{equation}
as $n \rightarrow \infty$, where $m_z({x}) = E \left( Y_i \mid Z_i = z, {X}_i = {x} \right)$, $\nu_z^2({x}) = Var \left( Y_i \mid  Z_i = z, {X}_i = {x} \right)$, 
$\xi_{1i} = (Y_i - m_1({X}_i))/\nu_1({X}_i)$, 
$\eta_z({x}) = E(\xi_{0} (\epsilon^2 - 1) \mid  Z_i = z, {X}_i = {x})$
for $z=0,1$, and
$\lambda^2({x}) = E \left\{ (\epsilon_i^2 - 1)^2 \mid {X}_i = {x} \right\}$.
\end{lemma}

\begin{proof}
To solve the unconstrained optimization problem
\begin{equation*}
\begin{split}
(\hat{\mu}_0, \hat{{\zeta}}_0, \hat{\mu}_1,\hat{{\zeta}}_1) = &
\arg\min_{\mu_0,{\zeta}_0,\mu_1,{\zeta}_1}
\sum_{i=1}^n Z_i \lVert Y_i - \mu_1-{\zeta}_1^\top({X}_i-{x}) \rVert^2 {K}\left({H}_3^{-1}({X}_i-{x})\right)\\
& \qquad\qquad + 
\sum_{i=1}^n (1-Z_i) \lVert Y_i - \mu_0-{\zeta}_0^\top({X}_i-{x}) \rVert^2 {K}\left({H}_4^{-1}({X}_i-{x})\right),
\end{split}
\end{equation*}
where $H_3 = h_3 {I}_{d}$ and $H_4 = h_4 {I}_{d}$,
it is equivalent to solve the following two problems separately,
\begin{equation}\label{eq_Z=1}
(\hat{\mu}_1,\hat{{\zeta}}_1^\top) = 
\arg\min_{\mu_1,{\zeta}_1}
\sum_{i=1}^n Z_i \lVert Y_i - \mu_1-{\zeta}_1^\top({X}_i-{x}) \rVert^2 {K}\left({H}_3^{-1}({X}_i-{x})\right),
\end{equation}
and
\begin{equation}\label{eq_Z=0}
(\hat{\mu}_0, \hat{{\zeta}}_0^\top) =  \arg\min_{\mu_0,{\zeta}_0}
\sum_{i=1}^n (1-Z_i) \lVert Y_i - \mu_0-{\zeta}_0^\top({X}_i-{x}) \rVert^2 {K}\left({H}_4^{-1}({X}_i-{x})\right).
\end{equation}

Denote ${e}_1 = (1, 0, \dots, 0)^\top \in \mathbb{R}^{d+1}$, %${\gamma}_z = (\mu_z, {\zeta}_z^\top)^\top \in \mathbb{R}^{d+1}$, for $z = 0,1$, 
$${\Omega}_1({x}) = \text{diag} \left(Z_1 {K}\left({H}_3^{-1}({X}_1-{x})\right), \dots, Z_n {K}\left({H}_3^{-1}({X}_n-{x})\right)  \right) \in \mathbb{R}^{n \times n},$$
$${\Omega}_0({x}) = \text{diag} \left((1-Z_1) {K}\left({H}_4^{-1}({X}_1-{x})\right), \dots, (1-Z_n) {K}\left({H}_4^{-1}({X}_n-{x})\right) \right)\in \mathbb{R}^{n \times n},$$
and
\begin{equation*}
    {\Gamma}({x}) = \left( \begin{array}{cc}
         1 & ({X}_1 - {x})^\top  \\
         \dots & \dots \\
         1 & ({X}_n - {x})^\top
    \end{array} \right) \in \mathbb{R}^{n \times (d+1)}.
\end{equation*}

Let's look at \eqref{eq_Z=1} first. 
The standard procedure gives us the solution to \eqref{eq_Z=1}:
\begin{equation*}
    (\hat{\mu}_1,\hat{{\zeta}}_1^\top)^\top = \left[ {\Gamma}({x})^\top {\Omega}_1({x}) {\Gamma}({x})  \right]^{-1} {\Gamma}({x})^\top {\Omega}_1({x}) {Y},
    % \hat{{\gamma}}_1({x}) = \left[ {\Gamma}({x})^\top {\Omega}_1({x}) {\Gamma}({x})  \right]^{-1} {\Gamma}({x})^\top {\Omega}_1({x}) {Y},
\end{equation*}
where ${Y} = (Y_1, \dots, Y_n)^\top \in \mathbb{R}^n.$
And we will be specifically interested in
\begin{equation*}
    \widehat{m_1}({x}) = \hat{\mu}_1 = {e}_1^\top \left[ {\Gamma}({x})^\top {\Omega}_1({x}) {\Gamma}({x})  \right]^{-1} {\Gamma}({x})^\top {\Omega}_1({x}) {Y}.
\end{equation*}
Apply Taylor expansion to $m_1({X}_i)$ at ${x}$ for each $i$,
% \begin{equation*}
% \begin{split}
%     Y_i &= m_1({X}_i) + \nu_1({X}_i) \xi_{1i} \\
%     &= m_1({x}) + \Dot{m}_1({x})^\top ({X}_i-{x}) + \frac{1}{2} ({X}_i-{x})^\top \Ddot{m}_1({x}) ({X}_i-{x}) \\
%     &+ o(\lVert {X}_i-{x} \rVert^2) + \nu_1({X}_i) \xi_{1i},
% \end{split}
% \end{equation*}
% where $\Dot{m}_1({x})$ and $\Ddot{m}_1({x})$ are the first and second derivatives of $m({x})$, respectively.
% If we further denote
% \begin{equation*}
%     o(\lVert {X}-{x} \rVert^2) = \left( o(\lVert {X}_1-{x} \rVert^2), \dots, o(\lVert {X}_n-{x} \rVert^2) \right)^\top,
% \end{equation*}
the vector ${Y}$ can now be expressed as
\begin{equation*}
\begin{split}
    {Y} &= {\Gamma}({x}) \left( m_1({x}), \Dot{m}_1({x})^\top \right)^\top + \frac{1}{2}
    \left( ({X}_i-{x})^\top \Ddot{m}_1({x}) ({X}_i-{x}) \right)_{n \times 1} \\
    &+ \left( o(\lVert {X}_i-{x} \rVert^2) + \nu_1({X}_i) \xi_{1i} \right)_{n \times 1}\\
    &:= {\Gamma}({x}) \left( m_1({x}), \Dot{m}_1({x})^\top \right)^\top + \frac{1}{2} {Q}({x})
    + {R}({x}) + \left( \nu_1({X}_i) \xi_{1i} \right)_{n \times 1},
\end{split}
\end{equation*}
in which a notation of the form $(a_i)_{n \times 1}$ stands for a column vector
$(a_1, \dots. a_n) \in \mathbb{R}^n.$
Since 
\begin{equation*}
    m_1({x}) = {e}_1\top \left[ {\Gamma}({x})^\top {\Omega}_1({x}) {\Gamma}({x}) \right]^{-1} \left[ {\Gamma}({x})^\top {\Omega}_1({x}) {\Gamma}({x}) \right] \left( m_1({x}), \Dot{m}_1({x})^\top \right)^\top,
\end{equation*}
we now are interested in 
\begin{equation*}
\begin{split}
    \widehat{m_1}({x}) - m_1({x}) = {e}_1\top \left[ {\Gamma}({x})^\top {\Omega}_1({x}) {\Gamma}({x}) \right]^{-1} {\Gamma}({x})^\top {\Omega}_1({x}) 
    \left( \frac{1}{2}{Q}({x}) + {R}({x}) + (\nu_1({X}_i) \xi_{1i})_{n \times 1} \right).
\end{split}
\end{equation*}
    
Using standard results from density estimation, 
\begin{equation*}
\begin{split}
    \frac{1}{n h_3^d} {\Gamma}({x})^\top {\Omega}_1({x}) {\Gamma}({x}) 
    % =& \left(
    % \begin{array}{cc}
    % \frac{1}{n h_3^d}\sum Z_i {K}\left({H}_3^{-1}({X}_1-{x})\right) & \frac{1}{n h_3^d}\sum Z_i {K}\left({H}_3^{-1}({X}_1-{x})\right) ({X}_i-{x})^\top \\
    % \frac{1}{n h_3^d}\sum Z_i {K}\left({H}_3^{-1}({X}_1-{x})\right) ({X}_i-{x}) & \frac{1}{n h_3^d}\sum Z_i {K}\left({H}_3^{-1}({X}_1-{x})\right) ({X}_i-{x}) ({X}_i-{x})^\top
    % \end{array} \right) \\
    = \left(
    \begin{array}{cc}
    \pi(x) f_1({x}) + o_p(1) & \pi(x) h_3^2 \sigma_k^2 \dot{f_1}({x})^\top + o_p(h^2) {1}_d^\top  \\
        \pi(x) h_3^2 \sigma_k^2 \dot{f_1}({x}) + o_p(h^2) {1}_d & \pi(x) h_3^2 \sigma_k^2  f_1({x}) {I}_{d} + o_p(h^2) {I}_{d}
    \end{array} \right).
\end{split}
\end{equation*}
It follows from this that
\begin{equation}\label{inverse}
    \begin{split}
        &\left[ \frac{1}{n h_3^d} {\Gamma}({x})^\top {\Omega}_1({x}) {\Gamma}({x}) \right]^{-1} \\ 
    =& \left(
    \begin{array}{cc}
    (\pi(x) f_1({x}))^{-1} + o_p(1) & - \pi(x)^{-1} (f_1({x}))^{-2}\dot{f_1}({x})^\top + o_p(1) {1}_d^\top  \\
    - \pi(x)^{-1} (f_1({x}))^{-2} \dot{f_1}({x}) + o_p(1) {1}_d & (\pi(x) h_3^2 \sigma_k^2 f_1({x}) )^{-1} {I}_{d} + o_p(h_3^{-2}) {I}_{d}
    \end{array} \right).
    \end{split}
\end{equation}

We then look at the ${\Gamma}({x})^\top {\Omega}_1({x}) 
    \left( \frac{1}{2}{Q}({x}) + {R}({x}) + (\nu_1({X}_i) \xi_{1i})_{n \times 1} \right)$ part.
Firstly, we have
\begin{equation}\label{bias}
    \begin{split}
        &\frac{1}{n h_3^d} {\Gamma}({x})^\top {\Omega}_1({x}) {Q}({x})\\
        =& \frac{1}{n h_3^d}
        \left(
        \begin{array}{c}
        \sum_{i=1}^n Z_i {K}\left({H}_3^{-1}({X}_i-{x})\right) ({X}_i-{x})^\top \Ddot{m}_1({x}) ({X}_i-{x}) \\
        \sum_{i=1}^n Z_i        {K}\left({H}_3^{-1}({X}_i-{x})\right) ({X}_i - {x}) ({X}_i-{x})^\top \Ddot{m}_1({x}) ({X}_i-{x})
    \end{array} \right).
    \end{split}
\end{equation}
In \eqref{bias},
\begin{equation*}
    \begin{split}
         &\frac{1}{n h_3^d} \sum_{i=1}^n Z_i {K}\left({H}_3^{-1}({X}_i-{x})\right) ({X}_i-{x})^\top \Ddot{m}_1({x}) ({X}_i-{x}) \\
         \rightarrow& \pi(x) E_{X \mid Z = 1} \left[ \frac{1}{h_3^d} {K}\left({H}_3^{-1}({X}_i-{x})\right) ({X}_i-{x})^\top \Ddot{m}_1({x}) ({X}_i-{x}) \right] \\
         =& \pi(x) \int \frac{1}{h_3^d} {K}\left({H}_3^{-1}({y}-{x})\right) ({y}-{x})^\top \Ddot{m}_1({x}) ({y}-{x}) f_1({y}) d {y} \\
         % =& \pi(x) \int \frac{1}{h_3^d} {K}\left({u}\right) h_3^2 {u}^\top \Ddot{m}_1({x}) {u} f_1(h_3 {u} + {x}) h_3^d d {u} \\
         =& \pi(x) h_3^2 \int {K}\left({u}\right) {u}^\top \Ddot{m}_1({x}) {u} f_1(h_3 {u} + {x}) d {u}\\
         =& \pi(x) h_3^2 \int {K}\left({u}\right) {u}^\top \Ddot{m}_1({x}) {u} \left( f_1({x}) + h_3 \dot{f}_1 ({x}) {u} + o(h_3) \right) d {u} \\
         =& \pi(x) h_3^2 \sigma_k^2 tr(\Ddot{m}({x})) f_1({x}) + o(h_3^3).
    \end{split}
\end{equation*}
Similarly, since $\int u^4 K(u) du$ is assumed to be finite, 
\begin{equation*}
    \begin{split}
         &\frac{1}{n h_3^d} \sum_{i=1}^n Z_i {K}\left({H}_3^{-1}({X}_i-{x})\right) ({X}_i-{x}) ({X}_i-{x})^\top \Ddot{m}_1({x}) ({X}_i-{x}) \\
         \rightarrow& \pi(x) E_{X \mid Z = 1} \left[ \frac{1}{h_3^d} {K}\left({H}_3^{-1}({X}_i-{x})\right) ({X}_i-{x}) ({X}_i-{x})^\top \Ddot{m}_1({x}) ({X}_i-{x}) \right] \\
         =& \pi(x) \int \frac{1}{h_3^d} {K}\left({H}_3^{-1}({y}-{x})\right) ({y}-{x}) ({y}-{x})^\top \Ddot{m}_1({x}) ({y}-{x}) f_1({y}) d {y} \\
         % =& \pi(x) \int \frac{1}{h_3^d} {K}\left({u}\right) h_3 {u} h_3^2 {u}^\top \Ddot{m}_1({x}) {u} f_1(h_3 {u} + {x}) h_3^d d {u} \\
         % =& \pi(x) h_3^3 \int {K}\left({u}\right) {u} {u}^\top \Ddot{m}_1({x}) {u} f_1(h_3 {u} + {x}) d {u}\\
         % =& \pi(x) h_3^3 \int {K}\left({u}\right) {u} {u}^\top \Ddot{m}_1({x}) {u} \left( f_1({x}) + h_3 \dot{f}_1 ({x}) {u} + o(h_3) \right) d {u} \\
         =& o(h_3^3) {1}_{d}.
    \end{split}
\end{equation*}
Thus, \eqref{bias} becomes
\begin{equation}\label{bias2}
    \begin{split}
        \frac{1}{n h_3^d} {\Gamma}({x})^\top {\Omega}_1({x}) {Q}({x})
        % =& \frac{1}{n h_3^d}
        % \left(
        % \begin{array}{c}
        % \sum_{i=1}^n Z_i {K}\left({H}_3^{-1}({X}_i-{x})\right) ({X}_i-{x})^\top \Ddot{m}_1({x}) ({X}_i-{x}) \\
        % \sum_{i=1}^n Z_i        {K}\left({H}_3^{-1}({X}_i-{x})\right) ({X}_i - {x}) ({X}_i-{x})^\top \Ddot{m}_1({x}) ({X}_i-{x})
    % \end{array} \right) \\
        =
        \left(
        \begin{array}{c}
        ph_3^2 \sigma_k^2 tr(\Ddot{m}_1({x})) f_1({x}) + o(h_3^3)  \\
        o(h_3^3) {1}_{d}
    \end{array} \right) .
    \end{split}
\end{equation}
Secondly,
\begin{equation}\label{main.term}
    \begin{split}
        &\frac{1}{n h_3^d} {\Gamma}({x})^\top {\Omega}_1({x}) \left( \nu_1({X}_i) \xi_{1i} \right)_{n \times 1}\\
        =& \frac{1}{n h_3^d}
        \left(
        \begin{array}{c}
        \sum_{i=1}^n Z_i {K}\left({H}_3^{-1}({X}_i-{x})\right) \nu_1({X}_i) \xi_{1i} \\
        \sum_{i=1}^n Z_i        {K}\left({H}_3^{-1}({X}_i-{x})\right) ({X}_i - {x}) \nu_1({X}_i) \xi_{1i}
    \end{array} \right).
    \end{split}
\end{equation}
For \eqref{main.term}, first recall that $E \left(\xi_{1i} \mid  Z_i = 1, {X}_i \right) = 0$ 
and $Var \left(\xi_{1i} \mid  Z_i = 1, {X}_i \right) = 1$. 
So, 
\begin{equation}
    \begin{split}
        &E \left[ Z_i {K}\left({H}_3^{-1}({X}_i-{x})\right) \nu_1({X}_i) \xi_{1i} \right] \\
        =& \pi(x) E_{X \mid Z = 1} \left[ {K}\left({H}_3^{-1}({X}_i-{x})\right) \nu_1({X}_i) E( \xi_{1i} \mid  Z_i = 1, {X}_i) \right] = 0,
    \end{split}
\end{equation}
and, denoting $\theta_{K}^d = \int K^2(u) du$,
\begin{equation*}
    \begin{split}
        &E \left[ ( Z_i {K}\left({H}_3^{-1}({X}_i-{x})\right) \nu_1({X}_i) \xi_{1i} )^2 \right] \\
        % =& \pi(x) E_{X \mid Z = 1} \left[ {K}^2\left({H}_3^{-1}({X}_i-{x})\right) \nu_1^2({X}_i) E( \xi_{1i}^2 \mid {X}_i, Z_i=1 ) \right] \\
        =& \pi(x) E_{X \mid Z = 1} \left[ {K}^2\left({H}_3^{-1}({X}_i-{x})\right) \nu_1^2({X}_i) \right] \\
        =& \pi(x) \int {K}^2\left({H}_3^{-1}({y}-{x})\right) \nu_1^2({y}) f_1({y}) d {y} \\
        % =& \pi(x) \int {K}^2\left( {u} \right) \nu_1^2(h_3 {u} + {x}) f_1(h_3 {u} + {x}) h_3^d d {u} \\
        % =& \pi(x) h_3^d \int {K}^2\left( {u} \right) \left( \nu_1^2({x}) + h_3 \dot{(\nu_1^2)} ({x}) {u} + o(h_3) \right) \left( f_1({x}) + h_3 \dot{f}_1 ({x}) {u} + o(h_3) \right) d {u} \\
        =& \pi(x) h_3^d \left( \theta_{K}^d \nu_1^2({x}) f_1({x}) + o(1) \right).
    \end{split}
\end{equation*}
We also have
\begin{equation*}
    \begin{split}
        &E \left[ Z_i {K}\left({H}_3^{-1}({X}_i-{x})\right) ({X}_i - {x}) \nu_1({X}_i) \xi_{1i} \right] \\
        =& \pi(x) E_{X \mid Z = 1} \left[ {K}\left({H}_3^{-1}({X}_i-{x})\right) ({X}_i - {x}) \nu_1({X}_i) E( \xi_{1i} \mid  Z_i = 1, {X}_i) \right] = {0}_{d},
    \end{split}
\end{equation*}
and, for a vector $a$, denoting $a^2 = a^\top a$,
\begin{equation*}
    \begin{split}
        &E \left[ ( Z_i {K}\left({H}_3^{-1}({X}_i-{x})\right) ({X}_i - {x})  \nu_1({X}_i) \xi_{1i} )^2 \right] \\
        % =& \pi(x) E_{X \mid Z = 1} \left[ {K}^2\left({H}_3^{-1}({X}_i-{x})\right) ({X}_i - {x})^2 \nu_1^2({X}_i) E( \xi_{1i}^2 \mid {X}_i, Z_i=1 ) \right] \\
        =& \pi(x) E_{X \mid Z = 1} \left[ {K}^2\left({H}_3^{-1}({X}_i-{x})\right) ({X}_i - {x})^2 \nu_1^2({X}_i) \right] \\
        =& \pi(x) \int {K}^2\left({H}_3^{-1}({y}-{x})\right) ({y} - {x})^2 \nu_1^2({y}) f_1({y}) d {y} \\
        % =& \pi(x) \int {K}^2\left( {u} \right) h_3^2 {u}^2 \nu_1^2(h_3 {u} + {x}) f_1(h_3 {u} + {x}) h_3^d d {u} \\
        % =& \pi(x) h_3^{d+2} \int {K}^2\left( {u} \right) {u}^2 \left( \nu_1^2({x}) + h_3 \dot{(\nu_1^2)} ({x}) {u} + o(h_3) \right) \left( f_1({x}) + h_3 \dot{f}_1 ({x}) {u} + o(h_3) \right) d {u} \\
        =& h_3^{d} o(h_3) {1}_{d} .
    \end{split}
\end{equation*}
Applying the central limit theorem to \eqref{main.term}, as $n \rightarrow \infty$, we get
\begin{equation}\label{main.term2.1}
    \begin{split}
        \sqrt{n h_3^d} \left( \frac{1}{n h_3^d} \sum_{i=1}^n Z_i {K}\left({H}_3^{-1}({X}_i-{x})\right) \nu_1({X}_i) \xi_{1i} \right) \overset{d}{\to} N \left(0,  \theta_{K}^d \nu_1^2({x}) f_1({x})\right),
    \end{split}
\end{equation}
and
\begin{equation}\label{main.term2.2}
    \begin{split}
        \sqrt{n h_3^d} \left( \frac{1}{n h_3^d} \sum_{i=1}^n Z_i {K}\left({H}_3^{-1}({X}_i-{x})\right)({X}_i-{x}) \nu_1({X}_i) \xi_{1i} \right) \overset{p}{\to} {0}_d,
    \end{split}
\end{equation}
since the asymptotic variance is $0$.

As for the remainder term ${R}({x})$,
\begin{equation}\label{remainder}
    \begin{split}
        & \frac{1}{n h_3^d}{\Gamma}({x})^\top {\Omega}_1({x})  {R}({x}) \\
        = & \frac{1}{n h_3^d}
        \left(
        \begin{array}{c}
        \sum_{i=1}^n Z_i {K}\left({H}_3^{-1}({X}_i-{x})\right) o(\lVert {X}_i - {x} \rVert^2) \\
        \sum_{i=1}^n Z_i        {K}\left({H}_3^{-1}({X}_i-{x})\right) ({X}_i - {x}) o(\lVert {X}_i - {x} \rVert^2) 
    \end{array} \right)\\
    =& o(h_3^2) {1}_{d+1}.
    \end{split}
\end{equation}

Combining \eqref{inverse}, \eqref{bias2} and \eqref{remainder},
\begin{equation*}
\begin{split}
     \widehat{m_1}({x}) - m_1({x}) =& {e}_1\top \left[  {\Gamma}({x})^\top {\Omega}_1({x}) {\Gamma}({x}) \right]^{-1} {\Gamma}({x})^\top {\Omega}_1({x}) 
    \left( \frac{1}{2}{Q}({x}) + {R}({x}) + (\nu_1({X}_i) \xi_{1i})_{n \times 1} \right)\\
    % =&  \left( pf_1({x}))^{-1} + o_p(1)  \right) 
    % \left( \frac{1}{2} ph_3^2 \sigma_k^2 tr(\Ddot{m}_1({x})) f_1({x}) + o(h_3^2)   \right) \\
    % +& \frac{1}{n h_3^d} \left( pf_1({x}))^{-1} + o_p(1)  \right) 
    % \left(  \sum_{i=1}^n Z_i {K}\left({H}_3^{-1}({X}_i-{x})\right) \nu_1({X}_i) \xi_{1i} \right) \\
    % +&  \left( - ^{-1} (f_1({x}))^{-2} \left( \frac{\partial f_1({x})}{\partial {x}}\right)^\top + o_p(1) {1}_d^\top  \right) 
    % \left( o(h_3^2) {1}_{d} \right) \\
    % +& \frac{1}{n h_3^d} \left(  - ^{-1} (f_1({x}))^{-2} \left( \frac{\partial f_1({x})}{\partial {x}}\right)^\top + o_p(1) {1}_d^\top \right)\\
    % \times&
    % \left(  \sum_{i=1}^n Z_i {K}\left({H}_3^{-1}({X}_i-{x})\right)({X}_i-{x}) \nu_1({X}_i) \xi_{1i} \right) \\
    =& \frac{1}{n h_3^d \pi(x) f_1({x})}  
    \left(  \sum_{i=1}^n Z_i {K}\left({H}_3^{-1}({X}_i-{x})\right) \nu_1({X}_i) \xi_{1i} \right) \\
    +& \frac{1}{n h_3^d  f_1^2({x})} \left( \frac{\partial f_1({x})}{\partial {x}}\right)^\top 
    \left(  \sum_{i=1}^n Z_i {K}\left({H}_3^{-1}({X}_i-{x})\right)({X}_i-{x}) \nu_1({X}_i) \xi_{1i} \right) \\
    +& \frac{1}{2} h_3^2 \sigma_k^2 tr(\Ddot{m}_1({x})) + o_p(h_3^2).
\end{split}
\end{equation*}
Further, together with \eqref{main.term2.1} and \eqref{main.term2.2},
\begin{equation*}
\begin{split}
    \sqrt{n h_3^d} \left( \widehat{m_1}({x}) - m_1({x}) - \frac{1}{2} h_3^2 \sigma_k^2 tr(\Ddot{m}_1({x})) + o_p(h_3^2) \right) 
    % =& \frac{\sqrt{n h_3^d}}{n h_3^d pf_1({x})}  
    % \left(  \sum_{i=1}^n Z_i {K}\left({H}_3^{-1}({X}_i-{x})\right) \nu_1({X}_i) \xi_{1i} \right) \\
    % +& \frac{\sqrt{n h_3^d}s}{n h_3^d  f_1^2({x})} \left( \frac{\partial f_1({x})}{\partial {x}}\right)^\top 
    % \left(  \sum_{i=1}^n Z_i {K}\left({H}_3^{-1}({X}_i-{x})\right)({X}_i-{x}) \nu_1({X}_i) \xi_{1i} \right) \\
    \overset{d}{\to} N \left(0, \frac{\theta_{K}^d \nu_1^2({x})}{ f_1({x})} \right) \text{, as } n \rightarrow \infty.
\end{split}
\end{equation*}

Similarly, for the solution $\widehat{m_0}({x}) := \hat{\mu}_0$ to the problem \eqref{eq_Z=0}, we have the following asymptotic distribution:
\begin{equation*}
\begin{split}
    \sqrt{n h_4^d} \left( \widehat{m_0}({x}) - m_0({x}) - \frac{1}{2} h_4^2 \sigma_k^2 tr(\Ddot{m}_0({x})) + o_p(h_4^2) \right) 
    \overset{d}{\to} N \left(0, \frac{\theta_{K}^d \nu_0^2({x})}{(1-\pi(x)) f_0({x})} \right) \text{, as } n \rightarrow \infty.
\end{split}
\end{equation*}

Moreover, if we look at the covariance between $\widehat{m_0}({x})$ and $\widehat{m_1}({x})$, since the covariance between the main terms are zero:
\begin{equation*}
    \begin{split}
        &cov \left( Z_i {K}\left({H}_3^{-1}({X}_i-{x})\right) \nu_1({X}_i) \xi_{1i}, (1-Z_i) {K}\left({H}_4^{-1}({X}_i-{x})\right) \nu_0({X}_i) \xi_{0i} \right) \\
        =& E \left( Z_i {K}\left({H}_3^{-1}({X}_i-{x})\right) \nu_1({X}_i) \xi_{1i} \times (1-Z_i) {K}\left({H}_4^{-1}({X}_i-{x})\right) \nu_0({X}_i) \xi_{0i} \right) \\
        -& E \left( Z_i {K}\left({H}_3^{-1}({X}_i-{x})\right) \nu_1({X}_i) \xi_{1i} \right) E \left( (1-Z_i) {K}\left({H}_4^{-1}({X}_i-{x})\right) \nu_0({X}_i) \xi_{0i} \right) \\
        =& 0,
    \end{split}
\end{equation*}
it follows that, asymptotically,
\begin{equation*}
    cov(\widehat{m_0}({x}), \widehat{m_1}({x})) = 0.
\end{equation*}

Fan and Yao (1998) derived the asymptotic distribution of $\widehat{\sigma^2}({x})$ in \eqref{eq_sigma2x}:
\begin{equation}
    \begin{split}
        \sqrt{n h_2^d} \left( \widehat{\sigma^2}({x}) - \sigma^2({x}) - \frac{1}{2} h_2^2 \sigma_k^2 tr(\Ddot{(\sigma^2)}({x})) + o_p(h_1^2+h_2^2)  \right) 
        \overset{d}{\to} N \left(0, \frac{\theta_{K}^d \sigma^4({x}) \lambda^2({x}) }{f({x})} \right),
    \end{split}
\end{equation}
as $n \rightarrow \infty$.

Besides the asymptotic distribution of $\widehat{\sigma^2}({x})$ itself, we are further interested in the asymptotic joint distribution of it and the other two estimators.
According to the proof in Fan and Yao (1998), the donimating term of $\widehat{\sigma^2}({x})$ is
\begin{equation*}
    \frac{1}{n h_2^d f(x)} \sum_{i=1}^n  {K}\left({H}_2^{-1}({X}_i-{x})\right) \sigma^2({X}_i) (\epsilon_i^2 - 1).
\end{equation*}
Now we calculate the covariances. Firstly,
\begin{equation*}
    \begin{split}
        &cov \left( Z_i {K}\left({H}_3^{-1}({X}_i-{x})\right) \nu_1({X}_i) \xi_{1i}, {K}\left({H}_2^{-1}({X}_i-{x})\right) \sigma^2({X}_i) (\epsilon_i^2 - 1) \right) \\
        =& E \left( Z_i {K}\left({H}_3^{-1}({X}_i-{x})\right) \nu_1({X}_i) \xi_{1i} \times {K}\left({H}_2^{-1}({X}_i-{x})\right) \sigma^2({X}_i) (\epsilon_i^2 - 1) \right) 
        - 0 \\
        =&  E_{X\mid Z=1} \left[  {K}\left({H}_3^{-1}({X}_i-{x})\right) {K}\left({H}_2^{-1}({X}_i-{x})\right) \nu_1({X}_i) \sigma^2({X}_i) E (\xi_{1i} (\epsilon_i^2 - 1) \mid  Z_i = 1, {X}_i ) \right] \\
        =&  \int {K}\left({H}_3^{-1}({y}-{x})\right) {K}\left({H}_2^{-1}({y}-{x})\right) \nu_1({y}) \sigma^2({y}) E (\xi_{1} (\epsilon^2 - 1) \mid Z_i = 1, {X}_i = {y}  ) f_1({y}) d{y} \\
        % =&  \int {K}\left({u}\right) {K}\left( \frac{h_2}{h_3}{u} \right) \nu_1(h_3 {u} + {x}) \sigma^2(h_3 {u} + {x}) E (\xi_{1} (\epsilon^2 - 1) \mid {X}_i= h_3 {u} + {x},  Z_i = 1 ) f_1(h_3 {u} + {x}) h_3^d d{u} \\
        =&  h_3^d \left( \nu_1({x}) \sigma^2({x}) \eta_1({x}) f_1({x}) \left( \int K({u}) K\left(\frac{h_2}{h_3}{u}\right) d {u} \right)^d + o(1) \right),
    \end{split}
\end{equation*}
where $\eta_1({x}) = E(\xi_{1i} (\epsilon_i^2 - 1) \mid  Z_i = 1, {X}_i = {x})$.
Similarly we have
\begin{equation}
    \begin{split}
        &cov \left( (1-Z_i) {K}\left({H}_4^{-1}({X}_i-{x})\right) \nu_0({X}_i) \xi_{0i}, {K}\left({H}_2^{-1}({X}_i-{x})\right) \sigma^2({X}_i) (\epsilon_i^2 - 1) \right) \\
        =& (1-) h_4^d \left( \nu_0({x}) \sigma^2({x}) \eta_0({x}) f_0({x}) \left( \int K({u}) K\left(\frac{h_2}{h_4}{u}\right) {u} \right)^d + o(1) \right),
    \end{split}
\end{equation}
where $\eta_0({x}) = E(\xi_{0} (\epsilon^2 - 1) \mid  Z_i = 0, {X}_i = {x})$.

Now we have proved the lemma.
\end{proof}

In order to prove Theorem \ref{thm_largesample}, we also need the next two lemmas.
Note that the constraint is equivalent to $\widehat{\Delta}(x) \geq 0$ and $\hat{\beta}_C^2(x) \geq \hat{s}(x) \equiv \frac{\widehat{\sigma^2}(x) - \widehat{\sigma_0^2}(x)}{\widehat{\pi}(x)(1-\widehat{\pi}(x))}$.

\begin{lemma}
    For a strictly convex function $f(x)$ if its  global minima $x^\star\not\in S$ where $S$ is a union of two closed half spaces, then $x^{\star\star}$ which minimizes $f(x)$ subject to $x\in S$ satisfies $x^{\star\star}\in \partial S$.
    ($\partial S$ denotes the boundary of $S$.)
\end{lemma}
\begin{proof}
    If possible suppose $x^{\star\star} \in S^o$, the interior of $S$. Then, $f(x) > f(x^{\star\star})$ for $x$ in a small open ball around $x^{\star\star}$.  Thus, $x^{\star\star}$ is a local minima. But, as $f$ is strictly convex, this implies that $x^{\star\star}$ must be a global minima. This produces a contradiction.
\end{proof}
\begin{lemma}
Suppose $\hat{\beta}^2_U(x) < \hat{s}(x)$. Then $\hat{\beta}^2_C(x) = \hat{s}(x)$, i.e., $\widehat{\Delta^2}(x)=0$.
\end{lemma}
\begin{proof}
    Because our optimization function is convex, the result follows from the above lemma.
\end{proof}

Hence we have shown that, $\widehat{\Delta^2}(x)=0$ if and only if $\hat{\beta}_U^2(x) \leq \hat{s}(x)$,
otherwise $\hat{\beta}_U(x) = \hat{\beta}_C(x)$.

\begin{proof}[Proof of Theorem \ref{thm_largesample}]
We prove part {(b)} first. 
When $E(Y_i(0)\mid Z_i=1, X_i=x) \neq E(Y_i(0)\mid Z_i=0, X_i=x)$, i.e., $\Delta^2(x) > 0$, by consistency of the estimators $\hat{\beta}_U(x)^2$ and $\hat{s}(x)$, there is a small enough $a>0$ so that
$$\Pr(\hat{\beta}_U(x)^2 > \hat{s}(x)+a) \longrightarrow 1.$$
Call the event inside the above probability as $A_n$ noting that $A_n$ says 
$\widehat{\Delta^2}(x) > a$ and hence $\Pr(A_n)\rightarrow 1$, and under $A_n$, $\hat{\beta}_C(x) = \hat{\beta}_U(x)$.

Now, for any function $f$ and set $B$ write $X_{nC} \equiv \sqrt{nh^d}( f(\hat{\beta}_C(x), \hat{s}(x))$ and $X_{nU} \equiv \sqrt{nh^d}( f(\hat{\beta}_U(x), \hat{s}(x))$.
\begin{align*}
    \Pr(X_{nC} \in B) & = \Pr(X_{nC}\in B\mid A_n)\Pr(A_n) +  \Pr(X_{nC}\in B\mid A_n^c)\Pr(A_{n}^c)\\
    & = \Pr(X_{nU}\in B\mid A_n)\Pr(A_n) +  \Pr(X_{nC}\in B\mid A_n^c)\Pr(A_{n}^c)\\
    & = \Pr(X_{nU}\in B) - \Pr(X_{nU}\in B\mid A_n^c)\Pr(A_n^c) +  \Pr(X_{nC}\in B\mid A_n^c)\Pr(A_{n}^c)
\end{align*}
We get,
\begin{align*}
    \Pr(X_{nC} \in B) - \Pr(X_{nU} \in B) | &\leq |(\Pr(X_{nC}\in B\mid A_n^c)-\Pr(X_{nU}\in B\mid A_n^c))\Pr(A_n^c) | \\ &\leq 2 \Pr(A_{n}^c).
\end{align*}| 
These calculations imply that when $\Delta^2(x) > 0$, 
$$\lim_{n\rightarrow\infty}  \Pr( \sqrt{nh^d}( f(\hat{\beta}_C(x), \hat{s}(x)) \in B) = 
\lim_{n\rightarrow\infty}  \Pr( \sqrt{nh^d}( f(\hat{\beta}_U(x), \hat{s}(x)) \in B),$$
for all measurable function $f$ and set $B$.

Then, using Delta-method and Lemma \ref{lemma_jointCLT}, we get the asymptotic distributions of both $\widehat{\Delta^2}(x)$ and $\widehat{\tau_\pm}(x)$.

Now we prove part (a). When $E(Y_i(0)\mid Z_i=1, X_i=x) = E(Y_i(0)\mid Z_i=0, X_i=x)$, i.e., $\Delta^2(x) = 0$, 
recall the fact that $\widehat{\Delta^2}(x)=0$ if and only if $\hat{\beta}_U^2(x) \leq \hat{s}(x)$,
otherwise $\hat{\beta}_U(x) = \hat{\beta}_C(x)$.

First, 
\begin{align*}
    \Pr(\sqrt{nh^d}\widehat{\Delta^2}(x) \leq 0) & = \Pr(\sqrt{nh^d}\widehat{\Delta^2}(x) = 0)\\
& = \Pr(\hat{\beta}^2_U(x) \leq \hat{s}(x)) = \Pr(\sqrt{nh^d}\widehat{\Delta^2}_U(x) \leq 0),
\end{align*}
where $\widehat{\Delta^2}_U(x) = \hat{\beta}_U^2(x) - \hat{s}(x)$.

Next, for $a>0$,
$$\Pr(\sqrt{nh^d}\widehat{\Delta^2}(x) \in (0, a)) = \Pr(\sqrt{nh^d}\widehat{\Delta^2}_U(x) \in (0, a)).$$
Finally, note that under $\Delta^2(x)=0$, using Lemma \ref{lemma_jointCLT} and the Delta method
$$\sqrt{nh^d}(\widehat{\Delta^2}_U(x) + O(h^2)) \longrightarrow N(0, v_{\Delta^2(x)}).$$ 
Hence, under the assumption that $nh^{d+4} \rightarrow 0$,
\begin{align*}
    \Pr(\sqrt{nh^d}\widehat{\Delta^2}(x) \leq a) & = \Pr(\sqrt{nh^d}\widehat{\Delta^2}(x) = 0) + \Pr(\sqrt{nh^d}\widehat{\Delta^2}(x) \in (0, a])\\ &
    =  \Pr(\sqrt{nh^d} (\widehat{\Delta^2}_U(x) + O(h^2)) \leq O(\sqrt{nh^{d+4}})) \\ &
    + \Pr(\sqrt{nh^d} (\widehat{\Delta^2}_U(x) + O(h^2)) \in (O(\sqrt{nh^{d+4}}), a+O(\sqrt{nh^{d+4}})])\\ &
    \longrightarrow  \frac12 + \left\{(\Phi(a/\sqrt{v_{\Delta^2(x)}}) - \frac12\right\}.
\end{align*}

Thus, assuming $nh^{d+4} \rightarrow 0$,
$$\sqrt{nh^d}\widehat{\Delta^2}(x) \longrightarrow \frac12 \delta_0 + \frac12 | N(0, v_{\Delta^2(x)})|.$$
Thus the proof is complete.\end{proof}

\begin{proof}[Proof of Theorem \ref{thm_ci}]
Note that, it is enough to show i) $\Pr( \widehat{\Delta^2}(x)  > \frac{\widehat{v_\Delta}(x)}{(nh^d)^{.5(1-\delta)}}) \rightarrow  0$ when $\Delta^2(x) = 0$ and 
ii) $\Pr( \widehat{\Delta^2}(x)  > \frac{\widehat{v_\Delta}(x)}{(nh^d)^{.5(1-\delta)}}) \rightarrow  1$ when $\Delta^2(x) > 0$.

For (i), when $\Delta^2(x) = 0$ we use part (a) of Theorem \ref{thm_largesample}. Rewrite,
$$\Pr( \widehat{\Delta^2}(x)  > \frac{\widehat{v_\Delta}(x)}{(nh^d)^{.5(1-\delta)}})= \Pr( \sqrt{nh^d} \widehat{\Delta^2}(x)/\widehat{v_\Delta}(x)  > (nh^d)^{\delta/2}).$$
Since $\delta >0$, $(nh^d)^{\delta/2}\rightarrow \infty$. Then, as  $\widehat{v_\Delta}(x) $ is a consistent estimator and $\sqrt{nh^d}h^2 \rightarrow 0$ we get the above probability goes in limit to the probability of a standard normal being infinitely large, which goes to 0.

For (ii), when $\Delta^2(x) > 0$ we use part (b) of Theorem \ref{thm_largesample}. Rewrite,
\begin{align*}
    &\Pr( \widehat{\Delta^2}(x) > \frac{\widehat{v_\Delta}(x)}{(nh^d)^{.5(1-\delta)}})\\ =& \Pr( \sqrt{nh^d} \{\widehat{\Delta^2}(x)-\Delta^2(x)\}/\widehat{v_\Delta}(x)  > (nh^d)^{\delta/2} - \sqrt{nh^d}\Delta^2(x)/\widehat{v_\Delta}(x)).
\end{align*}
Since $\delta < 1$, $(nh^d)^{\delta/2} - \sqrt{nh^d}\Delta^2(x)/\widehat{v_\Delta}(x)\rightarrow -\infty$ in probability. Then, as  $\widehat{v_\Delta}(x) $ is a consistent estimator and $\sqrt{nh^d}h^2 \rightarrow 0$ we get the above probability goes in limit to the probability of a standard normal being larger than negative infinity, which goes to 1.
\end{proof}

\section{Proofs of Example \ref{exam_1} and Remark \ref{rem_se}}\label{sec_se_consistency}

\begin{proof}[Proof of Example \ref{exam_1}]
Given $X_i=0$, then $\Pr(Z_i=1\mid X_i + U_i, X_i=0) = \Phi(U_i)$.
So $Pr(Z_i=1 \mid U_i=u) f(u) = \phi(u) \Phi(u)$, $Pr(Z_i=1) = 1/2$ and $f(u \mid Z_i=1) = 2\phi(u) \Phi(u)$.
Similarly $f(u \mid Z_i=0) = 2\phi(u) (1-\Phi(u))$.
Then $E(\lvert U_i \rvert \mid Z_i = 1, X_i = 0) - E(\lvert U_i \rvert \mid Z_i = 0, X_i = 0) = \int_{-\infty}^{\infty} 2 \lvert u \rvert \phi(u) (2\Phi(u) - 1) du = 0$ (integral of an odd function). 
%But, for example, if $Y_i(0) = X_i + U_i + \epsilon_i$, then $\beta(0) \neq 0$.

In general, given $X_i=x$, $\Pr(Z_i=1\mid X_i + U_i, X_i=x) = \Phi(U_i+x)$.
Then $Pr(Z_i=1 \mid X_i + U_i, X_i=x) f(u) = \phi(u) \Phi(u+x)$, $Pr(Z_i=1) = \Phi(x/\sqrt{2})$ and $f(u \mid Z_i=1) = \phi(u) \Phi(u+x) / \Phi(x/\sqrt{2})$.
Similarly, $f(u \mid Z_i=0) = \phi(u) (1-\Phi(u+x)) / (1-\Phi(x/\sqrt{2}))$. 
Let $V$ be a standard normal random variable that is independent of $U$.
\begin{align*}
    &\Phi(x/\sqrt{2}) E(\lvert U_i \rvert \mid Z_i = 1, X_i = x)  \\
    = & \int_{-\infty}^{\infty} |u| \phi(u) \Phi(u+x) du \\
    = & E_U (|U| \Phi(U+x)) \\
    = & E_{U, V} (|U| \mathbf{1}\{V \leq U+x\}) \\
    % = & \int_{-\infty}^{\infty} \int_{-\infty}^{u+x} |u| \phi(u) \phi(v) dv du \\
    = & \int_{-\infty}^{\infty} \int_{v-x}^{\infty} |u| \phi(u) \phi(v) du dv.
\end{align*}
Since $\int_{v-x}^{\infty} |u| \phi(u) du = \phi(v-x)$ if $v-x > 0$ and 
$\int_{v-x}^{\infty} |u| \phi(u) du = \sqrt{2/\pi} - \phi(v-x)$ if $v-x < 0$,
% and 
% $\int_{-\infty}^{\infty} \phi(v-x) \phi(v) dv = \phi(x/\sqrt{2}) / \sqrt{2}$,
% suppose for now $x > 0$, 
\begin{align*}
&\int_{-\infty}^{\infty} \int_{v-x}^{\infty} |u| \phi(u) \phi(v) du dv \\
=& \int_{x}^{\infty} \phi(v-x) \phi(v) dv + \int_{-\infty}^{x} (\sqrt{\frac{2}{\pi}} - \phi(v-x)) \phi(v) dv \\
=& \frac{1}{4 \sqrt{\pi}} exp(-\frac{x^2}{4}) erf(\frac{x}{2}) + \sqrt{\frac{2}{\pi}} - \frac{1}{4 \sqrt{\pi}} exp(-\frac{x^2}{4}) (erf(\frac{x}{2}) + 1) \\
=& \sqrt{\frac{2}{\pi}} \Phi(x) - \frac{1}{2 \sqrt{\pi}} exp(-\frac{x^2}{4}) erf(\frac{x}{2}).
\end{align*}

Similarly, $$(1-\Phi(x/\sqrt{2})) E(\lvert U_i \rvert \mid Z_i = 0, X_i = x) = \sqrt{\frac{2}{\pi}}(1-\Phi(x)) + \frac{1}{2 \sqrt{\pi}} exp(-\frac{x^2}{4}) erf(\frac{x}{2}).$$
Thus,
\begin{align*}
&\Delta(x) = E(\lvert U_i \rvert \mid Z_i = 1, X_i = x) - E(\lvert U_i \rvert \mid Z_i = 0, X_i = x) \\
= & - \frac{exp(-\frac{x^2}{4}) erf(\frac{x}{2})}{2 \sqrt{\pi} \Phi(x/\sqrt{2}) (1-\Phi(x/\sqrt{2}))}.
\end{align*}

\end{proof}

\begin{proof}[Proof of the consistency of \eqref{NW1}]
According to the standard kernel estimation theory, the denominator of \eqref{NW1}, as $n \rightarrow \infty$,
\begin{equation}\label{FourthMomentDenom}
    \frac{1}{nh_v^d}\sum_{i=1}^{n} {K}({H}_{v}^{-1}({X}_i - {x})) \overset{p}{\to} f({x}).
\end{equation}

As for the numerator of \eqref{NW1},
\begin{equation}\label{decomp}
    \begin{split}
        \frac{1}{nh_v^d}\sum_{i=1}^{n} {K}({H}_{v}^{-1}({X}_i - {x})) e_i^4 &= \frac{1}{nh_v^d}\sum_{i=1}^{n} {K}({H}_{v}^{-1}({X}_i - {x})) (Y_i - \hat{m}({X}_i))^4 \\
        &= \frac{1}{nh_v^d}\sum_{i=1}^{n} {K}({H}_{v}^{-1}({X}_i - {x})) (m({X}_i) - \hat{m}({X}_i) + \sigma({X}_i) \epsilon_i)^4 \\
        &= \frac{1}{nh_v^d}\sum_{i=1}^{n} {K}({H}_{v}^{-1}({X}_i - {x})) (m({X}_i) - \hat{m}({X}_i))^4 \\
        &+ \frac{1}{nh_v^d}\sum_{i=1}^{n} {K}({H}_{v}^{-1}({X}_i - {x})) 3(m({X}_i) - \hat{m}({X}_i))^3 (\sigma({X}_i) \epsilon_i) \\
        &+ \frac{1}{nh_v^d}\sum_{i=1}^{n} {K}({H}_{v}^{-1}({X}_i - {x})) 6(m({X}_i) - \hat{m}({X}_i))^2 (\sigma({X}_i) \epsilon_i)^2 \\
        &+ \frac{1}{nh_v^d}\sum_{i=1}^{n} {K}({H}_{v}^{-1}({X}_i - {x})) 3(m({X}_i) - \hat{m}({X}_i))(\sigma({X}_i) \epsilon_i)^3 \\
        &+ \frac{1}{nh_v^d}\sum_{i=1}^{n} {K}({H}_{v}^{-1}({X}_i - {x})) (\sigma({X}_i) \epsilon_i)^4.
    \end{split}
\end{equation}

Look at the last term first.
Denote $g({x}) = E((\sigma({X}_i) \epsilon_i)^4 \mid {X}_i = {x})$ and $G({x}) = E((\sigma({X}_i) \epsilon_i)^8 \mid {X}_i = {x})$.
Standard techniques gives the following results:
\begin{equation*}
    \begin{split}
        E\left(\frac{1}{h_v^d} {K}({H}_{v}^{-1}({X}_i - {x})) (\sigma({X}_i) \epsilon_i)^4\right) &= E\left( \frac{1}{h_v^d} {K}({H}_{v}^{-1}({X}_i - {x})) g({X}_i)\right) \\
        &= f({x}) g({x}) + h_v^2 \sigma_k^2 f({x}) \left(\frac{1}{2} tr(\Ddot{g}({x})) + tr(\Dot{f}^\top({x}) \Dot{g}({x}))\right) \\
        &+ o(h_v^2),
    \end{split}
\end{equation*}
and
\begin{equation*}
    \begin{split}
        E\left(\frac{1}{h_v^d} {K}({H}_{v}^{-1}({X}_i - {x})) (\sigma({X}_i) \epsilon_i)^4\right)^2 &= E\left( \frac{1}{h_v^{2d}} {K}^2({H}_{v}^{-1}({X}_i - {x})) G({X}_i)\right) \\
        &= \frac{1}{h_v^d} \theta_K f({x}) G({x}) + o(\frac{1}{h_v^d}).
    \end{split}
\end{equation*}
Hence, if as $n \rightarrow \infty$, $nh_v^d \rightarrow \infty$, we have the following results:
\begin{equation*}
    \begin{split}
        Var\left( \frac{1}{nh_v^d}\sum_{i=1}^{n} {K}({H}_{v}^{-1}({X}_i - {x})) (\sigma({X}_i) \epsilon_i)^4 \right) \rightarrow 0,
    \end{split}
\end{equation*}
and thus
\begin{equation*}
    \frac{1}{nh_v^d}\sum_{i=1}^{n} {K}({H}_{v}^{-1}({X}_i - {x})) (\sigma({X}_i) \epsilon_i)^4 \overset{p}{\to} f({x}) g({x}) = f({x}) E((\sigma({X}_i) \epsilon_i)^4 \mid {X}_i = {x}).
\end{equation*}

For the first term in \eqref{decomp}, we will show that 
\begin{equation*}
    \frac{1}{nh_v^d}\sum_{i=1}^{n} {K}({H}_{v}^{-1}({X}_i - {x})) (m({X}_i) - \hat{m}({X}_i))^4 \overset{p}{\to} 0.
\end{equation*}
By Markov Inequality, we only need to show that, as $n \rightarrow \infty$,
\begin{equation}\label{FourthMoment}
    E\left[\frac{1}{nh_v^d}\sum_{i=1}^{n} {K}({H}_{v}^{-1}({X}_i - {x})) (m({X}_i) - \hat{m}({X}_i))^4 \right] \to 0.
\end{equation}
Then, repeatedly using the Cauchy-Schwarz inequality, 
it is straightforward to show that the middle three terms are also of order $o_p(1)$. Hence the numerator
\begin{equation*}
    \frac{1}{nh_v^d}\sum_{i=1}^{n} {K}({H}_{v}^{-1}({X}_i - {x})) e_i^4 \overset{p}{\to} f({x}) E((\sigma({X}_i) \epsilon_i)^4 \mid {X}_i = {x}).
\end{equation*}
Combining \eqref{FourthMomentDenom} and applying Slutsky's theorem, it is proved that the proposed estimator \eqref{NW1} is a consistent estimator of $E \left\{\sigma^4({X}_i) \epsilon_i^4 \mid {X}_i = {x} \right\}$.
\end{proof}

Before proving \eqref{FourthMoment}, we make the following regularity assumptions:
\begin{enumerate}
    \item[(A1)] The univariate kernel function $k(u)$ is compactly supported on $[-1, 1]$. This means the $d$-variate kernel function ${K}({u})$ is compactly supported on the $d$-dimensional cubic $[-1,1]^d$, which is contained in the $d$-dimensional ball of radius $\sqrt{d}$.
    \item[(A2)] The univariate kernel function $k(u)$ is bounded above by $k_{m}$.
    \item[(A3)] $f({x})$, the density function of ${X}$ satisfies that,  for all ${x}$, there exists some compact set $I_{{x}}$ of which ${x}$ in the interior, and some finite constant $M_1 > 0$ such that,
    \begin{equation}
        \sup_{{u} \in I_{{x}}} \lVert (f({u})^{-1}, - f({u})^{-2} \dot{f}(u)^\top ) \rVert \leq M_1.
    \end{equation}
    \item[(A4)] The conditional expectation $m({x})$ is Lipschitz in ${x}$, i.e., there exists $L > 0$ such that, for all ${x}$ and ${y}$ in the support of ${X}$, 
    $$\lvert m({x}) - m({y}) \rvert \leq L \lVert {x} - {y} \rVert.$$
    \item[(A5)] For all ${x}$, there exists some compact set $I_{{x}}$ of which ${x}$ in the interior, and some finite constant $M_2 > 0$ such that, 
    \begin{equation}
        \sup_{{u} \in I_{{x}}} \lvert E[(\sigma({X}_i)) \epsilon_i)^k \mid {X}_i = {u}] \rvert \leq M_2,
    \end{equation}
    for $k = 2,3,$ and $4$.
    \item[(A6)] $ Pr(\lVert {X}_{1} - {x} \rVert \leq \sqrt{d} (h_2+h_v)) = (h_v h_2^{4-4/d})^\eta$ for some $\eta > d/5$.
    When $h_2=h_v=h$, this simplifies to $ Pr(\lVert {X}_{1} - {x} \rVert \leq 2 \sqrt{d} h) = h^{\eta'}$ for some $\eta' > d-4/5$.
\end{enumerate}

\begin{proof}[Proof of \eqref{FourthMoment}]
    First, write
    \begin{equation}\label{eq_weights}
        \widehat{m}({x}) = {e}_1^\top \left[ {\Gamma}({x})^\top {\Omega}({x}) {\Gamma}({x})  \right]^{-1} {\Gamma}({x})^\top {\Omega}({x}) {Y} =: \sum_{i=1}^n w_i({x}) Y_i.
    \end{equation}
    Since ${e}_1 = (1, 0, \dots, 0)^\top \in \mathbb{R}^{d+1}$, 
    the vector $(w_1({x}), \dots, w_n({x}))$ is the product of the first row vector of 
    $\left[ {\Gamma}({x})^\top {\Omega}({x}) {\Gamma}({x})  \right]^{-1}$ and the matrix
    ${\Gamma}({x})^\top {\Omega}({x})$.
    Asymptotically,
    \begin{equation*}
        \begin{split}
            &\left[ \frac{1}{n h_2^d} {\Gamma}({x})^\top {\Omega}({x}) {\Gamma}({x}) \right]^{-1} \\ 
            =& \left(
            \begin{array}{cc}
            f({x})^{-1} + o_p(1) & - f({x})^{-2}\dot{f}(u)^\top + o_p(1) {1}_d^\top  \\
            - f({x})^{-2} \dot{f}(u) + o_p(1) {1}_d & (h_2^2 \sigma_k^2 f({x}) )^{-1} {I}_{d} + o_p(h_2^{-2}) {I}_{d}
            \end{array} \right).
        \end{split}
    \end{equation*}
    Denoting ${\gamma}_i({x})^\top = (1, ({X}_i - {x})^\top)$ for $i=1,\dots,n$,
    then we can express
    \begin{equation*}
        \frac{1}{n}{\Gamma}({x})^\top {\Omega}({x}) = \frac{1}{nh_2^d}({K}({H}_{2}^{-1}({X}_1 - {x})) {\gamma}_1({x}), \dots, {K}({H}_{2}^{-1}({X}_n - {x})) {\gamma}_n({x})).
    \end{equation*}
    Thus, for $i = 1, \dots, n$,
    \begin{equation*}
        w_i({x}) = \frac{1}{nh_2^d} {K}({H}_{2}^{-1}({X}_i - {x})) \left\{ \left(f({x})^{-1}, - f({x})^{-2} \dot{f}(u)^\top \right) {\gamma}_i({x}) + o_p({1}^\top {\gamma}_i({x}) )\right\}
    \end{equation*}
    Under assumptions (A1) - (A3),
    \begin{equation*}
        \lvert w_i({x}) \rvert \leq \frac{K_m^d}{n h_2^d} 1\{\lVert {X}_i - {x} \rVert \leq \sqrt{d} h_2 \} 
        \sqrt{dh_2^2 + 1} (M_1 + o_p(1)).
    \end{equation*}

    Further, by setting $Y_i = 1, i=1,\dots,n$, 
    $\sum_{i=1}^n w_i({x})$ becomes the first regression coefficient from the local polynomial regression of $(1, \dots, 1)$ on ${X}_i$'s. Hence, $\sum_{i=1}^n w_i({x}) = 1$. 

    Now we prove \eqref{FourthMoment}.
    \begin{equation*}
    \begin{split}
        &E\left[\frac{1}{nh_v^d}\sum_{i=1}^{n} {K}({H}_{v}^{-1}({X}_i - {x})) (m({X}_i) - \hat{m}({X}_i))^4 \right]\\
        =& E\left[\frac{1}{h_v^d} {K}({H}_{v}^{-1}({X}_1 - {x})) (m({X}_1) - \hat{m}({X}_1))^4 \right]\\
        =& E\left[\frac{1}{h_v^d} {K}({H}_{v}^{-1}({X}_1 - {x})) E[(m({X}_1) - \hat{m}({X}_1))^4 \mid {X}_1,\dots,{X}_n] \right],
    \end{split}
    \end{equation*}
    in which
    \begin{equation*}
    \begin{split}
        &E[(m({X}_1) - \hat{m}({X}_1))^4 \mid {X}_1,\dots,{X}_n] \\
        =& E[(\sum_{i=1}^n w_i({x}) (Y_i - m({X}_1)) )^4 \mid {X}_1,\dots,{X}_n] \\
        =& \sum_{j_1=1}^n\sum_{j_2=1}^n\sum_{j_3=1}^n\sum_{j_4=1}^n E\left[\prod_{l=1}^4 w_{j_l}({x}) (Y_{j_l} - m({X}_1)) \mid {X}_1,\dots,{X}_n \right].
    \end{split}
    \end{equation*}
    Furthermore, under assumptions (A4) and (A5), we could show that,  as $n \rightarrow \infty$,
    \begin{equation*}
        E\left[\prod_{l=1}^4 w_{j_l}({x}) (Y_{j_l} - m({X}_1)) \mid {X}_1,\dots,{X}_n \right] \lesssim \prod_{l=1}^4 \lvert w_{j_l}({x}) \rvert.
    \end{equation*}
    Thus, as $n \rightarrow \infty$,
    \begin{equation*}
    \begin{split}
        &E[(m({X}_1) - \hat{m}({X}_1))^4 \mid {X}_1,\dots,{X}_n] \\
        \lesssim& \sum_{j_1=1}^n\sum_{j_2=1}^n\sum_{j_3=1}^n\sum_{j_4=1}^n \prod_{l=1}^4 \lvert w_{j_l}({x}) \rvert \\
        \lesssim& \left( \frac{1}{nh_2^{d-1}} \right)^4 \sum_{j_1=1}^n\sum_{j_2=1}^n\sum_{j_3=1}^n\sum_{j_4=1}^n \prod_{l=1}^4 1\{\lVert {X}_{j_l} - {X}_1 \rVert \leq \sqrt{d} h_2 \},
    \end{split}
    \end{equation*}
    and
    \begin{equation*}
    \begin{split}
        &\frac{1}{h_v^d} {K}({H}_{v}^{-1}({X}_1 - {x})) E[(m({X}_1) - \hat{m}({X}_1))^4 \mid {X}_1,\dots,{X}_n] \\
        \lesssim& \frac{1}{h_v^d} \left( \frac{1}{nh_2^{d-1}} \right)^4  1\{\lVert {X}_{1} - {x} \rVert \leq \sqrt{d} h_v \} \sum_{j_1=1}^n\sum_{j_2=1}^n\sum_{j_3=1}^n\sum_{j_4=1}^n \prod_{l=1}^4 1\{\lVert {X}_{j_l} - {X}_1 \rVert \leq \sqrt{d} h_2 \} \\
        =& \frac{1}{h_v^d} \left( \frac{1}{nh_2^{d-1}} \right)^4 \sum_{j_1=1}^n\sum_{j_2=1}^n\sum_{j_3=1}^n\sum_{j_4=1}^n \prod_{l=1}^4 1\{\lVert {X}_{j_l} - {X}_1 \rVert \leq \sqrt{d} h_2 \} 1\{\lVert {X}_{1} - {x} \rVert \leq \sqrt{d} h_v \} \\
        \leq& \frac{1}{h_v^d} \left( \frac{1}{nh_2^{d-1}} \right)^4 \sum_{j_1=1}^n\sum_{j_2=1}^n\sum_{j_3=1}^n\sum_{j_4=1}^n \prod_{l=1}^4 1\{\lVert {X}_{j_l} - {x} \rVert \leq \sqrt{d} (h_2 + h_v) \} 1\{\lVert {X}_{1} - {x} \rVert \leq \sqrt{d} h_2 \} \\
        =& \frac{1}{h_v^d} \left( \frac{1}{h_2^{d-1}} \right)^4 1\{\lVert {X}_{1} - {x} \rVert \leq \sqrt{d} h_2 \} \left(\frac{1}{n} \sum_{i=1}^n 1\{\lVert {X}_{i} - {x} \rVert \leq \sqrt{d} (h_2+h_v) \} \right)^4 \\
        \leq& \frac{1}{h_v^d} \left( \frac{1}{h_2^{d-1}} \right)^4 1\{\lVert {X}_{1} - {x} \rVert \leq \sqrt{d} (h_2+h_v) \} \left(\frac{1}{n} \sum_{i=1}^n 1\{\lVert {X}_{i} - {x} \rVert \leq \sqrt{d} (h_2+h_v) \} \right)^4,
    \end{split}
    \end{equation*}
    in which
    \begin{equation*}
    \begin{split}
        &E\left[ 1\{\lVert {X}_{1} - {x} \rVert \leq \sqrt{d} (h_2+h_v) \} \left(\frac{1}{n} \sum_{i=1}^n 1\{\lVert {X}_{i} - {x} \rVert \leq \sqrt{d} (h_2+h_v) \} \right)^4 \right] \\
        =& \left( Pr(\lVert {X}_{1} - {x} \rVert \leq \sqrt{d} (h_2+h_v)) \right)^5 + o(1),
    \end{split}
    \end{equation*}
    since the number of terms with repeated indexes are of a smaller order than $n^4$.
    % in which $\frac{1}{n} \sum_{i=1}^n 1\{\lVert {X}_{i} - {x} \rVert \leq \sqrt{d} (h_2+h_v) \} \overset{a.s.}{\to} Pr(\lVert {X}_{1} - {x} \rVert \leq \sqrt{d} (h_2+h_v))$, as $n \to \infty$.
    % {\color{red} Is this right since the $h$ inside also varies as $n$ goes to infinity?}
    Then we take the further expectation of the above and get
    \begin{equation*}
    \begin{split}
        &E\left[\frac{1}{nh_v^d}\sum_{i=1}^{n} {K}({H}_{v}^{-1}({X}_i - {x})) (m({X}_i) - \hat{m}({X}_i))^4 \right]\\
        \lesssim& \frac{1}{h_v^d} \left( \frac{1}{h_2^{d-1}} \right)^4 \left( Pr(\lVert {X}_{1} - {x} \rVert \leq \sqrt{d} (h_2+h_v)) \right)^4 Pr\left( \lVert {X}_{1} - {x} \rVert \leq \sqrt{d} h_2 \right) \\
        \leq& \frac{1}{h_v^d} \left( \frac{1}{h_2^{d-1}} \right)^4 \left( Pr(\lVert {X}_{1} - {x} \rVert \leq \sqrt{d} (h_2+h_v)) \right)^5.
    \end{split}
    \end{equation*}   
    Finally, assuming (A6), the above converges to $0$ as $n \to \infty$ and \eqref{FourthMoment} is proved.
\end{proof}

\begin{proof}[Proof of the consistency of \eqref{NW2}]
Using the same techniques as in the proof of \eqref{NW1}, it can be shown that the proposed estimator \eqref{NW2} is also consistent under corresponding regularity conditions.
Specifically, assumptions corresponding to (A3) - (A6) need to be made being conditional on $Z_i$.
\end{proof}

\section{Proof of parametric rate of convergence for ATT estimation}
\label{sec:parametric.rate}

We noted in Section \ref{sec_ate} that when the sign of the bias does interact with $X$, we can provide a two-point identification of ATT. Consequently, we can estimate $ATT_\pm$ by $n^{-1}\sum_i \widehat{\tau_\pm}(X_i)$. In this section, we show that this estimator is $\sqrt{n}$ consistent. 

Let $q(x) = m_1(x)-m_0(x) + [ \{m_1(x)-m_0(x)\}^2 - \{\sigma^2(x) -\sigma_0^2(x)\}/\{\pi(x)(1-\pi(x))\} ]^{1/2}$.
%and $g(a,b,c,d) = a - b + [ (a-b)^2 - c/\{d(1-d)\}]^{1/2}$. So, $q(x) = g(m_1(x), m_0(x), \sigma^2(x)-\sigma_0^2(x), \pi(x))$.
Assume that $\pi$ is estimated by a local linear regression of $Z$ on $x$ with kernel ${K}$ and bandwidth $h_4$; ${H}_4=h_4{I}_d$. Let $\delta(x) = ( \widehat{m_1}(x)-m_1(x), \widehat{m_0}(x)-m_0(x), \widehat{\sigma^2}(x)-\widehat{\sigma^2_0}(x) - {\sigma^2}(x) + {\sigma^2_0}(x), \widehat{\pi}(x)-\pi(x))^\top$. $\widehat{q}(x)$ is the plug-in estimator of $q(x)$. We are interested in the asymptotic distribution of 
$$\sqrt{n} \sum_i \{\widehat{\tau_+}(X_i) - \tau_+(X_i)\} = \sqrt{n}\frac1n \sum_i \{\widehat{q}(X_i) - q(X_i)\}.$$

We define the residuals in estimating our functions below. Let $e_l$ denote a $d$ unit vector with $1$ in coordinate $l$ and $0$ elsewhere.

Based on the linear approximation of the kernel estimators, write 
$$R_1(x) \equiv e_1^\top\delta(x) - \frac{1}{n h_3^d \pi(x) f_1({x})}  
    \left(  \sum_{i=1}^n Z_i {K}\left({H}_3^{-1}({X}_i-{x})\right) \nu_1({X}_i) \xi_{1i} \right), $$
$$R_2(x) \equiv e_2^\top\delta(x) - \frac{1}{n h_4^d (1-)f_0({x})}  
    \left(  \sum_{i=1}^n (1-Z_i) {K}\left({H}_4^{-1}({X}_i-{x})\right) \nu_0({X}_i) \xi_{0i} \right), $$
$$R_3(x) \equiv e_3^\top\delta(x) - \frac{1}{n h_2^d f(x)} \sum_{i=1}^n  {K}\left({H}_2^{-1}({X}_i-{x})\right) \sigma^2({X}_i) (\epsilon_i^2 - 1).$$
Similarly, 
$$R_4(x)  \equiv e_4^\top\delta(x) - \frac{1}{n h_5^d f({x})}  
    \left(  \sum_{i=1}^n {K}\left({H}_5^{-1}({X}_i-{x})\right) (Z_i-\pi({X}_i)) \right).$$

% \noindent {\it Assumption: } $\sum_i R_l(X_i)^2 = o_p(1)$. 
% $E \| f'(X) \|^4 < \infty$ for $l=1,2,3,4$ and $f''()$ is bounded. $\Pr(|| X_i-X_j||\leq h)=o(\sqrt{n}h^{2d})$, $E_j\{\Pr_i(|| X_i-X_j||\leq h)\}^2 = O(h^{2d})$ and $\sqrt{n}h^{2d}\rightarrow \infty$. $f()$, $\pi()$ and $1-\pi()$ are bounded away from 0. $\max\{E(\sigma^4(X_i)(\epsilon_i-1)^4), E(\nu_z^4(X_i)\xi_{zi}^4), \} < \infty$, .  $m_z()$, $\pi()$ are Lipschitz. $K()$ is compactly supported on $[-1/\sqrt{d},1/\sqrt{d}]^d$ and bounded.

Let $\zeta_{1i} = Z_iY_i - m_1(X_i)$ and $\zeta_{0i} = (1-Z_i)Y_i - m_0(X_i)$. Let $E_i(\cdot)$ denote expectation with respect to unit $i$'s data and $\Pr_i(A) = E_i1(A)$.

\begin{assumption}\label{assump_ateclt} (assumptions for asymptotic normality for $\widehat{ATT_\pm}$)
\begin{enumerate}
    \item[(a)] $\sum_i R_l(X_i)^2 = o_p(1)$ for $l=1,\ldots,4$.
    \item[(b)] $E \| q'(X) \|^4 < \infty$ and the largest eigenvalue of $q''(x)$ is bounded.
    \item[(c)] $h_l=h$ for all $l=1,\ldots,5$ and $nh^{2d}\rightarrow\infty$
    \item[(d)] $\max\{E(\sigma^4(X_i)(\epsilon_i-1)^4), E(\nu_1^4(X_i)\xi_{1i}^4), E(\nu_0^4(X_i)\xi_{0i}^4), E(\zeta_{1i}^4),  E(\zeta_{0i}^4)\} < \infty$
    \item[(e)] $f(\cdot)$, $\pi(\cdot)$ and $1-\pi(\cdot)$ are bounded away from 0 and $\sigma(\cdot)$ is bounded. $k(\cdot)$ is compactly supported on $[-1/\sqrt{d},1/\sqrt{d}]^d$ and bounded.
    \item[(f)] $\Pr(\|X_i-X_j\| \leq h)=o(nh^{4d})$ and \\
    $E_i \{{\Pr}_j(\|X_i-X_j\| \leq 2h)\}^2=o(\min\{\sqrt{n}h^{4-2d}), h^{4d-6}/\sqrt{n}, n^{7/2}h^{6d}\})$.
    \item[(g)] $m_1(\cdot), m_0(\cdot), \sigma(\cdot)$ and $\pi(\cdot)$ are H\"{o}lder continuous of degree 2.
\end{enumerate}
\end{assumption}

%\section{Asymptotic normality of $\widehat{ATT_\pm}$}
\begin{theorem}\label{thm_ateclt}
Under Assumption \ref{assump_ateclt}, with iid data,  $\sqrt{n}(n^{-1}\sum_i\widehat{\tau_+}(X_i)-{ATT_+})$ goes to a normal distribution with mean 0.
\end{theorem}
% \noindent{\bf Proof (of Theorem \ref{thm_ateclt}).}
\begin{proof}[Proof of Theorem \ref{thm_ateclt}.]

Using Taylor series expansion
\begin{align*}
    \sqrt{n}\frac1n \sum_i \{\widehat{q}(X_i) - q(X_i)\} 
    = \sqrt{n}\frac1n \sum_i q'(X_i)^\top \delta(X_i) +  
    \sqrt{n}\frac1n \sum_i \delta(X_i)^\top q''(X_i^\star)\delta(X_i)
\end{align*}

Consider first $\sqrt{n}\frac1n \sum_i q'(X_i)^\top \delta(X_i) = \sqrt{n}\frac1n \sum_i \sum_{l=1}^4 e_l^\top q'(X_i)e_l^\top\delta(X_i)$. We will show that this term is $O_p(1)$. We show that $\sqrt{n}\frac1n \sum_i e_1^\top q'(X_i)e_1^\top \delta(X_i) = O_p(1)$ and the same argument will work for the other three terms.
\begin{align*}
&\sqrt{n}\frac1n \sum_i e_1^\top q'(X_i)e_1^\top \delta(X_i) \\
=& \sqrt{n}\frac1n \sum_{i=1}^n \frac{e_1^\top q'(X_i)}{n h_3^d \pi(x) f_1({X}_i)}  
    \left(  \sum_{j=1}^n Z_j {K}\left({H}_3^{-1}({X}_j-{X}_i)\right) \nu_1({X}_j) \xi_{1j} \right) + \sqrt{n}\frac1n \sum_i e_1^\top f'(X_i) R_1(X_i).
\end{align*}
By Cauchy-Schwarz inequality, $$|\sqrt{n}\frac1n \sum_i e_1^\top q'(X_i) R_1(X_i)| 
\leq \{ \frac1n \sum_i (e_1^\top q'(X_i))^2 \}^{1/2} \{\sum_i  R_1(X_i)^2 \}^{1/2}.$$ Hence the second term in the above expression is $o_p(1)$ since $\sum_i  R_1(X_i)^2=o_p(1)$ and $E\|q'(X_i)\|^4<\infty$.

Throughout the rest of the paper, we use $C$ and $L$ as generic constants which may vary from line to line.

Next, let $W_i=(X_i,Z_i,Y_i)$. Rewrite the fist term as 
$\sqrt{n}\frac1{n^2} \sum_{i=1}^n \sum_{j=1}^n u_1(W_i,W_j),$ where
$$u_1(W_i, W_j) = \frac{e_1^\top q'(X_i)}{h_3^d \pi(x) f_1({X}_i)} Z_j {K}\left({H}_3^{-1}({X}_j-{X}_i)\right) \nu_1({X}_j) \xi_{1j}.$$

Using the fact that $E\{u_1(W_i,W_j)\mid W_i\} = 0$, by $V$-statistic projection we get 
\begin{align*}\sqrt{n}\frac1{n^2} \sum_{i=1}^n \sum_{j=1}^n [u_1(W_i,W_j) -& E\{u_1(W_i,W)\mid W\}]\\ &= O_p(\sqrt{n} [ \{E u_1(W_i,W)^2\}^{1/2} + \{E u_1(W_i,W_i)^2\}^{1/2} ]/n),
\end{align*}
where $W$ is an independent copy of $W_i$. 

We now show that $E u(W_i,W)^2=o(n)$ and $E u(W_i,W_i)^2=o(n)$. Using the facts that, $f$, $\pi$ and $1-\pi$ are bounded away from 0 and $k$ is bounded, we get that, for $j\neq i$
\begin{align*}
E u_1(W_i,W_j)^2 & \leq C E\{ h_3^{-d} 1(\|X_i-X_j\|\leq h_3) e_1^\top q'(X_i) \nu_1({X}_j) \xi_{1j}\}^2\\
& \leq C \{ E h_3^{-4d} 1(\|X_i-X_j\|\leq h_3)\}^{1/2}\{ E\{e_1^\top q'(X_i) \nu_1({X}_j) \xi_{1j}\}^4\}^{1/2}\\
& = C \{ E h_3^{-4d} 1(\|X_i-X_j\|\leq h_3)\}^{1/2} \|e_1^\top q'(X_i)\|_4^{2} \|\nu_1({X}_j) \xi_{1j}\|_4^{2}.
\end{align*}
Since, $\|q'(X_i)\|_4<\infty$ and $\|\nu_1({X}_j) \xi_{1j}\|_4<\infty$, we have $E u_1(W_i,W)^2=E u_1(W_i,W_j)^2 = o(n)$ by the fact that $nh_3^{2d}\rightarrow \infty$.
%$\Pr(\|X_i-X_j\|<h_3) = o(nh^{2d})$.

Next, 
\begin{align*}
E u_1(W_i,W_i)^2 & \leq C E\{ h_3^{-d} e_1^\top q'(X_i) \nu_1({X}_i) \xi_{1j}\}^2\\
&  \leq C h_3^{-2d} \|e_1^\top q'(X_i)\|_4^{2} \|\nu_1({X}_i) \xi_{1i}\|_4^{2}.
\end{align*}
Since, $\|q'(X_i)\|_4<\infty$ and $\|\nu_1({X}_i) \xi_{1i}\|_4<\infty$, we have $E u_1(W_i,W)^2=E u_1(W_i,W_j)^2 = o(n)$ by the fact that 
$nh_3^{2d}\rightarrow\infty$.

By parallel calculations for the other three terms, 
\begin{align*}
%&\sqrt{n}\frac1n \sum_i q'(X_i)^\top \delta(X_i)\\
& = \sqrt{n}\frac1n \sum_i \sum_{l=1}^4 e_l^\top q'(X_i)e_l^\top\delta(X_i)\\
&= \sqrt{n}\frac1n \sum_{i=1}^nE\{u_1(W_i,W) + u_2(W_i,W)+u_3(W_i,W)+u_4(W_i,W)\mid W\}+o_p(1),
\end{align*}
Which goes to a normal distribution by the CLT for average of iid random variables with bounded fourth moment and Slutsky's theorem.

The remaining term is 
$\sqrt{n}\frac1n \sum_i \delta(X_i)^\top q''(X_i^\star)\delta(X_i)$. Since 
$q''$ is uniformly bounded, we will only need to show that 
$\sum_i \delta(X_i)^\top\delta(X_i) = o_p(\sqrt{n})$, or $\sum_i \{e_l^\top\delta(X_i)\}^2 = o_p(\sqrt{n})$ for $l=1,2,3,4$. By Markov inequality, it suffices to show
$\sqrt{n}E\{e_l^\top\delta(X_i)\}^2\rightarrow 0$ for each $l=1,...,4$.

The calculations for $l=1,2$ and $4$ are similar. Thus, we only show the calculations for $\sqrt{n}E\{e_1^\top\delta(X_i)\}^2 = \sqrt{n}E\{\widehat{m_1}(X_i)-m(X_i)\}^2$ and $\sqrt{n}E\{e_3^\top\delta(X_i)\}^2=\sqrt{n}E\{\widehat{\sigma^2}(X_i)-\sigma^2(X_i)\}^2$.

Recall the definition of $w_i(x)$ from \eqref{eq_weights} in the proof of Remark \ref{rem_se} and the facts that $\sum_j w_i(x) = 1$ and $\sum_j w_j(x)(X_i-x)=0$. We use the fact that $|w_j(x)| \leq C (nh^d)^{-1}1(\|X_j-x\|\leq h)$.

Recall $E(Z_iY_i\mid X_i, Z_i=1) = m_1(X_i)$ and $\zeta_{1i} = Z_iY_i - m_1(X_i)$. Since $\widehat{m}_1(x) = \sum_j w_i(x) Z_iY_i$,  
using the H\"{o}lder continuity property of $m_1(\cdot)$, for some $d\times d$  matrices $L_{ij}^{m_1}$ which have uniformly bounded maximum eigenvalues
\begin{align*}
&E\{\widehat{m_1}(X_i)-m_1(X_i)\}^2  \\
= & E[ \sum_j w_j(X_i)\{Z_jY_j -m_1(X_i)\}]^2\\
= & E[ \sum_j w_j(X_i)\{m_1(X_j) + \zeta_{1j}-m_1(X_i)\}]^2\\
= & E[ \sum_j w_j(X_i) \{ (X_i-X_j)^\top L_{ij}^{m_1}(X_i-X_j) + \zeta_{1j}\} ]^2\\
\leq &  E\{ \sum_j |w_j(X_i)| L\|X_i-X_j\|^2 \}^2 + \sum_j E w_j(X_i)^2\zeta_{1j}^2\\
\leq & L^2h_3^4 E\{ \sum_j |w_j(X_i)| \}^2 + \sum_j E w_j(X_i)^2\zeta_{1j}^2\\
\leq & CL^2h_3^4 \frac{1}{n^2h_3^{2d}}E\{ \sum_j 1( \|X_i-X_j\| \leq h_3 ) \}^2 
+ \frac{C}{n^2h_3^{2d}} \sum_j E 1( \|X_i-X_j\| \leq h_3 )\zeta_{1j}^2 \\
\leq & CL^2h_3^{4-2d} \left[ \frac{n-1}{n}E_i \{{\Pr}_j(\|X_i-X_j\| \leq h_3)\}^2 + \frac{1}{n}\Pr(\|X_i-X_j\| \leq h_3) \right] \\
+ & \frac{C}{nh_3^{2d}} \{E \zeta_{1j}^4\}^{1/2}\{ \Pr(\|X_i-X_j\| \leq h_3) \}^{1/2}.
\end{align*}
Since $E \zeta_{1j}^4<\infty$, $\Pr(\|X_i-X_j\| \leq h_3)=o(nh_3^{4d})$ and
$E_i \{{\Pr}_j(\|X_i-X_j\| \leq h_3)\}^2=o(\sqrt{n}h_3^{4-2d})$, we get
$\sqrt{n}E\{\widehat{m_1}(X_i)-m_1(X_i)\}^2\rightarrow 0$.

Finally, $\sqrt{n}E\{e_1^\top\delta(X_i)\}^2= \sqrt{n}E\{\widehat{\sigma^2}(X_i)-\sigma^2(X_i)\}^2
= \sqrt{n}E[\sum_j w_j(X_i)\{e_j^2 - \sigma^2(X_i)\} ]^2$. Recall that $h_l=h$ for all $l=1,\ldots, 5$.
 Now, by H\"{o}lder continuity of $m$, and properties of $w_i(\cdot)$, for some $d\times d$ matrices $L_{ij}^{m}$  which have uniformly bounded maximum eigenvalues
\begin{align*}
    &E\{e_j^2\mid X_1,...,X_n\} \\
    =& E\left[ \sum_k w_k(X_j)((X_i-X_j)^\top L_{ij}^{m}(X_i-X_j) + \{\sigma(X_k)\epsilon_k-\sigma(X_j)\epsilon_j\} )\mid X_1,...,X_n\right]^2\\
    =&  \sigma^2(X_j) +  \sum_k w_k(X_j)^2\sigma^2(X_k) + 
    \{\sum_k w_k(X_j)(X_i-X_j)^\top L_{ij}(X_i-X_j) \}^2 \\
%    &  + 2E\left[ \sum_k w_k(X_j)(L\|X_k-X_j\|^2\{\sigma(X_k)\epsilon_k-\sigma(X_j)\epsilon_j\}\mid X_1,...,X_n\right]\\
   % & = \sigma^2(X_j) +  \sum_k w_k(X_j)^2\sigma^2(X_k) + 
%    \{\sum_k w_k(X_j)L\|X_k-X_j\|^2 \}^2\\
    =&  \sigma^2(X_j) +  \psi_j \quad (say).
\end{align*}
Let $\phi_j= e_i^2 - E\{e_j^2\mid X_1,...,X_n\}$. Then, for some $d\times d$ matrices $L_{ij}^{\sigma^2}$  which have uniformly bounded maximum eigenvalues
\begin{align*}
& E[\sum_j w_j(X_i)\{e_j^2 - \sigma^2(X_i)\} ]^2\\
= & E[\sum_j w_j(X_i)\{\psi_j + \phi_j + (X_i-X_j)^\top L_{ij}^{\sigma^2}(X_i-X_j)\} ]^2\\
= & E[\sum_j w_j(X_i)\psi_j]^2 + E[\sum_j w_j(X_i)\phi_j]^2 + E\left[\sum_j w_j(X_i)\{(X_i-X_j)^\top L_{ij}^{\sigma^2}(X_i-X_j)\} \right]^2 \\
+& 2 E\left[\sum_j\sum_k w_j(X_i)(X_i-X_j)^\top L_{ij}^{\sigma^2}(X_i-X_j)w_k(X_i)\psi_k\right]\\
\leq & E[\sum_j w_j(X_i)\psi_j]^2 + E[\sum_j w_j(X_i)\phi_j]^2 + E\left[\sum_j |w_j(X_i)|\{L\|X_j-X_i\|^2\} \right]^2 \\
+& 2 E\left[\sum_j\sum_k |w_j(X_i)|L\|X_j-X_i\|^2|w_k(X_i)||\psi_k|\right].
\end{align*}
From here, 

1) As shown in our previous calculations, $\sqrt{n}E[\sum_j w_j(X_i)\|X_j-X_i\|^2 ]^2\rightarrow 0$ because 
$E_i \{{\Pr}_j(\|X_i-X_j\| \leq h)\}^2=o(\sqrt{n}h^{4-2d})$.

2) $\sqrt{n}E[\sum_j w_j(X_i)\phi_j]^2 \leq C(\sqrt{n}h_2^{2d})^{-1}\{E \phi_j^4\}^{1/2}\{ \Pr(\|X_i-X_j\| \leq h) \}^{1/2}$. Which goes to 0 since $E \phi_j^4<\infty$ and $\Pr(\|X_i-X_j\| \leq h)=o(nh^{4d})$.

For the following, we use the boundedness of $\sigma^2(\cdot)$.

3) By our bound on $|w_j(x)|$,
\begin{align*}
    & E\left[\sum_j\sum_k |w_j(X_i)|L\|X_j-X_i\|^2|w_k(X_i)||\psi_k|\right]\\
    \le & q CLh_2^2\frac1{n^2h^{2d}}E\{\sum_j\sum_k 1(\|X_i-X_j\|\leq h)1(\|X_k-X_j\|\leq h)|\psi_k|\}\\
    = & CLh_2^2\frac1{h^{2d}}E\{1(\|X_i-X_j\|\leq h)1(\|X_k-X_j\|\leq h)|\psi_k|\}\\
    \leq & CLh_2^2\frac1{nh^{4d}}E\{1(\|X_i-X_j\|\leq h)1(\|X_k-X_j\|\leq h)1(\|X_l-X_j\|\leq h)\\
    + & CL^2h_2^2\frac{h^4}{n^2h^{4d}}E\{1(\|X_i-X_j\|\leq h)1(\|X_k-X_j\|\leq h)\{\sum_l 1(\|X_l-X_j\|\leq h)\}^2\\
    \leq & CLh_2^2\frac{1+h^4+nh^4}{nh^{4d}} E_i\{{\Pr}_j(\|X_i-X_j\|\leq 2h)\}^2.
\end{align*}
Thus, $\sqrt{n}E\left[\sum_j\sum_k |w_j(X_i)|L\|X_j-X_i\|^2|w_k(X_i)||\psi_k|\right]\rightarrow 0$ since
$E_i\{{\Pr}_j(\|X_i-X_j\|\leq 2h)\}^2 = o(h^{4d-6}/\sqrt{n})$.

4) $E[\sum_j |w_j(X_i)||\psi_j|]^2
\leq C h^{-2d} E\{ 1(\|X_i-X_j\| \leq h)1(\|X_i-X_k\| \leq h)|\psi_j||\psi_k|\}$.

Similar to the calculations in (3), $\sqrt{n}E[\sum_j |w_j(X_i)||\psi_j|]^2\rightarrow 0$ since
$E_i\{{\Pr}_j(\|X_i-X_j\|\leq 2h)\}^2 = o(n^{7/2}h^{6d})$.

Thus, we have established the desired asymptotic normality of $\sqrt{n}(n^{-1}\sum_i\widehat{\tau_+}(X_i)-{ATT_+})$.
% \hfill$\square$
\end{proof}
\end{document}